\documentclass[pra,floatfix,twocolumn,amsmath,amssymb]{revtex4-2}

\usepackage{graphicx}
\graphicspath{ {./images/} }
\usepackage{color}
\usepackage{amsmath,amssymb,mathrsfs,ulem}
\usepackage{braket}
\usepackage{bbm}
\usepackage{lipsum}
\usepackage{float}
\usepackage{tikz}
\usepackage{subfiles}

\usepackage[colorlinks = true,linkcolor = blue,urlcolor  = blue,     citecolor = blue,anchorcolor = blue]{hyperref}

\definecolor{lime}{HTML}{A6CE39}
\DeclareRobustCommand{\orcidicon}{
	\begin{tikzpicture}
	\draw[lime, fill=lime] (0,0) 
	circle [radius=0.16] 
	node[white] {{\fontfamily{qag}\selectfont \tiny ID}};
	\draw[white, fill=white] (-0.0625,0.095) 
	circle [radius=0.007];
	\end{tikzpicture}
	\hspace{-3mm}
}

\date{\today}

\begin{document}

\title{Diagrammatic Approach to Scattering of Multi-Photon States in Waveguide QED}

\author{Kiryl Piasotski\href{https://orcid.org/0000-0002-1265-4277}{\orcidicon}}
\email[Email: ]{kiryl.piastoski@rwth-aachen.de}
\affiliation{Institut f\"ur Theorie der Statistischen Physik, RWTH Aachen, 
52056 Aachen, Germany \\ and JARA - Fundamentals of Future Information Technology, 52056 Aachen, Germany}

\author{Mikhail Pletyukhov\href{https://orcid.org/0000-0002-7712-6520}{\orcidicon}}
\affiliation{Institut f\"ur Theorie der Statistischen Physik, RWTH Aachen, 
52056 Aachen, Germany \\ and JARA - Fundamentals of Future Information Technology, 52056 Aachen, Germany}

\begin{abstract}
We give an exposure to diagrammatic techniques in waveguide QED systems. A particular emphasis is placed on the systems with delayed coherent quantum feedback. Specifically, we show that the $N$-photon scattering matrices in single-qubit waveguide QED systems, within the rotating wave approximation, admit for a parametrization in terms of $N-1$-photon effective vertex functions and provide a detailed derivation of a closed hierarchy of generalized Bethe-Salpeter equations satisfied by these vertex functions. The advantage of this method is that the above mentioned integral equations hold independently of the number of radiation channels, their bandwidth, the dispersion of the modes they are supporting, and the structure of the radiation-qubit coupling interaction, thus enabling one to study multi-photon scattering problems beyond the Born-Markov approximation. Further, we generalize the diagrammatic techniques to the systems containing more than a single emitter by presenting an exact set of equations governing the generic two and three-photon scattering operators. The above described theoretical machinery is then showcased on the example of a three-photon scattering on a giant acoustic atom, recently studied experimentally [\href{https://www.nature.com/articles/s41567-019-0605-6}{Nat. Phys. {\bf 15}, 1123 (2019)}].

\end{abstract}

\maketitle

\section{Introduction} 
\label{sec: Intro}
Waveguide quantum electrodynamics (QED) is a modern field of research focused on the study of light-matter interactions in one dimension. The confinement of electromagnetic radiation to a single spatial dimension allows one to achieve a significant enhancement of the coupling between atoms and fields, as well as to attain a better matching between the spatial modes of the emitting and absorbing atoms \cite{Lalumiere_PRA_2013, Arcari_2014}. Apart from the purely academic interest in the study of strong light-matter coupling, a great deal of motivation in this field of research comes from the technological sector, namely from the quantum computing \cite{Angelakis_2007, Kockum_PRL_2018, Paulisch_2016, Vermersch_2017, Xu_2018, Zheng_2013}. One of the most prominent examples is the so-called quantum network: A system of quantum processors interconnected by quantum radiation channels propagating quantum information and entanglement between them \cite{Kimble_2008}. 

Although photons are the excellent carriers of quantum information, capable of high fidelity entanglement and information transfer \cite{Kimble_2015, Mitsch_2014, Sipahigil_2017}, recent advancement in quantum electronics offers a large variety of alternatives. Most commonly, nowadays, waveguide QED setups are realized in the experiments with superconducting quantum circuits, where superconducting transmission lines act as quantum radiation channels, whereas the Josephson-junction-based superconducting quantum bits play the role of quantum emitters \cite{Astafiev_Sci_2010, Hoi_2015, Loo_2013}. Other examples include superconducting qubits coupled to propagating phonons (surface acoustic waves) employing the piezoelectric effect \cite{Chu_2017, Gustafsson_2014, Manenti_2017,Andersson_Nat_Phys_2019}, and surface plasmons coupled to quantum dots \cite{Akimov_2007}.

Theoretically, the atom-field interactions in waveguide QED can often be accurately treated in the so-called Markovian limit $\gamma\tau\ll1$, where $\gamma$ and $\tau$ stand for the characteristic decay rate and time delay in the system \cite{Guimond_QST_2017}. In the course of the last few decades, a significant number of theoretical approaches allowing one to tackle the waveguide QED problems within this limit was proposed. In particular, the Markovian waveguide QED systems are commonly examined theoretically with help of master equation based-approaches \cite{Lalumiere_PRA_2013, Lehmberg_1970_1, Lehmberg_1970_2, You_2018, Mirza_2016, Li_2009, Lin_2019}. Indeed, within the framework of the associated input-output formalism \cite{Lalumiere_PRA_2013, Fan_2010, Xu_2015}, Lindblad-type equation approaches allow one to study transmission and reflection characteristics, as well as the photonic correlations for arbitrary initial photon states, including coherent, thermal, and Fock ones. Moreover, at a rather modest expense, a substantial variety of theoretical tools are available for a derivation of master equations:  equation of motion-based methods \cite{Lalumiere_PRA_2013, Lehmberg_1970_1, Lehmberg_1970_2}, path integral techniques \cite{Shi_2015}, SLH formalism \cite{Combes_2017}, to mention just a few. Recently master equation-based approach was generalized to capture the effects of temporal modulation of the system parameters \cite{PPG_2017,Liao_2020}, such as the light-matter coupling strength and the transition frequencies of quantum emitters. 
\par
Another frequenter method of studying waveguide QED systems within Markov approximation is the coordinate space Bethe ansatz \cite{Shen_2015, Yudson_2008, Yudson_1998, Zheng_2010, Shen_2005, Tsoi_2008, Cheng_2017, Liao_2016}. Within this approach, one is able to determine the exact stationary eigenstates of the system's Hamiltonian in a subspace of a given excitation number, which, in turn, enables one to calculate stationary observables as well as photonic correlation functions exactly. Moreover, the Bethe ansatz technique was shown to be applicable to the studies of real-time dynamics of few-photon states \cite{Yudson_1998, Chen_2011, Dinc_2019}, systems where the photon-photon bound states occur \cite{Facchi_2016, Zheng_2010}, and systems with delayed coherent feedback \cite{Calajo_2019}. Despite its scalability to the cases of multiple emitters and emitters with complicated level structures, this approach is known to be strongly limited by the number of excitations in the system due to the rapid increase of complexity of the resulting Bethe wavefunctions \cite{Fang_2014, Yudson_1998, Shen_2015, Fang_2017}. 
\par
Another group of systematic methods of studying the waveguide QED systems in the limit of the negligible delay times is comprised of the field theory-based approaches. For example, in \cite{Shi_2009, Shi_2011, Shi_2013}, by representing the atomic operators in terms of the slave fermions, the authors were able to employ the path integral representation of the field correlation functions, from which the $S$-matrix may be established by means of the celebrated Lehmann-Symanzik-Zimmermann reduction formula. Another field theoretical method to be mentioned is the so-called diagrammatic resummation theory developed in \cite{Pletyukhov_New_J_Phys_2012, Pletyukhov_2015, Pedersen_PRA_2017, Tian_Feng_2017}. Within the framework of resummation theory, one sums the perturbative series of Feynmann diagrams for the matrix elements of the transition operator to infinite orders, which, in turn, allows one to determine the exact $S$-matrix, with the help of which all of the stationary observables may be calculated \cite{Pletyukhov_2015}.
\par
Although physics behind waveguide QED systems is easily accessible in the Markovian limit using a large variety of theoretical tools, there exists a number of problems forcing one to go beyond this approximation \cite{Guimond_QST_2017, Rivas_2014, Breuer_2016, Sinha_prl_2020, Sinha_pra_2020}. As it is well known, in the single excitation subspace, all of the dynamical and stationary information about the system may be easily obtained (either analytically or numerically) by means of Bethe ansatz for a system of arbitrary complexity \cite{Guo_2020, Guo_Phys_Rev_2017, Dinc-2020, Dorner_2002}. Primarily, this assertion has to do with the fact that it is relatively easy to conceive a closed system of (delay) differential equations governing the evolution of the system. Even though it is possible to proceed in the same manner in higher excitation subspaces \cite{Redchenko_2014}, the calculations become much more cumbersome and lack systematicity. In the course of the last decade, a number of theoretical approaches allowing one to overcome these difficulties were put forward. On the numerical side, for example, significant progress in the dynamical simulation of the non-Markovian 1D quantum systems was achieved within the framework of matrix product state ansatz \cite{Grimsm_2015, Pichler_2016, Fin_2020, Wang_2017}. Despite the complexity associated with the non-Markovian waveguide QED systems, a few analytical methods were also recently suggested. In particular, it is a common practice to generalize the Lindblad-type equation approaches to the non-Markovian realm \cite{Shi_2015, Chen_2018, Wu_2010, Tan_2011}. Although generalized master equations can only provide exact results in the case of linear scatterers, they allow for systematic approximate treatment of systems with nonlinearities such as qubits. 
\par
Another common approach to waveguide QED problems with delayed coherent quantum feedback is based on the diagrammatic resummation theory. In recent years resummation ideas were successfully applied to solve the two-photon scattering and dynamics in the systems with two distant qubits \cite{Laakso_PRL_2014}, a single qubit in front of a mirror \cite{Guimond_QST_2017}, a giant acoustic atom \cite{Guo_Phys_Rev_2017}, and a qubit coupled to a resonator array \cite{Koc_2016}. 
\par
In this paper, we present a systematic generalizattion of the diagrammatic approach to scattering of multiphoton states in waveguide QED. Our approach is based on the exact resummation of the perturbation theory for the transition operator which turns out to be possible due to the conservation of excitation number guaranteed by the rotating wave approximation. The advantage of our approach is its insensitivity to the form of light-matter coupling constants, thus, allowing one to potentially examine any kinds of waveguide QED systems, including the systems with delayed coherent quantum feedback. We start by making an exposure of the method by the direct example of $1$-qubit waveguide QED. This framework lays down a basis for further calculations, in particular general qubit number two and three-photon scattering theory. We then apply the developed theory to a weakly coherent pulse scattering on a giant acoustic atom, an intrinsically non-Markovian system. In particular, we consider the scattering of a coherent state with small enough coherence parameter $|\alpha|\ll1$ chosen such that the terms of order $|\alpha|^{4}$ can be ignored. With the help of the general methods developed in Section \ref{sec: Scat_Theor} we compute an exact output state of radiation and study the particle correlations in it with the help of the theory of optical coherence. In particular, we compute the first, second, and third-order coherence functions to the lowest order in $|\alpha|$ and discuss the impact of the non-Markovianity of the scatterer on the observable quantities.

\section{Scattering Theory}
\label{sec: Scat_Theor}

In this section, we first set up the notations used throughout the paper. Next, we introduce the general formalism in the framework of the single-qubit waveguide QED. In particular, we extend the findings of the preceding papers \cite{Guimond_QST_2017, Pletyukhov_New_J_Phys_2012, Pedersen_PRA_2017, Guo_Phys_Rev_2017} to the realm of non-Markovian models by allowing for arbitrary momentum dependence of waveguide modes and radiation-qubit couplings. This development, in turn, allows one to study multi-photon scattering problems in systems with non-linear dispersion of the modes supported by the radiation channels, as well as the systems with artificial feedback loops, such as a qubit placed in front of a mirror or a giant acoustic atom,  for example. 
\par
Further on, we generalize the scattering formalism to the systems with a higher number of qubits, where the quantum feedback loops a naturally present due to the finiteness of time required for a photon to propagate between a given pair of distant emitters. Specifically, we discuss two and three-photon scattering problems on an arbitrary number of emitters, extending the approach of \cite{Laakso_PRL_2014}. 
\par
Alongside this, we discuss several practical issues, such as the separation of elastic contributions to the scattering matrices and the generalized cluster decomposition.

\subsection{Hamiltonian, generalized summation convention, and the $S$ -matrix}
\label{sec: Conventions}

Let us consider a collection of $N_{q}$ qubits coupled to a waveguide with $N_{c}$ radiation channels. The Hamiltonian of such a system assumes the following form $\mathcal{H}=\mathcal{H}_{0}+\mathcal{V}$, where
\begin{align}
\label{eq: Bare_Hamiltonian}
\mathcal{H}_{0}=&\sum_{n=1}^{N_{q}}\Omega_{n}\sigma_{+}^{(n)}\sigma_{-}^{(n)}+\sum_{\mu=1}^{N_{c}}\int_{B_{\mu}}dk\omega_{\mu}(k)a^{\dagger}_{\mu}(k)a_{\mu}(k),
\end{align}
is the free Hamiltonian, and
\begin{align}
\label{eq: inter}
\mathcal{V}=&\sum_{\mu=1}^{N_{c}}\sum_{n=1}^{N_{q}}\int_{B_{\mu}}dk[g_{\mu, n}(k)a^{\dagger}_{\mu}(k)\sigma_{-}^{(n)} + \text{h.c.} ]
\end{align}
is the dipole light-matter interaction in the rotating wave approximation (RWA). In the above expression, $\Omega_{n}$ is the transition frequency of the $n^{\text{th}}$ qubit, the dispersion relation $\omega_{\mu}(k)$ and the bandwidth $B_{\mu}$ characterize the radiation channel $\mu$, while $a^{\dagger}_{\mu}(k)$ and $a_{\mu}(k)$ stand for the creation and annihilation field operators of a photon with momentum $k$ and obey the standard bosonic commutation relations:
\begin{align}
\label{eq: commutator_boson_zero}
[a_{\mu}(k), a^{\dagger}_{\mu'}(k')]=&\delta_{\mu, \mu'}\delta(k-k'),\\
\label{eq: commutator_boson_non-zero}
[a_{\mu}^{\dagger}(k), a^{\dagger}_{\mu'}(k')]=&[a_{\mu}(k), a_{\mu'}(k')]=0.
\end{align}
The operators acting on the Hilbert space of the $n^{\text{th}}$ qubit, $\{\sigma_{3}^{(n)}, \sigma_{+}^{(n)}, \sigma_{-}^{(n)}\}$, are defined according to $\sigma^{(n)}_{l}=1_{\mathbb{C}^{2}}^{\otimes(n-1)}\otimes\sigma_{l}\otimes1_{\mathbb{C}^{2}}^{\otimes(N_{q}-n)}$, with the $\sigma$-matrices being chosen according to the standard convention
\begin{align}
\label{eq: Paulis}
\sigma_{3}=\begin{pmatrix}1 &0 \\ 0& -1 \end{pmatrix}, \quad \sigma_{+}=\sigma_{-}^{\dagger}=\begin{pmatrix} 0 & 1\\ 0& 0  \end{pmatrix}.
\end{align}
\par
The RWA is justified as long as the characteristic operational frequency $\omega_{0}$ is such that the condition $|g_{\mu}^{2}(\omega_{0})/\omega_{0}|\ll1$ is satisfied \cite{Mandel_Wolf}.
One of the main benefits of this approximation is the conservation of the total number of excitations in the system, i.e. an operator $\mathcal{N}=\sum_{n=1}^{N_q}\sigma_{+}^{(n)}\sigma_{-}^{(n)}+\sum_{\mu=1}^{N_c}\int_{B_{\mu}}dka_{\mu}^{\dagger}(k)a_{\mu}(k)$ commutes with the full Hamiltonian, $[\mathcal{N}, \mathcal{H}]=0$. This property allows us to simultaneously diagonalize $\mathcal{H}$ and $\mathcal{N}$. Moreover, the eigenspaces of the Hamiltonian labeled by the eigenvalues of $\mathcal{N}$ are certainly orthogonal, hence there exists a direct sum decomposition of the total Hilbert space of the system
\begin{equation}
\label{eq: Direct_sum_hilbert}
\mathscr{H}=\bigoplus_{N=0}^{\infty}\mathscr{H}_{N},
\end{equation}
where $\mathscr{H}_{N}$ is the $D(N, N_{q})=\sum_{l=0}^{\min(N_{q}, N)}\frac{N_{q}!}{l!(N_{q}-l)!}$ dimensional subspace of all possible states with $N$ excitations. Note that here $D(N, N_{q})$ does not stand for the dimension of $\mathscr{H}_{N}$ in the strict mathematical sense. Instead, it is a number of ways to distribute $N$ excitations in a system of $N_{q}$ qubits (i.e., each single-photon infinite-dimensional vector space adds unity to $D(N, N_{q})$). Analogous direct sum decompositions hold for the Hamiltonian, unitary evolution operator, and the $S$-matrix (to be introduced later). Moreover, such a decomposition considerably simplifies the problem since the calculations may be performed in all of the subspaces separately.
\par
To simplify our further analysis, it is useful to introduce compact notations. Thus we define a multi-index $s=(\mu, k)$ and the generalized summation convention: if two multi-indices are repeated, summation over the channel index $\mu$ and integration with respect to the momentum $k$ (over the relevant bandwidth $B_{\mu}$) is implied. We also introduce the following notation for the bare interaction vertex 
\begin{equation}
\label{eq: Bare_vertex_sc}
v_{s}=\sum_{n=1}^{N_{q}}g_{\mu, n}(k)\sigma^{(n)}_{-}.
\end{equation}
Using these conventions we rewrite the Hamiltonian as
\begin{align}
\label{eq: Bare_Hamiltonian_compact}
\mathcal{H}=&\mathcal{H}_{0}+\mathcal{V}, \quad \mathcal{H}_{0}=\omega_{s}a^{\dagger}_{s}a_{s}+\sum_{n=1}^{N_{q}}\Omega_{n}\sigma_{+}^{(n)}\sigma_{-}^{(n)}, \\
\label{eq: Potential_compact}
\mathcal{V}=&a^{\dagger}_{s}v_{s}+v_{s}^{\dagger}a_{s}.
\end{align}
\par
The scattering matrix, or the $S$-matrix, is the main object of interest in the present section. It can be generally defined via the so-called transition operator, or the $T$-matrix, in the following way:
\begin{align}
\label{eq: SMatrix_def}
\mathcal{S}=&1_{\mathscr{H}}-2\pi{i}\delta(\epsilon_{i}-\epsilon_{f})\mathcal{T}(\epsilon_{i}), \\
\label{eq: TMatrix_def}
\mathcal{T}(\epsilon)=&\mathcal{V}+\mathcal{V}\mathcal{G}(\epsilon)\mathcal{V},
\end{align}
where the $T$-matrix is put on-shell, i.e. $\epsilon=\epsilon_i$, and the energies  $\epsilon_{i}, \epsilon_{f}$, corresponding to initial and final states of the system, obey the energy conservation in the scattering processes, which is mathematically ensured by the delta-function. In (\ref{eq: TMatrix_def}) we have denoted by $\mathcal{G}(\epsilon)$ the retarded Green's operator defined according to
\begin{align}
\label{eq: Green_operator}
\mathcal{G}(\epsilon)=&\frac{1}{\epsilon-\mathcal{H}+i\eta}=\sum_{n=0}^{\infty}(\mathcal{G}_{0}(\epsilon)\mathcal{V})^n \mathcal{G}_{0}(\epsilon),\\
\label{eq: Bare_Green_operator}
\mathcal{G}_{0}(\epsilon)=&\frac{1}{\epsilon-\mathcal{H}_{0}+i\eta}, \quad \eta\rightarrow0^{+}.
\end{align}

\subsection{General properties of the scattering matrix in waveguide QED}
\label{sec: Properties}

Let us consider the scattering problem for the following initial state $\ket{N_{p}}\otimes\ket{g}$, where $\ket{N_{p}}$ is a $N_{p}$-photon state, and $\ket{g}=\ket{0}^{\otimes{N_{q}}}$ is the ground state of the scatterer. Due to the conservation of excitation-number operator, all of the possible $D(N_{p}, N_{q})$ scattering outcomes must contain $N_{p}$ excitations. Due to the fact that for any system of qubits the ground state $\ket{g}$ is the only non-decaying (subradiant) subspace, in the long-time limit (a priory assumed in scattering theory) all of the emitters will definitely decay into the continuum, leaving us with the only possibility for the system to end up in the state $\ket{N_{p}'}\otimes\ket{g}$  (here $\ket{N_{p}'}$ is again a $N_{p}$-photon state, with potentially redistributed momenta) [\onlinecite{pisotski}]. If one wishes to extend the scattering theory to the systems with metastable ground states, such as e.g. a $\Lambda$ three-level system, one then has to consider calculating more matrix elements of the transition operator \cite{Trivedi_2018,Xu_Phys_Rev_2017}.
\par
Since the only matrix elements, we are interested in are diagonal in both the photon and qubit space, due to the RWA, the only terms contributing to the perturbation expansion of the $T$-matrix are those containing an even number of interactions $\mathcal{V}$, thus reducing the number of diagrams by half. Another important feature to be mentioned is the nilpotency of the photon-qubit interaction vertex operator $v_{s_{1}}^{\dagger}...v_{s_{N_{q}+1}}^{\dagger}=v_{s_{1}}...v_{s_{N_{q}+1}}=0$, which along with the property $v_{s}\ket{g}=\bra{g}v_{s}^{\dagger}=0$ and the fact that the number of $v$'s and $v^{\dagger}$'s has to be equal in each graph contributing to the expansion, significantly reduces the number of non-zero diagrams at each order in perturbation theory. In fact, as we are going to see in this section, all of the diagrams contributing to the series for any fixed $N_{q}$ may be constructed out of a finite number of "clusters", in turn allowing, in principle for the exact resummation of the perturbation series. One can also organize the calculation differently, namely by fixing $N_{p}$ and allowing $N_{q}$ to vary instead. Calculation within this approach is facilitated in its turn by the fact that $a_{s_{1}}...a_{s_{N_{p}+1}}\ket{N_{p}}=\bra{N_{p}}a_{s_{1}}^{\dagger}...a_{s_{N_{p}+1}}^{\dagger}=0$ and the fact that the number of $a$'s and $a^{\dagger}$'s in each term of the perturbation series have to be equal. Although, these two approaches are clearly dual due to the structure of interaction potential in RWA. This approach is beneficial when studying few-photon scattering on a large number of qubits. This assertion has to do with the fact that once the solution of $N_{p}$ photon scattering problem on $N_{q}=N_{p}$ the scattering of $N_{p}$ particles on $N_{q}>N_{p}$ follows the same lines. Indeed, since $v_{s}$ ($v_{s}^{\dagger}$) by itself contains the sum of all of the single-qubit lowering (raising) operators and the highest number of qubits that the $N_{p}$ photon pulse can excite equals to $N_{p}$, the normal ordered $N_{p}$-photon $S$-matrix cannot contain projectors on subspaces with higher-excitation number than $N_{p}$. Using this fact in the following we are going to derive the generic two and three-photon scattering matrices by considering two and three particle scattering on two and three atoms respectively.

\subsection{Scattering theory in $N_{q}=1$ waveguide QED}
\label{sec:Q=1}

In this subsection, we would like to make a detailed exposition of our general method by considering the simplest imaginable scenario of a single qubit coupled to a waveguide. As it was anticipated in Subsection \ref{sec: Properties}, to solve the $N_{p}$-photon scattering problem, we shall determine the following matrix elements of the transition operator
\begin{align}
\label{eq: Cll1_1}
&\Braket{N_{p}', g|\mathcal{T}^{(1)}(\epsilon)|N_{p}, g}=\Braket{N_{p}', g|\mathcal{V}\mathcal{G}(\epsilon)\mathcal{V}|N_{p}, g}\\
\label{eq: Cll1_3}
&=\Braket{N_{p}', g|a^{\dagger}_{s_{1}'}v_{s_{1}'}\mathcal{G}(\epsilon)v^{\dagger}_{s_{1}}a_{s_{1}}|N_{p}, g}\\
\label{eq: Cll1_4}
&=\sum_{n=0}^{\infty}\Braket{N_{p}', g|a^{\dagger}_{s_{1}'}v_{s_{1}'}(\mathcal{G}_{0}(\epsilon)\mathcal{V})^{n}\mathcal{G}_{0}(\epsilon)v^{\dagger}_{s_{1}}a_{s_{1}}|N_{p}, g},
\end{align}
where the superscript $(1)$ refers to the $N_{q}=1$ waveguide QED and we have used the fact that $\Braket{N_{p}', g|\mathcal{V}|N_{p}, g}=0$. Note that in what follows, whenever an argument of an object is omitted, we understand that its argument is $\epsilon$, e.g $\mathcal{G}^{(1)}$ stands for $\mathcal{G}^{(1)}(\epsilon)$, etc, however, when the argument of a given operator or vertex function is of importance, it would be explicitly stated. As it was mentioned above, due to RWA, only even terms contribute to the above geometric series, so that
\begin{align}
\nonumber
&\Braket{N_{p}', g|\mathcal{T}^{(1)}|N_{p}, g}\\
\label{eq: Cll1_5}
&=\sum_{n=0}^{\infty}\Braket{N_{p}', g|a^{\dagger}_{s_{1}'}v_{s_{1}'}(\mathcal{G}_{0}\mathcal{V}\mathcal{G}_{0}\mathcal{V})^{n}\mathcal{G}_{0}v^{\dagger}_{s_{1}}a_{s_{1}}|N_{p}, g}.
\end{align}
Now, let us consider the $\mathcal{V}\mathcal{G}_{0}\mathcal{V}$ term in the brackets above
\begin{align}
\nonumber
\mathcal{V}\mathcal{G}_{0}\mathcal{V}=&a^{\dagger}_{s'}v_{s'}\mathcal{G}_{0}a^{\dagger}_{s}v_{s}+a^{\dagger}_{s'}v_{s'}\mathcal{G}_{0}v_{s}^{\dagger}a_{s}\\
\label{eq: Cll1_6}
+&v_{s'}^{\dagger}a_{s'}\mathcal{G}_{0}v_{s}^{\dagger}a_{s}+v_{s'}^{\dagger}a_{s'}\mathcal{G}_{0}a^{\dagger}_{s}v_{s}.
\end{align}
Clearly, the terms with two $v$s and two $v^{\dagger}$s do not contribute (they are in fact zero) by the single-qubit nilpotency condition $v^{2}=(v^{\dagger})^{2}=0$. Among the "diagonal" terms, the only non-zero contribution comes from the $v^{\dagger}v$ term since the term $vv^{\dagger}$ annihilates the state $v^{\dagger}\ket{g}$ and gives zero whenever it is multiplied with $v^{\dagger}v$ term. Hence, we arrive at
\begin{align}
\nonumber
&\Braket{N_{p}', g|\mathcal{T}^{(1)}|N_{p}, g}\\
\label{eq: Cll1_7}
&=\sum_{n=0}^{\infty}\bra{N_{p}', g}a^{\dagger}_{s_{1}'}v_{s_{1}'}(\mathcal{G}_{0}v_{s'}^{\dagger}a_{s'}\mathcal{G}_{0}a^{\dagger}_{s}v_{s})^{n}\mathcal{G}_{0}v^{\dagger}_{s_{1}}a_{s_{1}}\ket{N_{p}, g}.
\end{align}
Now, by using the commutation relations (\ref{eq: commutator_boson_zero}) and (\ref{eq: commutator_boson_non-zero}), one may easily establish the following intertwining property of bosonic operators \cite{Pletyukhov_New_J_Phys_2012}
\begin{align}
\label{eq: CII1_8}
a_{s}f(\mathcal{H}_{0})&=f(\mathcal{H}_{0}-\omega_{s})a_{s},\\
f(\mathcal{H}_{0})a_{s}^{\dagger}&=a_{s}^{\dagger}f(\mathcal{H}_{0}-\omega_{s}),
\end{align}
where $f$ is some function admitting for the Maclaurin expansion $f(z)=\sum_{n=0}^{\infty}f_{n}z^{n}$. So, we see that
\begin{align}
\nonumber
v_{s'}^{\dagger}a_{s'}\mathcal{G}_{0}(\epsilon)a^{\dagger}_{s}v_{s}&=v_{s}^{\dagger}\mathcal{G}_{0}(\epsilon-\omega_{s})v_{s}\\
\label{eq: CII1_9}
&+a^{\dagger}_{s'}v_{s}^{\dagger}\mathcal{G}_{0}(\epsilon-\omega_{s'}-\omega_{s})v_{s'}a_{s},
\end{align}
where in the last term of (\ref{eq: CII1_9}) the contraction of the indices $s$ and $s'$ is not assumed. By defining the self-energy operator $\Sigma^{(1)}$ and the effective potential energy operator $\mathcal{R}^{(1)}$ as
\begin{align}
\label{eq: CII1_10}
\Sigma^{(1)}(\epsilon)=&v_{s}^{\dagger}\mathcal{G}_{0}(\epsilon-\omega_{s})v_{s}, \quad \mathcal{R}^{(1)}(\epsilon)=a^{\dagger}_{s'}\mathcal{R}_{s', s}^{(1)}(\epsilon)a_{s}, \\
\label{eq: CII1_11}
\mathcal{R}_{s', s}^{(1)}(\epsilon)=&v_{s}^{\dagger}\mathcal{G}_{0}(\epsilon-\omega_{s'}-\omega_{s})v_{s'},
\end{align}
and resumming the perturbative series in (\ref{eq: Cll1_7}), we conclude that
\begin{align}
\nonumber
&\Braket{N_{p}', g|\mathcal{T}^{(1)}|N_{p}, g}\\
\label{eq: CII1_12}
&=\Bra{N_{p}', g}a^{\dagger}_{s_{1}'}v_{s_{1}'}\frac{1}{1-{\mathcal{G}}^{(1)}\mathcal{R}^{(1)}}{\mathcal{G}}^{(1)}v^{\dagger}_{s_{1}}a_{s_{1}}\ket{N_{p}, g},
\end{align}
where we have introduced the following Green's operator $(\mathcal{G}^{(1)})^{-1}(\epsilon)=\mathcal{G}_{0}^{-1}(\epsilon)-\Sigma^{(1)}(\epsilon)$. By defining the generating operator:
\begin{align}
\label{eq: CII1_13}
\mathcal{W}^{(1)}=\mathcal{R}^{(1)}+\mathcal{R}^{(1)}\mathcal{G}^{(1)}\mathcal{W}^{(1)},
\end{align}
we arrive at the following result:
\begin{align}
\nonumber
&\Braket{N_{p}', g|\mathcal{T}^{(1)}|N_{p}, g}\\
\nonumber
&=\Braket{N_{p}', g|a^{\dagger}_{s_{1}'}v_{s_{1}'}\mathcal{G}^{(1)}v^{\dagger}_{s_{1}}a_{s_{1}}|N_{p}, g}\\
\label{eq: CII1_14}
&+\Braket{N_{p}', g|a^{\dagger}_{s_{1}'}v_{s_{1}'}\mathcal{G}^{(1)}\mathcal{W}^{(1)}\mathcal{G}^{(1)}v^{\dagger}_{s_{1}}a_{s_{1}}|N_{p}, g}. \end{align}
By analyzing the structure of $\mathcal{R}^{(1)}$, we conclude that $\mathcal{W}^{(1)}$ admits for the following series representation:
\begin{align}
\label{eq: CII1_15}
\mathcal{W}^{(1)}=\sum_{n=1}^{\infty}a^{\dagger}_{s_{1}'}...a^{\dagger}_{s_{n}'}\mathcal{W}^{(1, n)}_{s_{1}'...s_{n}', s_{1}...s_{n}}a_{s_{n}}...a_{s_{1}},
\end{align}
where $\mathcal{W}^{(1, n)}$'s are the operator-valued functions, which depend only on $\epsilon-\mathcal{H}_{0}$ and $2n$ multi-indices $\{s_{1}', ..., s_{n}', s_{1}, ..., s_{n}\}$. 
\par
\begin{figure}[b]
                \includegraphics[width=\columnwidth]{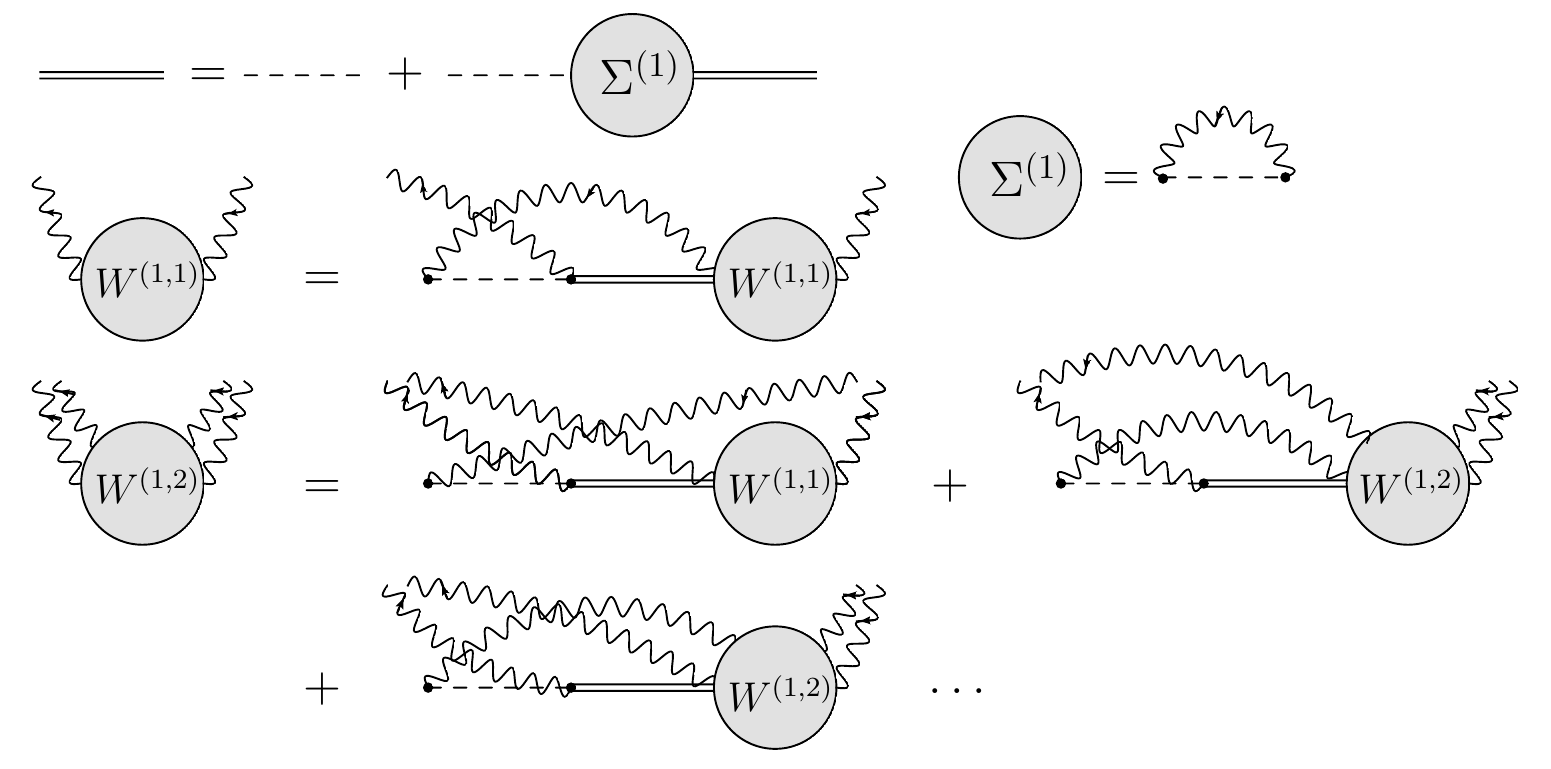}
                %\linewidth or \textwidth
                \caption{Hierarchy of integral equations governing effective multi-photon vertex functions in $N_{q}=1$ waveguide QED. Here the $\Sigma^{(1)}$ bubble corresponds to the self-energy in a subspace of a single excitation used to the define the dressed Green's function (represented by a double line) from the bare Green's function (depicted by the dashed line). The $W^{(1, n)}$ bubbles (bubbles with $2n$-amputated legs) represent effective $n$-photon vertex functions describing effective interaction between $n$-photons induced by non-linearity. Bare absorption $v^{\dagger}$ and emission $v$ vertices are represented by black dots with incoming and outgoing photon lines respectively. Diagrammatic rules are as stated in Sec. \ref{sec:Q=1}. }
                \label{fig: CII1_1}
\end{figure}

By inserting (\ref{eq: CII1_15}) in (\ref{eq: CII1_13}) and taking the projections onto the particle subspaces we arrive at the following hierarchy of integral equations:
\begin{widetext}
\begin{align}
\label{eq:W11_Integral}
W_{s_{1}', s_{1}}^{(1, 1)}(\epsilon)=&R_{s_{1}', s_{1}}^{(1)}(\epsilon)+R_{s_{1}', s}^{(1)}(\epsilon)G^{(1)}(\epsilon-\omega_{s})W_{s, s_{1}}^{(1, 1)}(\epsilon),\\
\nonumber
\vdots\\
\nonumber
W^{(1, n)}_{s_{1}'...s_{n}', s_{1}...s_{n}}(\epsilon)=&R_{s_{1}', s_{1}}^{(1)}\Bigg(\epsilon-\sum_{l=2}^{n}\omega_{s_{n}'}\Bigg)G^{(1)}\Bigg(\epsilon-\omega_{s_{1}}-\sum_{l=2}^{n}\omega_{s_{l}'}\Bigg)W^{(1, n-1)}_{s_{2}'...s_{n}', s_{2}...s_{n}}(\epsilon-\omega_{s_{1}})\\
\label{eq: CII1_18}
+&R_{s_{1}', s}^{(1)}\Bigg(\epsilon-\sum_{l=2}^{n}\omega_{s_{n}'}\Bigg)G^{(1)}\Bigg(\epsilon-\omega_{s}-\sum_{l=2}^{n}\omega_{s_{l}'}\Bigg)\Big[W^{(1, n)}_{ss_{2}'...s_{n}', s_{1}...s_{n}}(\epsilon)+...+W^{(1, n)}_{s_{2}'...s_{n}'s, s_{1}...s_{n}}(\epsilon)\Big]\\
\nonumber
\vdots
\end{align}
\end{widetext}
Here, we have used regular letters $W, G, R$ instead of calligraphic ones to indicate that these objects are not operators but are rather their projections on the photonic vacuum (note that $W^{(1, n)}$s are still acting as operators on the qubit space). The collection of the integral equations above may be conveniently represented diagrammatically, see Figure \ref{fig: CII1_1}. Diagrammatic rules may be formulated as follows. To each dotted line associate the bare propagator $G_{0}(\epsilon)$. To each wavy line with incoming (outgoing) arrow associate the bare absorption $v^{\dagger}$ (emission $v$) vertex. Whenever two wavy lines are contracted together, one has to integrate over all momentum and sum over all channels. When a given wavy line passes over the propagator, its argument has to be shifted by the frequency carried by this line (which is a direct consequence of the intertwining property (\ref{eq: CII1_8})). 
For example, the self-energy diagram in Figure \ref{fig: CII1_1} is translated as
\begin{align}
    \Sigma^{(1)}(\epsilon)=v_{s}^{\dagger}G_{0}(\epsilon-\omega_{s})v_{s}.
\end{align}
Now having identified the equations defining the components of $\mathcal{W}^{(1)}$, we may express the transition operator as follows
\begin{equation}
\label{eq: CII1_19}
\mathcal{T}^{(1)}(\epsilon)=\sum_{n=1}^{\infty}T^{(1, n)}_{s_{1}'...s_{n}', s_{1}...s_{n}}(\epsilon)a^{\dagger}_{s_{1}'}...a_{s_{n}'}^{\dagger}a_{s_{n}}...a_{s_{1}},
\end{equation}
where we have defined the following functions
\begin{align}
\label{eq:T11}
T^{(1, 1)}_{s_{1}', s_{1}}(\epsilon)=\Braket{g|v_{s_{1}'}G^{(1)}(\epsilon)v^{\dagger}_{s_{1}}|g},
\end{align}
\begin{widetext}
\begin{align}
\label{eq: CII1_21}
T^{(1, n>1)}_{s_{1}'...s_{n}', s_{1}...s_{n}}(\epsilon)=\Braket{g|v_{s_{1}'}G^{(1)}\Bigg(\epsilon-\sum_{l=2}^{n}\omega_{s_{l}'}\Bigg)W^{(1, n-1)}_{s_{2}'...s_{n}', s_{2}...s_{n}}(\epsilon)G^{(1)}\Bigg(\epsilon-\sum_{l=2}^{n}\omega_{s_{l}}\Bigg)v^{\dagger}_{s_{1}}|g}.
\end{align}
\end{widetext}
Note that we are working with the non-symmetrized forms of operators and perform the symmetrization only when doing actual calculations. Also note that the dependence of $\mathcal{T}_{s_{1}'...s_{n}', s_{1}...s_{n}}^{(1, n)}$ on $\omega_{s}a^{\dagger}_{s}a_{s}$ may be omitted taking into account normal ordering, when $S$-matrix is going to be finally contracted with the initial state, all of the $T$-matrix components are going to be eventually projected onto the photonic vacuum and energies are going to be put on-shell. Using the above representation of the transition operator we immediately deduce that the scattering operator takes the following form:
\begin{widetext}
\begin{equation}
\label{eq: CII1_22}
\mathcal{S}=1_{\mathscr{H}}-2\pi{i}\sum_{n=1}^{\infty}T^{(n)}_{s_{1}'...s_{n}', s_{1}...s_{n}}\Bigg(\sum_{l=1}^{n}\omega_{s_{l}}\Bigg)\delta\Bigg(\sum_{l=1}^{n}\omega_{s_{l}'}-\sum_{l=1}^{n}\omega_{s_{l}}\Bigg)a^{\dagger}_{s_{1}'}...a_{s_{n}'}^{\dagger}a_{s_{n}}...a_{s_{1}}.
\end{equation}
\end{widetext}
Practically, when one contracts the $S$-matrix with the initial $N_{p}$-photon state, only first $N_{p}$ terms in the above series contribute. It is, however, beneficial to represent the $S$-matrix in terms of the direct sum of $N$-body operators which act solely in the $N$-particle subspaces
\begin{equation}
\label{eq: CII1_23}
\mathcal{S}=1_{\mathscr{H}}\oplus\bigoplus_{n=1}^{\infty}\mathcal{S}_{n},
\end{equation}
where $\mathcal{S}_{n}$ may be written as
\begin{equation}
\label{eq:Sn}
\mathcal{S}_{n}=S^{(n)}_{s_{1}'...s_{n}', s_{1}...s_{n}}a^{\dagger}_{s_{1}'}...a_{s_{n}'}^{\dagger}a_{s_{n}}...a_{s_{1}},
\end{equation}
\begin{widetext}
\begin{align}
\label{eq:Sn_Matrix}
S^{(n)}_{s_{1}'...s_{n}', s_{1}...s_{n}}=\frac{1}{n!}\prod_{l=1}^{n}\delta_{s_{l}', s_{l}}-2\pi{i}\sum_{m=1}^{n}\frac{1}{(n-m)!}T^{(n)}_{s_{1}'...s_{m}', s_{1}...s_{m}}\Bigg(\sum_{l=1}^{m}\omega_{s_{l}}\Bigg)\delta\Bigg(\sum_{l=1}^{n}\omega_{s_{l}'}-\sum_{l=1}^{n}\omega_{s_{l}}\Bigg)\prod_{r=m+1}^{n}\delta_{s_{r}', s_{r}}.
\end{align}
\end{widetext}
Note that the combinatorial prefactors $\frac{1}{(n-m)!}, \quad m\in\{0, ..., n\}$ and additional $\delta$ functions $\delta_{s_{r}', s_{r}}$ are chosen to take into account the excess number of $(n-m)!$ Wick contractions.

\subsection{Generalized cluster decomposition}
\label{sec:GCD}

In this section, we would like to illustrate how the generalized cluster decomposition in waveguide QED, extensively studied in [\onlinecite{Xu_Phys_Rev_2017}-\onlinecite{Burillo_New_J_Phys_2018}], naturally follows from the results of the previous section. 
\par
The cluster decomposition is a way of separating the elastic and inelastic contributions to the $S$-matrix. The elastic contribution physically corresponds to the scattering channel in which all photons scatter coherently, i.e. conserve their energy individually in the scattering process. On the other hand, the inelastic contribution  corresponds to the incoherent scattering. Thereby the photons redistribute their initial energy between one another via effective photon-photon interaction which is mediated by their interaction with nonlinear scatterers (e.g., qubits). In contrast, in the case of a linear scatter (e.g., a cavity mode), the effective photon-photon interaction is not generated, and the $n$-photon $S$-matrix simply factors out into the product of $n$ single-particle $S$-matrices.
\par
The first step towards the cluster decomposition of the multi-photon $S$-matrices is the realization that the multi-photon vertex functions $W^{(1, n)}$ contain the disconnected components. These, in turn, arise from the projections of $R^{(1)}_{s', s}$ on the ground state of the scatterer thus resulting in the (quasi)elastic contributions $\propto\delta(\epsilon-...)$ to $W^{(1, n)}$'s. Not only the separation of these contributions is crucial for the cluster decomposition but is of evident importance for both numerical and analytical treatment of the integral equations (\ref{eq:W11_Integral}), (\ref{eq: CII1_18}). It is the easiest to define the connected parts of the multi-photon vertex functions recursively. First, we observe that
\begin{align}
\label{eq: CII2_1}
R_{s', s}^{(1)}(\epsilon)=&v_{s}^{\dagger}G_{0}(\epsilon-\omega_{s'}-\omega_{s})v_{s'}\\
=&-i\pi\delta(\epsilon-\omega_{s'}-\omega_{s})v_{s}^{\dagger}v_{s'}\\
+&v_{s}^{\dagger}P\frac{1}{\epsilon-\omega_{s'}-\omega_{s}}v_{s'},
\end{align}
where $P$ stands for the Cauchy principle value. Which helps us to define the connected part of one-particle vertex function as
\begin{align}
\label{eq: CII2_2}
W_{s_{1}', s_{1}}^{(1, 1, C)}(\epsilon)=&W_{s_{1}', s_{1}}^{(1, 1)}(\epsilon)+i\pi\delta(\epsilon-\omega_{s_{1}'}-\omega_{s_{1}})v_{s_{1}}^{\dagger}v_{s_{1}'}.
\end{align}
By then analyzing the structure of the general equation in the hierarchy (\ref{eq: CII1_18}) and bearing in mind the equation satisfied by $W_{s_{1}', s_{1}}^{(1, 1)}(\epsilon)$ one may easily deduce the connected part of $n$-photon effective vertex function may be defined in the following manner
\begin{widetext}
\begin{align}
\label{eq: CII2_3}
W^{(1, n, C)}_{s_{1}'...s_{n}', s_{1}...s_{n}}(\epsilon)=&W^{(1, n)}_{s_{1}'...s_{n}', s_{1}...s_{n}}(\epsilon)-W_{s_{1}', s_{1}}^{(1, 1)}\Bigg(\epsilon-\sum_{l=2}^{n}\omega_{s_{n}'}\Bigg)G^{(1)}\Bigg(\epsilon-\omega_{s_{1}}-\sum_{l=2}^{n}\omega_{s_{l}'}\Bigg)W^{(1, n-1)}_{s_{2}'...s_{n}', s_{2}...s_{n}}(\epsilon-\omega_{s_{1}}).
\end{align}
\end{widetext}
Now, with the help of the definition (\ref{eq: CII2_2}) one may immediately deduce the following decomposition of the two-body transition operator
\begin{align}
\nonumber
&T^{(1, 2)}_{s_{1}'s_{2}', s_{1}s_{2}}(\omega_{s_{1}}+\omega_{s_{2}})\\
\nonumber
&=-i\pi\delta(\omega_{s_{1}}-\omega_{s_{2}'})T^{(1, 1)}_{s_{1}', s_{1}}(\omega_{s_{1}})T^{(1, 1)}_{s_{2}', s_{2}}(\omega_{s_{2}})\\
\label{eq: CII2_4}
&+T^{(1, 2, C)}_{s_{1}'s_{2}', s_{1}s_{2}}(\omega_{s_{1}}+\omega_{s_{2}}),
\end{align}
where $T^{(1, 2, C)}$ is defined in precisely the same way as $T^{(1, 2)}$ but with $W^{(1, 1, C)}$ replacing $W^{(1, 1)}$. By plugging the representation (\ref{eq: CII2_4}) into the definition of the two-photon $S$-matrix (\ref{eq:Sn_Matrix}) we immediately arrive at the following cluster decomposition principle:
\begin{align}
\nonumber
\mathcal{S}_{2}=&\Bigg[\frac{1}{2}S^{(1)}_{s_{1}', s_{1}}S^{(1)}_{s_{2}', s_{2}}-2\pi{i}T^{(1, 2, C)}_{s_{1}'s_{2}', s_{1}s_{2}}(\omega_{s_{1}}+\omega_{s_{2}})\\
\label{eq: CII2_5}
\times&\delta(\omega_{s_{1}'}+\omega_{s_{2}'}-\omega_{s_{1}}-\omega_{s_{2}})\Bigg]a_{s_{1}'}^{\dagger}a_{s_{2}'}^{\dagger}a_{s_{2}}a_{s_{1}}.
\end{align}
The physical meaning of the above expression is rather clear. The first term, being a product of one-particle $S$-matrices, describes the coherent scattering of two particles, i.e. a scattering process in which there is no effective interactions between the photons. The second term, on the other hand, is completely connected and describes the incoherent scattering of photons in which the particles redistribute their initial energy by interaction. A slightly more involved, but otherwise completely analogous, calculation may be done in the three-photon sector. As a result, one arrives at the following cluster decomposition of the three-body $S$-matrix:
\begin{align}
\nonumber
\mathcal{S}_{3}=&\Bigg[\frac{1}{6}S^{(1)}_{s_{1}', s_{1}}S^{(1)}_{s_{2}', s_{2}}S^{(1)}_{s_{3}', s_{3}}-2\pi{i}T^{(1, 2, C)}_{s_{1}'s_{2}', s_{1}s_{2}}(\omega_{s_{1}}+\omega_{s_{2}})\\
\nonumber
\times&\delta(\omega_{s_{1}'}+\omega_{s_{2}'}-\omega_{s_{1}}-\omega_{s_{2}})S^{(1)}_{s_{3}', s_{3}}\\
\nonumber
-&2\pi{i}T^{(1, 3, C)}_{s_{1}'s_{2}'s_{3}', s_{1}s_{2}s_{3}}(\omega_{s_{1}}+\omega_{s_{2}}+\omega_{s_{3}})\\
\nonumber
\times&\delta(\omega_{s_{1}'}+\omega_{s_{2}'}+\omega_{s_{3}'}-\omega_{s_{1}}-\omega_{s_{2}}-\omega_{s_{3}})\Bigg]\\
\label{eq: STE41}
\times&{a}_{s_{1}'}^{\dagger}a_{s_{2}'}^{\dagger}a_{s_{3}'}^{\dagger}a_{s_{3}}a_{s_{2}}a_{s_{1}},
\end{align}
where the connected part of the three-photon $T$-matrix may be found in the Appendix \ref{3bodyScat}.

\subsection{Closed form solution in the Markovian limit}

\begin{figure}[b]
                \includegraphics[width=\columnwidth]{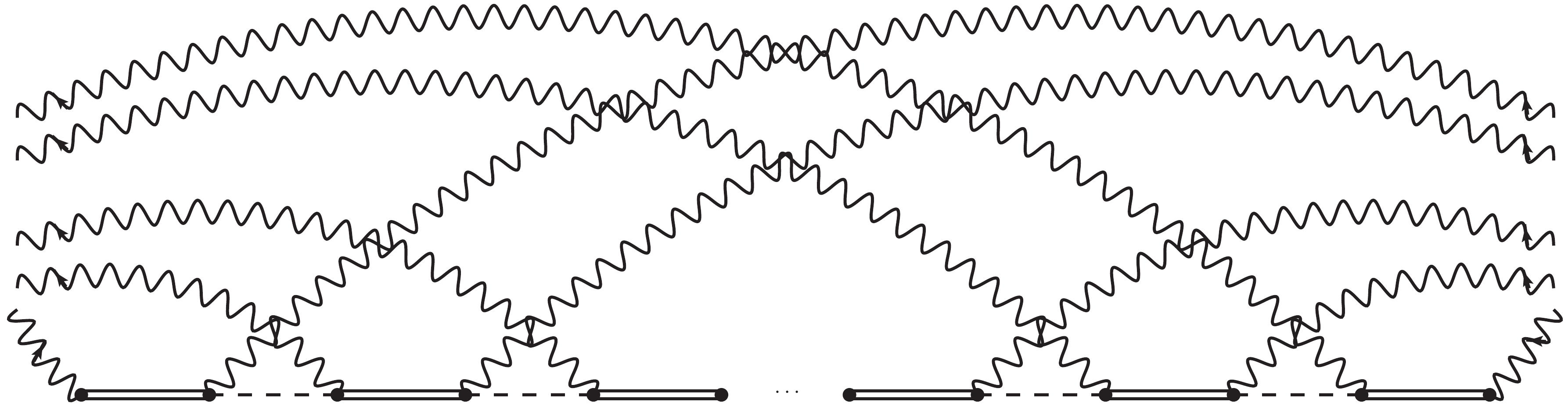}
                %\linewidth or \textwidth
                \caption{Exact diagrammatic representation of the $n$-particle $T$-matrix in the Markovian limit. The above diagrams also correspond to the non-crossing approximation discussed in section \ref{sec: AS}. As opposed to the exact $n$-photon $T$-matrix parameterized by the $n-1$-particle effective vertex function resuming an infinite number of emission and absorption processes in both direct and exchange channels, this approximation replaces the vertex by a sequence of $n-2$ and $n-1$ ($n\geq 2$) alternating excitations and deexcitations of the emitter. }
                \label{fig: CII3_1}
\end{figure}
Let us now consider the single-qubit scattering problem in the Markovian limit. Validity of Markovian approximation demands a number of assumptions: linear dispersion relation in all of the channels $\omega_{\mu}(k)=\omega_{0\mu}+v_{\mu}k$, broadband limit $B_{\mu}=\mathbb{R}, \forall\mu\in\{1, ..., N_{c}\}$ and local couplings (i.e. independent of frequency) $g_{\mu}(k)=\sqrt{\gamma_{\mu}/\pi}$. Within the above assumptions, the self-energy diagram reads as $\Sigma^{(1)}(\epsilon)=-i\sum_{\mu=1}^{N_{c}}\gamma_{\mu}/v_{\mu}\ket{1}\bra{1}$, which is independent of $\epsilon$ (here we have introduced the projector on the qubit's excited state $\ket{1}\bra{1}=\frac{1+\sigma_{3}}{2}$). By causality, all of the $W$ functions are analytic in momentum (energy) variables in the lower (upper) half of the complex plane. By exploiting this analyticity along with the momentum independence of both coupling constants and self-energy, we can close all of the integration contours in the lower half of the complex momentum plane to render all of the integrals zero, thus promoting the integral equations into algebraic ones. As a result, we recover the solution obtained in [\onlinecite{Pletyukhov_New_J_Phys_2012}]. Diagrammatically the above solution is equivalent to the so-called non-crossing approximation (which becomes exact in the Markovian limit) and it is represented in Figure \ref{fig: CII3_1}.

\subsection{Approximation strategies}

\label{sec: AS}
\begin{figure}[b]
                \includegraphics[width=\columnwidth]{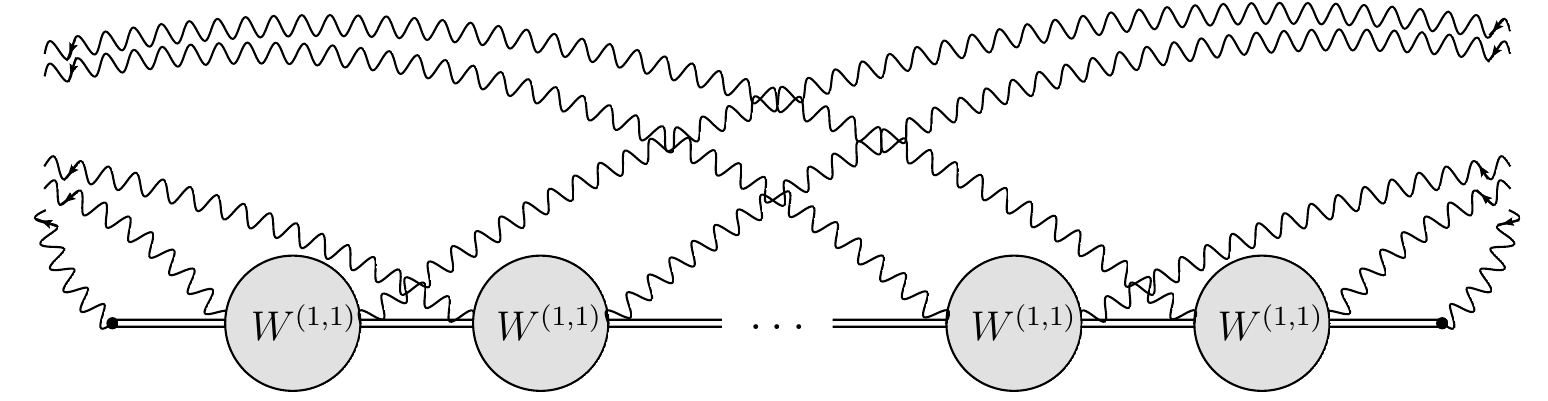}
                %\linewidth or \textwidth
                \caption{Diagrammatic representation of $n$-particle transition matrix within the weak correlation approximation. Here two-particle bubbles correspond to a single-photon effective vertex functions $W^{(1, 1)}$. Opposite to the quasi-Markovian approximation, this approximation resumms exactly all of the single particle processes, ignoring all the connected diagrams corresponding to multiple-particle scattering.}
                \label{fig: CII4_1}
\end{figure}
Although it is possible to write an entire hierarchy of exact equations for the multi-photon vertex functions, which in turn parametrize the exact $S$-matrix, their analytical solution is in general only possible in the Markovian limit. To make some progress in solving the scattering problem, one has to resort to some kind of approximation routine. In this section we propose a couple of physically motivated resummation approaches, allowing one to avoid (or partially avoid) the solution of the exact integral equations.
\par
The most basic approximation, which is accurate in the regime $\gamma\tau\lesssim1$, where $\tau$ is a typical time delay in the system due to photons' propagation between two nearby scatterers, and $\gamma$ is a typical decay rate of scatterers, is the so called quasi-Markovian approximation. While in the quantum master equation approach it has a long tradition (see, e.g., \cite{Lehmberg_1970_1,Lehmberg_1970_2}), it has been recently realised \cite{Pletyukhov_New_J_Phys_2012,Pedersen_PRA_2017} that in the diagrammatic approach it is mathematically equivalent to picking only non-crossing diagrams shown in Fig. \ref{fig: CII3_1} and thereby fully neglecting vertex corrections. To give a physical explanation of this equivalence, we view photons propagating in the waveguide as an effective reservoir. It is intuitively clear that a fast decay of time correlations between these photons is an essential feature of the Markovian regime (we add here the prefix quasi in order to indicate that the field values at positions of different scatterers can still differ from each other by a phase factor). On the other hand, these time correlations result from correlated virtual processes of emission and absorption of different photons at different scatterers, which are mathematically encoded in the vertex corrections. Thus, the neglect of vertex corrections is equivalent to making the quasi-Markovian approximation. It is also worth noting that this approximation for a single-photon scattering matrix coincides with its exact expression, since in the absence of other photons the correlations in questions are not generated.

Below we give a step-by-step prescription how to implement the quasi-Markovian approximation in our diagrammatic approach. The exact $n$-photon transition matrices are represented by a $n-1$-particle effective vertex function in between two dressed Green's functions followed by emission/absorption vertices (see equation (\ref{eq: CII1_21})). In this approximation, the exact multi-photon vertex functions are simply replaced with a sequence of $n-1$ bare and $n-2$ dressed Green's functions interspersed with $2n-2$ emission and absorption vertices. In particular, in the $n=2$ case we get $W^{(1, 1)}_{s_{1}', s_{1}}(\epsilon)\rightarrow v_{s_{1}}^{\dagger}G_{0}(\epsilon-\omega_{s_{1}'}-\omega_{s_{1}})v_{s_{1}'}$. In the $n=3$ case we obtain $W^{(1, 2)}_{s_{1}'s_{2}', s_{1}s_{2}}(\epsilon)\rightarrow v_{s_{1}}^{\dagger}G_{0}(\epsilon-\omega_{s_{1}}-\omega_{s_{1}'}-\omega_{s_{2}'})v_{s_{1}'}G^{(1)}(\epsilon-\omega_{s_{1}}-\omega_{s_{2}'})v_{s_{2}}^{\dagger}G_{0}(\epsilon-\omega_{s_{1}}-\omega_{s_{2}}-\omega_{s_{2}'})v_{s_{2'}}$, and so on. In this approximation, all of the non-trivial momentum dependence of the $S$-matrix, comes entirely from the frequency dependence of the self-energy diagram. Note that the single-photon $T$-matrix is completely determined by the dressed propagator $G^{(1)}(\epsilon)$ (see equation (\ref{eq:T11})). This observation explicitly confirms the statement that the quasi-Markovian approximation becomes exact for a single propagating photon.
\par
The second approximation routine, which is more accurate than the quasi-Markovian one for $\gamma \tau \gtrsim 1$, makes a partial account of the vertex corrections. In particular, this approximation is based on the resummation of the full single-photon vertex functions $W^{(1, 1)}_{s', s}(\epsilon)$. Diagrammatically, this approximation amounts to the replacement of the dotted lines in Figure \ref{fig: CII3_1} by the full single-particle bubbles, the resulting $n$-photon $T$-matrix is shown in Figure \ref{fig: CII4_1}. This approximation corresponds to resummation of all of the direct interaction diagrams (comping entirely from the single particle sector), completely ignoring the exchange processes (encompassed by the connected parts of many-particle vertices). In order to clarify the matters, let us consider an integral equation governing the two-particle effective vertex shown in Figure \ref{fig: CII1_1}. Using the diagrammatic rules we deduce
\begin{align}
\nonumber
    &W^{(1, 2)}_{s_{1}'s_{2}', s_{1}, s_{2}}(\epsilon)\\
    \nonumber
    &=R^{(1)}_{s_{1}', s_{1}}(\epsilon-\omega_{s_{2}'})G^{(1)}(\epsilon-\omega_{s_{1}}-\omega_{s_{2}'})W^{(1, 1)}_{s_{2}', s_{2}}(\epsilon-\omega_{1})\\
        \nonumber
    &+R^{(1)}_{s_{1}, s}(\epsilon-\omega_{s_{2}'})G^{(1)}(\epsilon-\omega_{s}-\omega_{s_{2}'})W^{(1, 2)}_{ss_{2}', s_{1}s_{2}}\\
    &+R^{(1)}_{s_{1}, s}(\epsilon-\omega_{s_{2}'})G^{(1)}(\epsilon-\omega_{s}-\omega_{s_{2}'})W^{(1, 2)}_{s_{2}'s, s_{1}s_{2}}.
    \label{eq:nn}
\end{align}
Note that the last line above corresponds to the exchange interaction between that particles (see also Figure \ref{fig: CII1_1}). If we ignore the the last term completely and take into account the equation satisfied by $W^{(1, 1)}$ it is easy to show that 
\begin{align}
\nonumber
    &W^{(1, 2)}_{s_{1}'s_{2}', s_{1}, s_{2}}(\epsilon)\\
    &=W^{(1, 1)}_{s_{1}', s_{1}}(\epsilon-\omega_{s_{2}'})G^{(1)}(\epsilon-\omega_{s_{1}}-\omega_{s_{2}'})W^{(1, 1)}_{s_{2}', s_{2}}(\epsilon-\omega_{1})
\end{align}
is an exact solution of (\ref{eq:nn}).

We expect this approximation to get worse as one increases the number of photons in the system since this approximation ignores larger and larger fraction of diagrams with the increase in the number of particles involved in a scattering process. This approximation is beneficial since an integral equation for $W^{(1, 1)}$ may be frequently solved analytically by means of the method developed in the supplementary material of [\onlinecite{Laakso_PRL_2014}]. Due to the non-trivial momentum dependence of $W^{(1, 1)}$, we expect this approximation to be better than a quasi-Markovian one. In what follows we refer to this approximation as to the weak correlation one. Note that as the quasi-Markovian approximation is exact in the single-particle sector, the weak correlation approximation is exact in the two-particle sector, thus in order to test its validity, one has to consider at least a three-photon scattering problem.

\subsection{Systems with more than one qubit}

In this section, we would like to present certain generalizations of the theory developed in Section \ref{sec:Q=1}. Specifically, we would like to demonstrate how the above-presented formalism may be extended to the systems with a larger number of qubits. In particular, we are going to focus our attention on the systems with two and three emitters. By deriving the equations governing two and three-photon scattering matrices in two and three-qubit waveguide QED systems respectively we present the most general two and three-particle equations, holding independently of the number of qubits (see the discussion in Section \ref{sec: Properties}).
\subsubsection{$N_{q}=2$ waveguide QED}
\begin{figure}[h]
                \includegraphics[width=0.8\columnwidth]{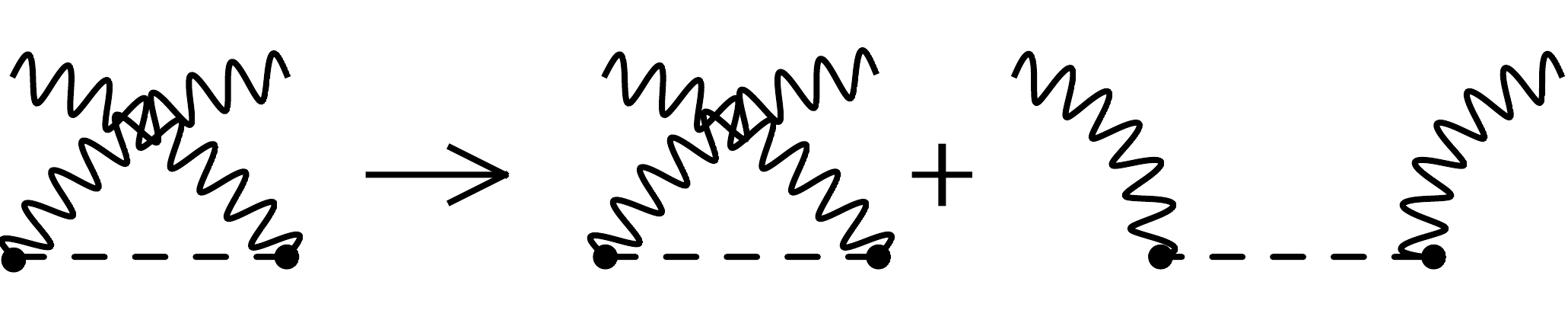}
                \caption{Diagrammatic representation of the replacement required to obtain $R^{(2)}_{s', s}$ from $R^{(1)}_{s', s}$. Note that the extra contribution to the effective potential energy vertex $R^{(2)}$ in two-qubit systems is not single-particle connected, i.e. it is possible to cut the intermediate propagator such that the diagram falls into two distinct pieces. This very fact makes the splitting (\ref{eq: DII1_8})-(\ref{eq: DII1_11}) into reducible and irreducible contributions possible.}
                \label{fig: replacement}
\end{figure}
First, we would like to focus on the case of two emitters coupled to a waveguide since the generalization of the single-qubit results is the most apparent in this case. The starting point in the analysis are equations (\ref{eq: Cll1_5}) and (\ref{eq: Cll1_6}). As opposed to the single-emitter case, now, clearly, both diagonal terms $v_{s'}^{\dagger}a_{s'}\mathcal{G}_{0}a^{\dagger}_{s}v_{s}$ and $a^{\dagger}_{s'}v_{s'}\mathcal{G}_{0}v_{s}^{\dagger}a_{s}$ contribute to the geometric series for the transition operator. Non-diagonal ones still give zero, since they both annihilate the state $v^{\dagger}\ket{g}$. By resumming the series and introducing the following set of objects
\begin{align}
\label{eq: DII1_1}
\mathcal{R}^{(2)}=&a_{s_{1}'}^{\dagger}\mathcal{R}^{(2)}_{s_{1}', s_{1}}a_{s_{1}},\\
\label{eq: DII1_2}
\mathcal{R}^{(2)}_{s_{1}', s_{1}}=&\mathcal{R}^{(1)}_{s_{1}', s_{1}}+v_{s_{1}'}\mathcal{G}_{0}v_{s_{1}}^{\dagger},\\
\label{eq: DII1_3}
\mathcal{W}^{(2)}=&\mathcal{R}^{(2)}+\mathcal{R}^{(2)}\mathcal{G}^{(1)}\mathcal{W}^{(2)},
\end{align}
we render the matrix elements of the transition operator in the following form
% \begin{widetext}
\begin{align}
\nonumber
&\Braket{N_{p}', g|\mathcal{T}^{(2)}|N_{p}, g}\\\nonumber&=\Braket{N_{p}', g|a^{\dagger}_{s_{1}'}v_{s_{1}'}\mathcal{G}^{(1)}v^{\dagger}_{s_{1}}a_{s_{1}}|N_{p}, g}\\\label{eq: DII1_4}&+\Braket{N_{p}', g|a^{\dagger}_{s_{1}'}v_{s_{1}'}\mathcal{G}^{(1)}\mathcal{W}^{(2)}\mathcal{G}^{(1)}v^{\dagger}_{s_{1}}a_{s_{1}}|N_{p}, g}.
\end{align}
% \end{widetext}
Since the lowest term in the expansion of $\mathcal{W}^{(2)}$ is again a single-photon operator
\begin{align}
\label{eq: DII1_5}
\mathcal{W}^{(2)}=\sum_{n=1}^{\infty}a^{\dagger}_{s_{1}'}...a^{\dagger}_{s_{n}'}\mathcal{W}^{(2, n)}_{s_{1}'...s_{n}', s_{1}...s_{n}}a_{s_{n}}...a_{s_{1}},
\end{align}
we see that the hierarchy of equations satisfied by the two-photon vertex functions is precisely the same as (\ref{eq:W11_Integral})-(\ref{eq: CII1_18}) with $R^{(1)}$ being replaced by $R^{(2)}$ (see Figure \ref{fig: replacement}). 
\par
In order to understand the difference between the single-qubit and the two-qubit theories, let us have a closer look at the equation defining $\mathcal{W}^{(2)}$
\begin{align}
\nonumber
\mathcal{W}^{(2)}=&\mathcal{R}^{(1)}+a_{s_{1}'}^{\dagger}v_{s_{1}'}\mathcal{G}_{0}v_{s_{1}}^{\dagger}a_{s_{1}}+\mathcal{R}^{(1)}\mathcal{G}^{(1)}\mathcal{W}^{(2)}\\
\label{eq: DII1_6}
+&a_{s_{1}'}^{\dagger}v_{s_{1}'}\mathcal{G}_{0}v_{s_{1}}^{\dagger}a_{s_{1}}\mathcal{G}^{(1)}\mathcal{W}^{(2)}.
\end{align}
Now let us perform the following separation $\mathcal{W}^{(2)}=\mathcal{W}^{(2, i)}+\mathcal{W}^{(2, r)}$, where the superscripts $i$ and $r$ stand for the irreducible and reducible contributions respectively, and $\mathcal{W}^{(2, i)}$ is chosen to satisfy the following equation (\ref{eq: CII1_13}). This leads one to the following equation satisfied by the reducible part
\begin{align}
\nonumber
\mathcal{W}^{(2, r)}=&a_{s_{1}'}^{\dagger}v_{s_{1}'}\mathcal{G}_{0}v_{s_{1}}^{\dagger}a_{s_{1}}+\mathcal{R}^{(1)}\mathcal{G}^{(1)}\mathcal{W}^{(2, r)}\\
\nonumber
+&a_{s_{1}'}^{\dagger}v_{s_{1}'}\mathcal{G}_{0}v_{s_{1}}^{\dagger}a_{s_{1}}\mathcal{G}^{(1)}\mathcal{W}^{(2, i)}\\
\label{eq: DII1_7}
+&a_{s_{1}'}^{\dagger}v_{s_{1}'}\mathcal{G}_{0}v_{s_{1}}^{\dagger}a_{s_{1}}\mathcal{G}^{(1)}\mathcal{W}^{(2, r)}.
\end{align}
The solution of equation (\ref{eq: DII1_7}) may then conveniently parametrized as $W^{(2,r)}(\epsilon)=\mathcal{V}^{(2)}(\epsilon)\mathcal{G}^{(2)}(\epsilon)\overline{\mathcal{V}}^{(2)}(\epsilon)$, where
\begin{align}
\label{eq: DII1_8}
\mathcal{G}^{(2)}=&\mathcal{G}_{0}+\mathcal{G}_{0}\Sigma^{(2)}\mathcal{G}^{(2)}, \\ 
\label{eq: DII1_9}
\Sigma^{(2)}=&a_{s}v_{s}^{\dagger}\mathcal{G}^{(1)}v_{s'}a^{\dagger}_{s'}+a_{s}v_{s}^{\dagger}\mathcal{G}^{(1)}\mathcal{W}^{(2, i)}\mathcal{G}^{(1)}v_{s'}a^{\dagger}_{s'},\\
\label{eq: DII1_10}
\overline{\mathcal{V}}^{(2)}=&v_{s}^{\dagger}a_{s}+v_{s}^{\dagger}a_{s}\mathcal{G}\mathcal{W}^{(2, i)}=:\sum_{n=1}^{\infty}\overline{V}^{(2, n)}_{s_{1}...s_{n}}a_{s_{1}}...a_{s_{n}}, \\
\label{eq: DII1_11}
\mathcal{V}^{(2)}=&a_{s}^{\dagger}v_{s}+\mathcal{W}^{(2, i)}\mathcal{G}v_{s}a_{s}^{\dagger}=:\sum_{n=1}^{\infty}{V}^{(2, n)}_{s_{1}...s_{n}}a_{s_{1}}^{\dagger}...a_{s_{n}}^{\dagger}.
\end{align}
The meaning of the above-defined objects is as follows. $\mathcal{G}^{(2)}(\epsilon)$ may be thought of as a Green's operator in the two-qubit excitation subspace since in practice, it is always projected there. $\Sigma^{(2)}(\epsilon)$ in its turn may be thought of as a self-energy operator in the two-qubit excitation subspace. $\mathcal{V}(\epsilon)$ and $\overline{\mathcal{V}}(\epsilon)$ may be understood as the renormalized absorption and emission vertex operators.
\par
\begin{figure}[b!]
\includegraphics[width=\columnwidth]{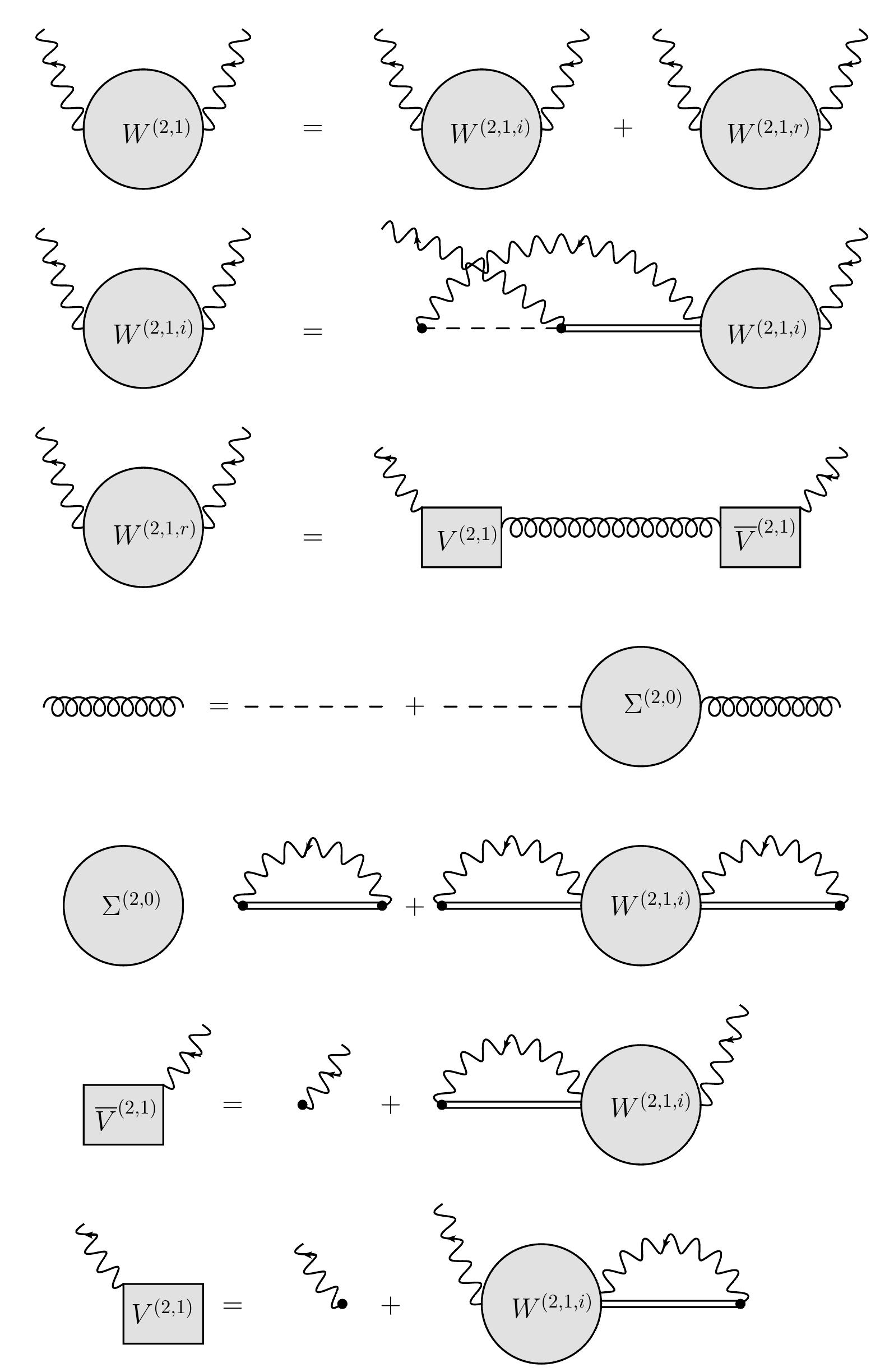}
\caption{Equations corresponding to the general two-photon scattering problem. Equation in the first line describes the splitting of the effective single-particle vertex function $W^{(2, 1)}$ in the generic two-photon scattering problem into its reducible $W^{(2, 1, r)}$ and irreducible parts $W^{(2, 1, i)}$. The irreducible part is generated from the fundamental process of the single-qubit waveguide QED $R^{(1)}$ (as shown in the second line) and thus satisfies the same integral equation as $W^{(1, 1)}$. The reducible part, steming from non-single-particle connectedness of the additional contribution to $R^{(2)}$ (see Figure \ref{fig: replacement}), is parametrized by the dressed Green's function $G^{(2, 0)}$ in two excitation subspace (shown as the curly line) as well as renormalized emission and absorption vertices (indicated by rectangles). As one can see the entire system of equations blows down to the solution of the equation defining the irreducible part of the vertex function, since its knowledge is sufficient to completely determine both the renormalized vertices $V^{(2, 1)}, \ \overline{V}^{(2, 1)}$ as well as the self energy $\Sigma^{(2, 0)}$ in the two excitation subspace.}
        \label{fig: DII1_2}
\end{figure}

As it was anticipated above, the equations governing a two-photon scattering on a pair of qubits hold in the case of a two-photon scattering for a general $N_{q}$ system. Bearing this in mind, in Figure \ref{fig: DII1_2} we present the system of exact integral equations governing the general two-photon scattering problem in waveguide QED (i.e. equations defining $W^{(2, 1)}$). These equations were first obtained in [\onlinecite{Laakso_PRL_2014}] in relation with the two-photon scattering problem on two distant qubits.
\subsubsection{$N_{q}=3$ waveguide QED}
\begin{figure}[b!]
\includegraphics[width=\columnwidth]{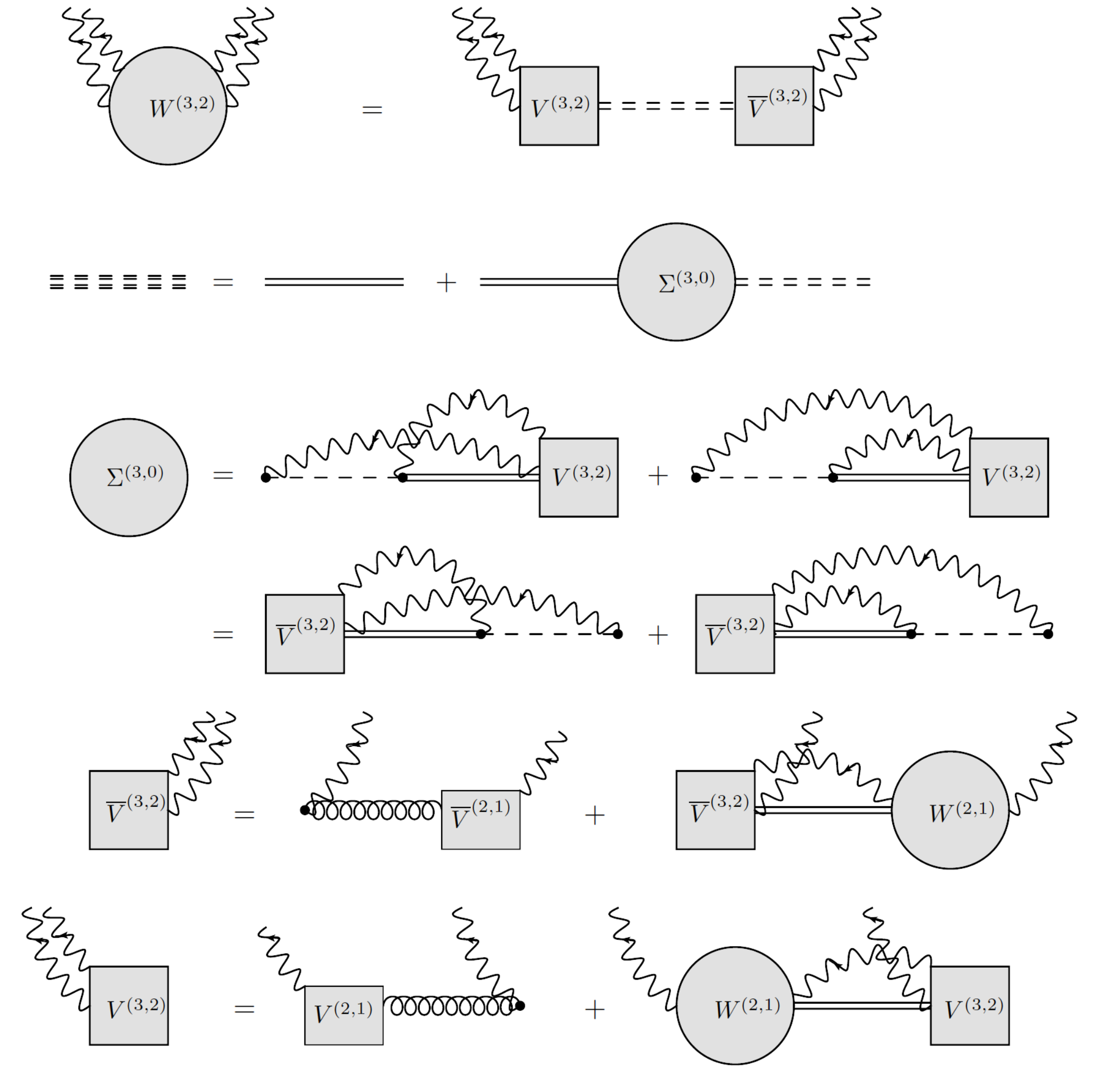}
\caption{Equations corresponding to the general three-photon scattering problem. Here the first line defines the effective two-body vertex function $W^{(3, 2)}$ in terms of the dressed Green's function in three excitation subspace shown as a double dashed line as well as effective two-photon absorption/emission vertices depicted by square boxes with a pair of incoming/outgoing amputated photon legs (note that these vertices correspond to the absorption/emission of photon pairs by a system as a whole). As it is shown in the last five lines the effective two-photon emission/absorption vertices as well as the self-energy bubble defining the Green's function in the three-excitation subspace satisfy a closed hierarchy of self-consistent equations, the validity of which is directly proven in Appendix \ref{ap: dressed}. Curly lines as well as the rectangular boxes with single amputated photon lines are precisely defined in Figure \ref{fig: DII1_2}. }
        \label{fig: DII2_3}
\end{figure}

Let us finally dedicate our attention to three-qubit waveguide QED systems. Due to the enormous increase in the computational complexity, we restrict ourselves to the first non-trivial subspace within such a setup - a three-excitation subspace. As before, the starting point of the analysis is the equation (\ref{eq: Cll1_5}). No doubt, both of the "diagonal" terms $\sim v^{\dagger}v$ and $\sim vv^{\dagger}$ give a non-zero contribution to the transition operator as it was the case in the previous section, however, one can clearly see that now the "off-diagonal" terms ($vv, \ v^{\dagger}v^{\dagger}$) have to be also taken into account. By defining the following operators
\begin{align}
\label{eq: DII2_1}
\mathcal{D}_{+}=a^{\dagger}_{s'}v_{s'}\mathcal{G}_{0}a^{\dagger}_{s}v_{s}, \quad \mathcal{D}_{-}=v_{s'}^{\dagger}a_{s'}\mathcal{G}_{0}v_{s}^{\dagger}a_{s},
\end{align}
bearing in mind the rotating wave approximation, and resuming the series (\ref{eq: Cll1_5}) to incorporate the diagonal terms and again making use of RWA, in the realm of $N_{q}=3$ waveguide QED, we obtain
\begin{align}
\nonumber
&\Braket{N_{p}', g|\mathcal{T}^{(3)}|N_{p}, g}\\
\label{eq: DII2_2}
&=\Braket{N_{p}', g|a^{\dagger}_{s_{1}'}v_{s_{1}'}\frac{1}{(\mathcal{G}^{(1)})^{-1}-\mathcal{R}^{(3)}}v^{\dagger}_{s_{1}}a_{s_{1}}|N_{p}, g},\\
\label{eq: DII2_4}
&\mathcal{R}^{(3)}=\mathcal{R}^{(2)}+\mathcal{D},\\
\nonumber
&\mathcal{D}=\mathcal{D}_{+}\frac{1}{(\mathcal{G}^{(1)})^{-1}-\mathcal{R}^{(2)}}\mathcal{D}_{-}\\
\label{eq: DII2_5}
&=\mathcal{R}^{(2)}+\mathcal{D}_{+}\mathcal{G}^{(1)}\mathcal{D}_{-}+\mathcal{D}_{+}\mathcal{G}^{(1)}\mathcal{W}^{(2)}\mathcal{G}^{(1)}\mathcal{D}_{-}.
\end{align}
As usual, we define the $\mathcal{W}^{(3)}$ operator according to 
\begin{align}
\label{eq: 3body_vertex_generator}
\mathcal{W}^{(3)}=\mathcal{D}+\mathcal{D}(\mathcal{G}^{(1)}+\mathcal{G}^{(1)}\mathcal{W}^{(2)}\mathcal{G}^{(1)})\mathcal{W}^{(3)}.
\end{align}
Which brings the transition operator to the following form
\begin{widetext}
\begin{align}
\label{eq: Tphoton_Transition_operator}
&\Braket{N_{p}', g|\mathcal{T}^{(3)}|N_{p}, g}=\Braket{N_{p}', g|\mathcal{T}^{(2)}+a^{\dagger}_{s_{1}'}v_{s_{1}'}\mathcal{G}^{(1)}(1+\mathcal{W}^{(2)}\mathcal{G}^{(1)})\mathcal{W}^{(3)}(\mathcal{G}^{(1)}\mathcal{W}^{(2)}+1)\mathcal{G}^{(1)}v^{\dagger}_{s_{1}}a_{s_{1}}|N_{p}, g}.
\end{align}
\end{widetext}
Upon the projection, onto the three-photon subspace, we obtain the following elegant set of equations governing the generic (see Section \ref{sec: Properties}) three-particle $T$-matrix (see Appendix \ref{ap: dressed} for the detailed derivation)
\begin{widetext}
\begin{align}
\nonumber
T^{(3, 3)}_{s_{1}'s_{2}'s_{3}', s_{1}s_{2}s_{3}}(\epsilon)=&\Braket{g|v_{s_{1}'}{G}^{(1)}(\epsilon-\omega_{s_{2}'}-\omega_{s_{3}'})[W^{(2, 2)}_{s_{2}'s_{3}', s_{2}s_{3}}(\epsilon)+V^{(3, 2)}_{s_{2}'s_{3}'}(\epsilon)G^{(3, 0)}(\epsilon)\overline{V}^{(3, 2)}_{s_{2}s_{3}}(\epsilon)]{G}^{(1)}(\epsilon-\omega_{s_{2}}-\omega_{s_{3}})v^{\dagger}_{s_{1}}|g}\\
=:&\Braket{g|v_{s_{1}'}{G}^{(1)}(\epsilon-\omega_{s_{2}'}-\omega_{s_{3}'})[W^{(2, 2)}_{s_{2}'s_{3}', s_{2}s_{3}}(\epsilon)+W^{(3, 2)}_{s_{2}'s_{3}', s_{2}s_{3}}(\epsilon)]{G}^{(1)}(\epsilon-\omega_{s_{2}}-\omega_{s_{3}})v^{\dagger}_{s_{1}}|g},\\
G^{(3, 0)}(\epsilon)=&G^{(1)}(\epsilon)+G^{(1)}(\epsilon)\Sigma^{(3, 0)}(\epsilon)G^{(3, 0)}(\epsilon),\\
\label{eq:3photon_self1}
\Sigma^{(3, 0)}(\epsilon)=&(v_{s}^{\dagger}G_{0}(\epsilon-\omega_{s})v_{s'}^{\dagger}+v_{s'}^{\dagger}G_{0}(\epsilon-\omega_{s'})v_{s}^{\dagger})G^{(1)}(\epsilon-\omega_{s}-\omega_{s'})V^{(3, 2)}_{ss'}(\epsilon)\\
\label{eq:3photon_self2}
\equiv&\overline{V}^{(3, 2)}_{ss'}(\epsilon)G^{(1)}(\epsilon-\omega_{s}-\omega_{s'})(v_{s'}G_{0}(\epsilon-\omega_{s})v_{s}+v_{s}G_{0}(\epsilon-\omega_{s'})v_{s'}),
\end{align}
\begin{align}
\label{eq: twophotonabsorb}
\overline{V}^{(3, 2)}_{s_{1}s_{2}}(\epsilon)=&v^{\dagger}_{s_{1}}G^{(2, 0)}(\epsilon)\overline{V}^{(2, 1)}_{s_{2}}(\epsilon)+\overline{V}^{(3, 2)}_{ss_{1}}(\epsilon)G^{(1)}(\epsilon-\omega_{s})W^{(2, 1)}_{s, s_{2}}(\epsilon),\\
\label{eq: twophotonemis}
V^{(3, 2)}_{s_{1}'s_{2}'}(\epsilon)=&{V}^{(2, 1)}_{s_{1}'}(\epsilon)G^{(2, 0)}(\epsilon)v^{\dagger}_{s_{2}'}+W^{(2, 1)}_{s_{1}', s}(\epsilon)G^{(1)}(\epsilon-\omega_{s}){V}^{(3, 2)}_{s_{2}'s}(\epsilon).
\end{align}
\end{widetext}
Here $G^{(2, 0)}(\epsilon)$ is the projection of (\ref{eq: DII1_8}) onto the photon vacuum state. As usual, the above equation may be compactly represented diagrammatically as shown in Figure \ref{fig: DII2_3}.

\section{Application: a giant acoustic atom}
\label{sec:GIANT_ATOM}

In this section, we would like to consider a practical application of the above-developed theory. In order to showcase our method in full glory, we would like to focus on the non-Markovian scattering setup - a giant acoustic atom, extensively  studied theoretically and experimentally in [\onlinecite{Andersson_Nat_Phys_2019}, \onlinecite{Guo_Phys_Rev_2017}, \onlinecite{Kockum_PRA_2014, Vadiraj_2020, Kannan_2020, Guo_2020, Cilluffo_2020}] (see also [\onlinecite{Kockum_2021}] for a detailed review).

\subsection{Model and conventions}
 A giant acoustic atom may be defined as a two-level system coupled to an acoustical waveguide, in which the radiation is carried by the surface acoustic waves (SAW), at two distant points $x=\pm R/2$. The Hamiltonian of such a system may be written as $\mathcal{H}=\mathcal{H_{0}}+\mathcal{V}$
 \begin{align}
    \mathcal{H}_{0}&=\Omega\sigma_{+}\sigma_{-}-iv\sum_{\mu=1, 2}\int{dx}c_{\mu}b^{\dagger}_{\mu}(x)\partial_{x}b_{\mu}(x),\\
     \mathcal{V}&=\sum_{\mu=1, 2}(\sqrt{\Gamma_{1}}b^{\dagger}_{\mu}(-R/2)+\sqrt{\Gamma_{2}}b^{\dagger}_{\mu}(R/2))\sigma_{-}+\text{h.c.}
 \end{align}
Here $b_{\mu}(x), b_{\mu}^{\dagger}(x)$ are the position space field operators of phonons, the index $\mu$ distinguishes beetween the right $\mu=1$ and left $\mu=2$ mooving fields, and $c_{\mu}=(-1)^{\mu-1}$. Note that we have made use of the common assumption of the mode dispersion being linear in the vicinity of the relevant energy scale $\sim\Omega$ and the the bandwidth be infinite. Defining the Fourier transformation as 
 \begin{align}
b_{\mu}(x)=\frac{1}{\sqrt{2\pi}}\int{dk}e^{ic_{\mu}kx}a_{\mu}(k),
 \end{align}
 we bring the Hamiltonian in to the form (\ref{eq: Bare_Hamiltonian}), (\ref{eq: inter}):
  \begin{align}
     \mathcal{H}_{0}&=\Omega\sigma_{+}\sigma_{-}+v\sum_{\mu=1, 2}\int{dk}ka^{\dagger}_{\mu}(k)a_{\mu}(k),\\
     \mathcal{V}&=\sum_{\mu=1, 2}\int{dk}{g}_{\mu}(k)a_{\mu}(k)\sigma_{-}+\text{h.c.},\\
     {g}_{\mu}(k)&=\sqrt{\frac{\Gamma_{1}}{2\pi}}e^{-ic_{\mu}kR/2}+\sqrt{\frac{\Gamma_{1}}{2\pi}}e^{ic_{\mu}kR/2}.
 \end{align}
 In the following subsections we are going to study the scattering of a coherent pulse in the form of a wavepacket centered around the frequency $vk_{0}$ (see Section \ref{sec:pf}), for that sake it is convenient to perform the following time-dependant gauge transformation: 
 \begin{align}
     \mathcal{U}(t)=\exp\Bigg(-ivk_{0}\Bigg[\int{dk}a^{\dagger}_{\mu}(k)a_{\mu}(k)+\sigma_{+}\sigma_{-}\Bigg]t\Bigg).
 \end{align}
 The Hamiltonian transforms as $\mathcal{H}\rightarrow\tilde{\mathcal{H}}=\mathcal{U}^{\dagger}(t)\mathcal{H}\mathcal{U}(t)-i\mathcal{U}^{\dagger}(t)\frac{d\mathcal{U}(t)}{dt}=:\tilde{\mathcal{H}}_{0}+\tilde{\mathcal{V}}$, resulting in final form of the Hamiltonian we are going to work with:
   \begin{align}
     \tilde{\mathcal{H}}_{0}&=-\Delta\sigma_{+}\sigma_{-}+v\sum_{\mu=1, 2}\int{dk}k\tilde{a}^{\dagger}_{\mu}(k)\tilde{a}_{\mu}(k),\\
     \tilde{\mathcal{V}}&=\sum_{\mu=1, 2}\int{dk}\tilde{g}_{\mu}(k)\tilde{a}_{\mu}(k)\sigma_{-}+\text{h.c.},\\
     \tilde{g}_{\mu}(k)&=\sqrt{\frac{\Gamma_{1}}{2\pi}}e^{-ic_{\mu}(k+k_{0})R/2}+\sqrt{\frac{\Gamma_{1}}{2\pi}}e^{ic_{\mu}(k+k_{0})R/2},\\
     \tilde{a}_{\mu}(k)&=a_{\mu}(k+k_{0}), \quad \Delta=\omega_{0}-\Omega,\quad \omega_{0}=vk_{0}.
 \end{align}
 In the following we adopt the system of units such that $v=1$, for simplicity we will also assume $\Gamma_{1}=\Gamma_{2}=\gamma/2$. Furthermore, we shall also drop the tilde symbols out of operators for notational convenience.

\subsection{Problem formulation}
\label{sec:pf}
In order to formulate the scattering problem, we define the following wave-packet operators
\begin{align}
A_{\mu}^{\dagger}=\int {dk}\varphi_{L}(k)a_{\mu}^{\dagger}(k),
\end{align}
where $\varphi_{L}(k)=\sqrt{\frac{2}{\pi L}}\frac{\sin(kL/2)}{k}$ is the Fourier transform of the rectangular pulse of length $L$, with the property $\varphi_{L}(k)\sim\sqrt{\frac{2\pi}{L}}\delta(k), \ L\rightarrow\infty$ (obviously, it is possible to choose any other nascent $\delta$-function for $\varphi_{L}(k)$). Note that since we are working in the frame of reference rotating with frequency $k_{0}$, peaking of $\varphi_{L}(k)$ at $k=0$ corresponds to a pulse centered at $k=k_{0}$ in the original one. Note that throughout this section we focus on the zero detuning setup $\Delta=0$, i.e. the initial $k_{0}$ is chosen to be equal to $\Omega$. The effect of non-zero detuning of atom from radiation is studied in Appendix \ref{ap:detun}. Definition of these wave-packet operators is crucial since the eigenstates of the bare Hamiltonian $\mathcal{H}_{0}$ are not normalizable. Hence, in order to avoid the ascent of the undefinable quantities such as $\delta(0), (\delta(0))^{2}, ...$ coming from the elastic clusters of the $S$-matrix in the calculation of the observables, one has to work with the normalizable states and take the plane-wave limit $L\rightarrow\infty$ at the very end. Having defined the Fock creation operators $A^{\dagger}_{\mu}$, we can define the coherent state as a displaced vacuum
\begin{align}
    \label{eq:Coherent_State}
    \ket{\alpha}_{\mu}=e^{-|\alpha|^{2}/2}e^{\alpha A^{\dagger}_{\mu}}\ket{\Omega}=e^{-|\alpha|^{2}/2}\sum_{n=0}^{\infty}\frac{\alpha^{n}}{\sqrt{n!}}\ket{\Phi^{(n)}_{\mu}},
\end{align}
where $\alpha\in\mathbb{C}$ is the so-called coherence parameter, defined such that $|\alpha|^{2}$ is the average photon number (power), and the normalized $n$-particle Fock states $\ket{\Phi^{(n)}_{\mu}}$ were defined according to
\begin{equation}
\label{eq:Fock_State}
\ket{\Phi^{(n)}_{\mu}}=\frac{1}{\sqrt{n!}}\Big(A^{\dagger}_{\mu}\Big)^{n}\ket{\Omega}.
\end{equation}
In what follows, we assume that $|\alpha|\ll1$ in such a way that the terms of order $|\alpha|^{4}$ are negligible
\begin{align}
\nonumber
\ket{\alpha}_{\mu}\approx&{e}^{-|\alpha|^{2}/2}\Bigg(\ket{\Omega}+\frac{\alpha}{\sqrt{1!}}\ket{\Phi^{(1)}_{\mu}}+\frac{\alpha^{2}}{\sqrt{2!}}\ket{\Phi^{(2)}_{\mu}}\\
\label{eq:Weak_Coherent_State}
+&\frac{\alpha^{3}}{\sqrt{3!}}\ket{\Phi^{(3)}_{\mu}}+\mathcal{O}(|\alpha|^{4})\Bigg)\equiv\ket{\alpha}_{\mu}^{(3)}.
\end{align}
The factor $e^{-|\alpha|^{2}/2}$ has to be retained until various overlaps of states are calculated, to ensure the normalization of both the initial and final states as well as the power conservation (here the normalization is again assumed in the power perturbative regime, that is up to order $\mathcal{O}(|\alpha|^{8})$). With this in hands, we formulate the scattering problem as follows. We assume that the initial state of the system is given by $\ket{\psi_{i}}=\ket{\alpha}_{1}^{(3)}\otimes\ket{0}$, that is a leftwards-propagating weakly coherent pulse $\ket{\alpha}_{1}^{(3)}$ is incident on a two-level system which is initially in its ground state $\ket{0}$. According to the theory developed above, the final state of the systems has the following form
\begin{align}
\nonumber
\ket{\psi_{f}}=&{e}^{-|\alpha|^{2}/2}\Bigg(\ket{\Omega}+\frac{\alpha}{\sqrt{1!}}{\mathcal{S}}_{1}\ket{\Phi^{(1)}_{1}}\\
\label{eq:Weak_Coherent_FState}
+&\frac{\alpha^{2}}{\sqrt{2!}}{\mathcal{S}}_{2}\ket{\Phi^{(2)}_{1}}+\frac{\alpha^{3}}{\sqrt{3!}}{\mathcal{S}}_{3}\ket{\Phi^{(3)}_{1}}\Bigg)\otimes\ket{0}.
\end{align}
\par
Inelastic contributions to the $n$-phonon $S$-matrices (discussed in Section \ref{sec:GCD}) introduce a non-trivial momentum redistribution of the incident phonons (note that owing to the linear dispersion relation and our choice of units energy and momentum may be used interchangeably), leading to the non-trivial phonon correlations in the final state. As it is well known, the phononic correlations may be conveniently examined with the help of the so-called coherence functions introduced by Glauber \cite{Glauber_PR_1963} in 1963. Defining the Fourier transform of the field operators according to
\begin{align}
\label{eq:Fourier_Operators}
a_{\mu}(\tau)=\int{dk}\frac{e^{ik\tau}}{\sqrt{2\pi}}a(k),
\end{align}
one constructs the first-order coherence function as
\begin{widetext}
\begin{align}
\nonumber
C_{\mu}^{(1)}(\tau)=&\braket{\psi_{f}|a^{\dagger}_{\mu}(\tau)a_{\mu}(0)|\psi_{f}}\\
\label{eq:1st_order_coherence}
=&\Bigg(1-|\alpha|^{2}+\frac{|\alpha|^{4}}{2}\Bigg)|\alpha|^{2}C^{(1, 1)}_{\mu}(\tau)+(1-|\alpha|^{2})\frac{|\alpha|^{4}}{2}C^{(1, 2)}_{\mu}(\tau)+\frac{|\alpha|^{6}}{6}C^{(1, 3)}_{\mu}(\tau)+\mathcal{O}(|\alpha|^{8}),
\end{align}
\end{widetext}
where we have introduced the $n$-particle Fock state first-order correlation functions as 
\begin{equation}
\label{eq:1st_order_coh_Fock}
C^{(1, n)}_{\mu}(\tau)=\Braket{\Phi^{(n)}_{1}|(\mathcal{S}_{n})^{\dagger}a^{\dagger}_{\mu}(\tau)a_{\mu}(0)\mathcal{S}_{n}|\Phi_{1}^{(n)}}.
\end{equation}
With the help of the first-order coherence function, one may define the spectral power density as the Fourier transform of (\ref{eq:1st_order_coherence}):
\begin{equation}
\label{eq:Spectral_Density_Def}
S_{\mu}(k)=\int d\tau\frac{e^{-ik\tau}}{2\pi}C^{(1)}_{\mu}(\tau).
\end{equation}
$S_{\mu}(k)$ is understood as a momentum space distribution of power in the scattered state of radiation, that is to a given mode $k$ (supported by the $\mu^{th}$ channel) it associates a certain power $S_{\mu}(k)$. The power conservation condition 
\begin{equation}
\label{eq:Power_Conservation}
\sum_{\mu=1, 2}\int dkS_{\mu}(k)\rightarrow\Phi, \quad \Phi=\frac{|\alpha|^{2}}{L},
\end{equation}
is automatically satisfied due to the unitarity of the $S$-matrix. In a linear system, where the $n$ body $S$-matrix factorises into the product of single-particle $S$-matrices, the first order coherence function is a constant, thus leading to the purely elastic power density $S_{\mu}(k)\propto\delta(k)$. Since a qubit is an intrinsically non-linear system, we are going to see that the spectral power density admits for the following decomposition $S(k)=S^{\text{el}}(k)+S^{\text{inel}}(k)$, where $S^{\text{el}}(k)\propto\delta(k)$ is the elastic contribution to the spectral density, and $S^{\text{inel}}(k)$, in turn, is the inelastic part of spectral power density with a non-trivial momentum dependence.
\par
Further, we define the normalized second and third-order coherence functions:
\begin{widetext}
\begin{align}
\label{eq:2d_coh_fun_def}
C^{(2)}_{\mu, \mu'}(\tau)=&\frac{\braket{\psi_{f}|a^{\dagger}_{\mu}(0)a^{\dagger}_{\mu'}(\tau)a_{\mu'}(\tau)a_{\mu}(0)|\psi_{f}}}{C^{(1)}_{\mu}(0)C^{(1)}_{\mu'}(0)}=\Bigg(1-\frac{|\alpha|^{2}}{2}\Bigg)\frac{1}{2}C^{(2, 2)}_{\mu, \mu'}(\tau)+\frac{|\alpha|^{2}}{6}C^{(2, 3)}_{\mu, \mu'}(\tau)+\mathcal{O}(|\alpha|^{4}),\\
\label{eq:3d_coh_fun_def}
C^{(3)}_{\mu, \mu', \mu''}(\tau, \ \tau')=&\frac{\braket{\psi_{f}|a^{\dagger}_{\mu}(0)a^{\dagger}_{\mu'}(\tau)a^{\dagger}_{\mu''}(\tau')a_{\mu''}(\tau')a_{\mu'}(\tau)a_{\mu}(0)|\psi_{f}}}{C^{(1)}_{\mu}(0)C^{(1)}_{\mu'}(0)C^{(1)}_{\mu''}(0)}=\frac{1}{6}C^{(3, 3)}_{\mu, \mu', \mu''}(\tau, \ \tau')+\mathcal{O}(|\alpha|^{2}),\\
C^{(m, n)}_{\mu_{1}, ..., \mu_{m}}(\tau_{1}, .., \tau_{m-1})=&\frac{\braket{\Phi^{(n)}_{1}|{\mathcal{S}}_{n}^{\dagger}a^{\dagger}_{\mu_{1}}(0)...a^{\dagger}_{\mu_{m}}(\tau_{m-1})a_{\mu_{m}}(\tau_{m-1})...a_{\mu_{1}}(0){\mathcal{S}}_{n}|\Phi^{(n)}_{1}}}{\prod_{l=1}^{m}C^{(1)}_{\mu_{l}}(0)}.
\end{align}
\end{widetext}
For pairs of phonons, the normalized second-order coherence function is defined as the arrival probability of the second particle as a function of the delay $\tau$ following the detection of the first one, normalized by the individual photon probabilities. Likewise, for particle triples, the normalized third-order coherence function is defined as the arrival probability of the third and second particle as a function of delays $\tau', \tau$ following the detection of the first one, normalized by the individual phonon probabilities. A perfectly coherent source is characterized by a uniform arrival probability, yielding the correlation functions of all orders equal to unity. Whenever correlation functions exceed unity, particle statistics is said to be super-Poissonian and the particles are said to be bunched together, whereas in the case of correlation functions falling below unity, the statistics of particles is said to be sub-Poissonian and particles are correspondingly said to be anti-bunched. 
\par
 In the following subsections, we are going to use the ideas developed in Section \ref{sec:Q=1} in order to compute the above-discussed observables to the lowest order in $|\alpha|$ exactly.

\subsection{Spectral power density}
\label{SPd}

Let us begin with the analysis of the power spectrum of the SAW scattered by the giant atom. We start by considering the Fock state first-order coherence functions defined via (\ref{eq:1st_order_coh_Fock}). In order to establish $C^{(1, 1)}_{\mu}(\tau)$ one first has to find the matrix elements of the single-particle $S$-matrix, which, in turn, demands the knowledge of the dressed propagator in the single-excitation subspace $G^{(1)}(\epsilon)$. The self-energy diagram may be evaluated straightforwardly to yield:
\begin{equation}
\label{eq:Self_Energy}
\Sigma^{(1)}(\epsilon)=-i\gamma(1+e^{i(\epsilon+k_{0})R})\sigma_{+}\sigma_{-},
\end{equation}
in accordance with reference [\onlinecite{Guo_Phys_Rev_2017}]. With this in hands, the single-phonon scattering operator may be easily established with the help of equations (\ref{eq:T11}), (\ref{eq:Sn}) and (\ref{eq:Sn_Matrix}):
\begin{align}
\label{eq:1Photon_S_operator}
\mathcal{S}_{1}=&\sum_{\mu', \mu}\int{dkdk'}S^{(1)}_{\mu' k', \mu k}a^{\dagger}_{\mu'}(k')a_{\mu}(k),\\
\label{eq:comp}
S^{(1)}_{\mu' k', \mu k}=&\delta(k-k')S^{(1)}_{\mu', \mu}(k)\sigma_{-}\sigma_{+}\\
\label{eq:1Photon_S_on-shell}
S^{(1)}_{\mu', \mu}(k)=&(\delta_{\mu, \mu'}-2\pi{i}g_{\mu}^{*}(k)g_{\mu'}(k)\tilde{G}^{(1)}(k)).
\end{align}
Here the tilde symbol denotes the projection onto the excited state in the qubit space, and ${G}^{(1)}(\epsilon)={G}_{0}^{-1}(\epsilon)-{\Sigma}^{(1)}(\epsilon)$ in accordance with the definition in Section \ref{sec:Q=1}. Performing a straightforward calculation, we obtain
\begin{align}
\nonumber
C^{(1)}_{\mu}(\tau)=&|\alpha|^{2}C^{(1, 1)}_{\mu}(\tau)+\mathcal{O}(|\alpha|^{4})\\
\label{eq:1Photon_C1}
=&\Phi|S^{(1)}_{\mu0, 10}|^{2}+\mathcal{O}(\Phi^{2})\\
\label{eq:1Photon_S1}
\implies& S_{\mu}(k)=\Phi|S^{(1)}_{\mu0, 10}|^{2}\delta(k)+\mathcal{O}(\Phi^{2}).
\end{align}
As we can see there exists no inelastic contribution to the spectral density in single-photon sector. In general the phenomenon of resonance fluorescence [\onlinecite{Mollow_1969}], leading to the inelastic power spectrum, is underpinned by the possibility of the two-level system to emit into the modes other than the incident one. This naturally allows the photons incident on the atom to exchange their energy between one another (whilst conserving the total energy) due to the higher order emission and absorption processes, leading to the inelastic power spectrum. In the case of the single-photon however, the particle is ought to conserve its energy individually (as mathematically prescribed by the $\delta$-function in equation (\ref{eq:comp})) leading to the purely elastic spectrum.
\par
In order to obtain the leading order inelastic contribution to the spectral power density, we thus have to consider the contribution of the multi-particle states in the expansion (\ref{eq:Weak_Coherent_FState}), so that to allow for the inter-particle interaction. In this case, the information about the inelastic scattering is entirely contained in the connected component of the two-phonon $T$-matrix (see Section \ref{sec:GCD}), which captures the nonlinear acoustic effects via the effective single-particle vertex function. Contracting the cluster decomposed $S$-matrix (\ref{eq: CII2_5}) with the two-particle Fock state we arrive at the following result
\begin{widetext}
\begin{align}
\label{eq: final_two_particle_state}
\sum_{\mu, \mu'}\Bigg(\frac{1}{\sqrt{2}}\int{dkdk'}\varphi(k)\varphi(k')S^{(1)}_{\mu, 1}(k)S^{(1)}_{\mu', 1}(k'){a}_{\mu}^{\dagger}(k)a_{\mu'}^{\dagger}(k')-\frac{8\pi^{2}i}{L\sqrt{2}}\int{dk}T^{(2, C)}_{\mu k, \mu'-k, 10, 10}(0){a}_{\mu}^{\dagger}(k)a_{\mu'}^{\dagger}(-k)\Bigg)\ket{\Omega}\otimes\ket{0}.
\end{align}
\end{widetext}
Again we would like to emphasize that at this point the plane-wave limit ($L\rightarrow\infty$) may only be taken in the term containing the connected component of the $S$-matrix since in this limit the first part of the state (\ref{eq: final_two_particle_state}) is clearly non-normalizable. With the help of the final two-particle state derived above one may easily establish
\begin{align}
\nonumber
C^{(1, 2)}_{\mu}(\tau)=&16\pi^{2}\sum_{\mu'}\text{Im}\{M_{\mu', \mu}(0)(S_{\mu'; 1}^{(1)}(0)S_{\mu; 1}^{(1)}(0))^{*}\}\\
\label{eq: first_order_coherence_2photon}
+&32\pi^{3}\sum_{\mu'}\int {dk}e^{ik\tau}|M_{\mu', \mu}(k)|^{2},
\end{align}
where $M_{\mu', \mu}(k)$ is the symmetric part of $T^{(2, C)}_{\mu k, \mu'-k, 10, 10}(0)$, i.e. $M_{\mu', \mu}(k)=M_{\mu, \mu'}(-k)=(T^{(2, C)}_{\mu' k, \mu-k, 10, 10}(0)+T^{(2, C)}_{\mu -k, \mu'k, 10, 10}(0))/2$. Performing the Fourier transform of (\ref{eq: first_order_coherence_2photon}) we arrive at the following expressions for the elastic $S^{\text{el}}(k)$ and inelastic $S^{\text{inel}}(k)$ spectral power densities valid to order $\Phi^{3}$:
\begin{align}
\nonumber
S_{\mu}^{\text{el}}(k)=&\delta(k)\Bigg(\Phi|S^{(1)}_{\mu; 1}(0)|^{2}+16\pi^{2}\Phi^{2}\\
\label{eq: Elastic_SPD}
\times&\sum_{\mu'}\text{Im}\{M_{\mu', \mu}(0)(S_{\mu'; 1}^{(1)}(0)S_{\mu; 1}^{(1)}(0))^{*}\}\Bigg),\\
\label{eq: Inelastic_SPD}
S_{\mu}^{\text{inel}}(k)=&32\pi^{3}\Phi^{2}\sum_{\mu'}|M_{\mu', \mu}(k)|^{2}.
\end{align}

\begin{figure}[t]
\includegraphics[width=\columnwidth]{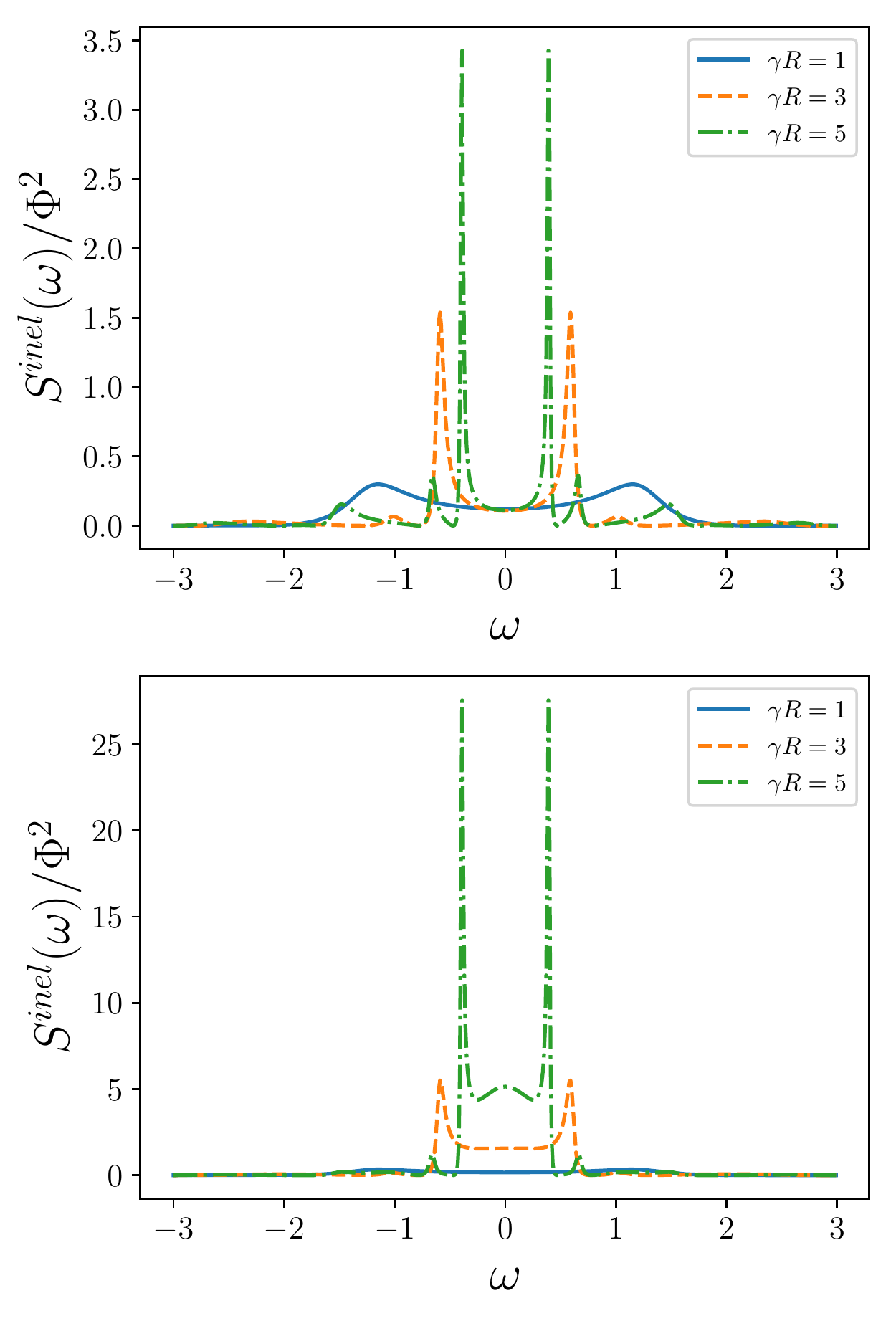}
\caption{Inelastic spectral power density (scaled by $\Phi^{2}$) of SAW scattered by a giant acoustic atom as a function of frequency $\omega=k$ for a variety of inter-leg separations $\gamma R=1, 3, 5$. Top panel: exact solution; Bottom panel: quasi-Markovian approximation. Other model parameters: $k_{0}R=\pi/4 \mod2\pi$, and $\Delta=0$. Here the limit $L\rightarrow\infty$ was taken.}
        \label{fig: Spectral_Density}
\end{figure}

The inelastic spectral densities are shown in Figure \ref{fig: Spectral_Density} for different inter-leg separations (R = 1, 3, 5). As one can see, the inelastic power spectrum develops a couple of sharp peaks near the origin of momentum space followed by infinitely many smaller side peaks. In order to interpret the nature of these peaks, we adopt the physical picture discussed in [\onlinecite{Chang_NJP_2012, Guimond_PRA_2016, Guimond_QST_2017}] for an "atom in front of a mirror" system. One can interpret this system as a leaky cavity formed by the two connection points of a giant atom. In this picture these peaks may be thought of as being located at the renormalized excitation frequencies of the effective cavity broadened by the renormalized decay rates. The resulting effect is the formation of sharp, bound-state like peaks in the intensity of the scattered phonons, corresponding to cavity resonances. As the delay $\gamma R$ increases, the peaks get closer to $\omega=0$, and get sharper, corresponding to a decrease of the effective cavity linewidth $\simeq1/(\gamma R)$, that is effective cavity resonances approach the resonance of the two-level system, thus making atomic connection points better and better mirrors and hence increasing the quality factor of the effective cavity (quality factor is a common physical measure of both the rate of excitation damping and the rate of energy loss of an oscillator or a cavity)

\par
Additionally, spectral densities based on the approximate solution of the scattering problem in the quasi-Markovian approximation are shown in Figure \ref{fig: Spectral_Density}. As it was mentioned in the section \ref{sec: AS}, quasi-Markovian results show a good agreement with an exact solution in $\gamma{R}\ll1$ parameter regime. As the delay time increases $\gamma{R}\geq1$, the discrepancy between the approximate and complete solutions becomes dramatic. One can clearly see the tendency of the quasi-Markovian theory to enhance the scattering into the incoming frequency states $\omega=0$. Indeed, in contrast to the exact solution, taking into account infinitely many excursions of phonons to the qubit,  the quasi-Markovian approximation is based on the assumption of a single scattering event after which each phonon immediately leaves the system, thus leading to a more elastic result.

\subsubsection{Second-order coherence}
\label{2d_order_coh}

\begin{figure*}[t]
\includegraphics[width=2\columnwidth]{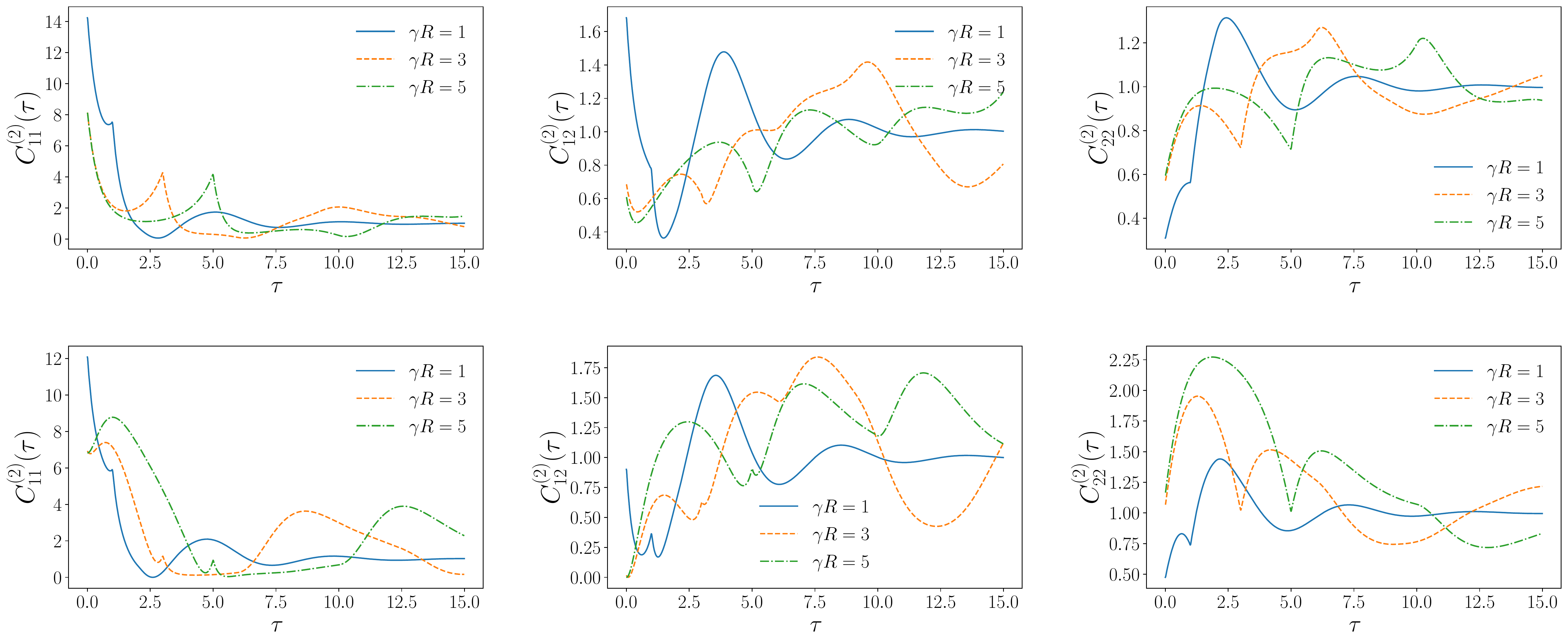} 
\caption{The figure demonstrates the independent components of the normalized second order coherence function $C^{(2)}_{11}, \ C^{(2)}_{12}, \ C^{(2)}_{22}$ for a various inter-leg separations $\gamma R=1, 3, 5$. As before the top panel corresponds to an exact solution, the bottom one to its quasi-Markovian approximation, and $k_{0}R=\pi/4 \mod2\pi$, $\Delta=0$. Here the limit $L\rightarrow\infty$ was taken and by definition $C^{(2)}$ is dimensionless.}
        \label{fig: 2d_Order_Coh}
\end{figure*}

Let us now consider the second-order coherence function to the lowest order in $\Phi$ - $\mathcal{O}(\Phi^{0})$. Performing the relevant Wick contractions we arrive at the following simple expression for the normalized second-order coherence 
\begin{align}
\label{eq: second_order_coherence}
C_{\mu', \mu}^{(2)}(\tau)
=\Bigg{|}1-\frac{4\pi{i}}{S^{(1)}_{\mu', 1}(0)S^{(1)}_{\mu, 1}(0)}\int{dk}e^{ik\tau}M_{\mu', \mu}(k)\Bigg{|}^{2}.
\end{align}
The independent components of second-order coherence function's components are presented in Figure \ref{fig: 2d_Order_Coh}.
\par
As it was mentioned above, the higher-order coherence functions provide the information about the information about the statistics of radiation scattered by a giant acoustic atom. Since $C^{(2)}_{1, 1}(0)>1$, one can clearly see that the statistics of back-scattered phonons is super-Posissonian, i.e. the particles tend to bunch together. Indeed, this result agrees with the physical expectation that at zero detuning $\Delta=0$ the power extinction $1-|S_{11}(0)|^{2}$ is enhanced even in the presence of pure dephasing\cite{Astafiev_Sci_2010, Kimble_PRA_1976}. This assertion also explains the anti-bunching of forwardly scattered photons $C^{(2)}_{2, 2}(0)$. Another clear feature of the second-order coherence function shown in Figure \ref{fig: 2d_Order_Coh} is presence of long-range quantum correlations of phonons, i.e. the components of $C^{(2)}$ do not decay to unity even for delays significantly exceeding the inter-leg separation $\tau\gg1$. Note that this correlation effect becomes more and more pronounced with the increase in $R$, where correlation functions exhibit slightly damped oscillations around unity.
\par
Another interesting feature of the second-order coherence is the presence of the non-differentiable peaks at natural multiples of the inter-leg separation $\tau_{n}=nR, \ n\in\mathbb{N}$. Physically, this property may again be understood with the help of the simple picture of a leaky cavity formed by the scatterer. Indeed, once a phonon is trapped between the legs of the atom, it may bounce off the cavity's walls back and forth multiple times, leading to the formation of the non-analytical structures present in the second-order coherence. The fact that the non-differentiable peaks are more pronounced at smaller values of $n$ is a direct manifestation of the fact that the quality factor of the effective cavity is not infinite. This property is indeed of interest since the second-order coherence function is an experimentally measurable quantity which makes these sharp peaks a potentially observable effect.
\par
Alongside the exact solution, the results based on the quasi-Markovian approximation are presented in Figure \ref{fig: 2d_Order_Coh} (lower panel). The discrepancy between the two is apparent. Another feature typical of this approximation is the tendency to overestimate the amplitudes of oscillation as was first pointed out in the supplemental material of reference [\onlinecite{Laakso_PRL_2014}].

\subsubsection{Third-order coherence}
\label{3d_order_coh}
\begin{figure*}[t]
\includegraphics[width=1.99\columnwidth]{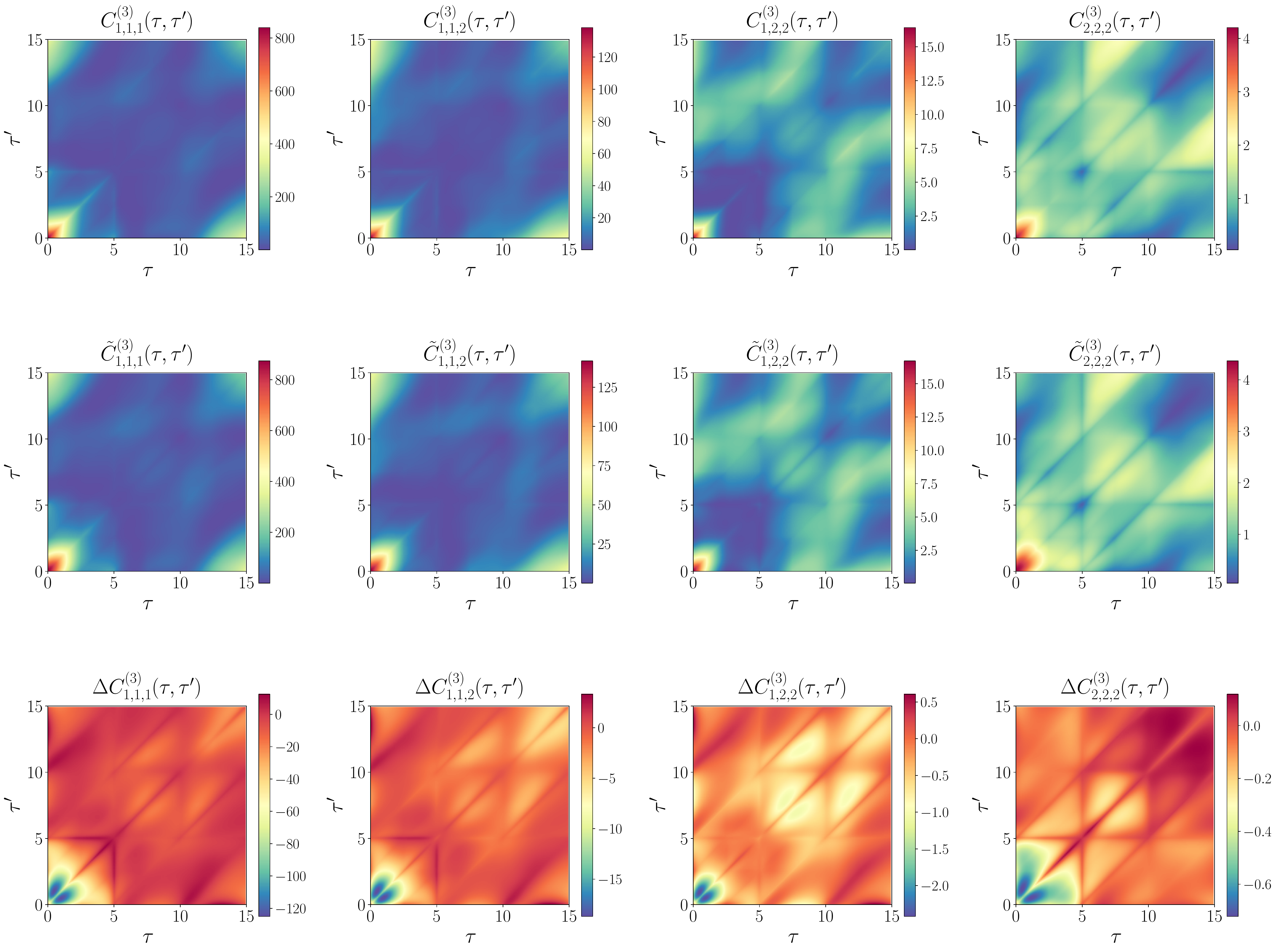} 
\caption{Independent components $(\mu, \mu', \mu'')=(1, 1, 1),\ (1, 1, 2),\ (1, 2, 2),\ (2, 2, 2)$ of the third-order coherence function of phonons scattered by a giant acoustic atom. Top, central, and bottom panels correspond to the exact solution $C^{(3)}_{\mu, \mu', \mu''}(\tau, \tau')$, weak correlation approximation $\tilde{C}^{(3)}_{\mu, \mu', \mu''}(\tau, \tau')$, and their difference $\Delta{C^{(3)}_{\mu, \mu', \mu''}(\tau, \tau')}=C^{(3)}_{\mu, \mu', \mu''}(\tau, \tau')-\tilde{C}^{(3)}_{\mu, \mu', \mu''}(\tau, \tau')$ respectively. Here $\gamma R=5$, $k_{0}R=\pi/4 \mod2\pi$, and $\Delta=0$. Here the limit $L\rightarrow\infty$ was taken and by definition $C^{(2)}$ is dimensionless.}
        \label{fig: 3d_Order_Coh_res}
\end{figure*}
Let us finally consider the third-order coherence function. The full expression for the third-order coherence function $C^{(3)}_{\mu, \mu', \mu''}(\tau, \tau')$ in terms of the symmetrized components of two and three-particle transition matrices may be found in Appendix \ref{ap: 3ocf_exp}. Although in general, $C^{(3)}_{\mu, \mu', \mu''}$ has $8$ independent components in the case of two radiation channels, due to our particular choice of the coupling $\Gamma_{1}=\Gamma_{2}=\gamma/2$, only four independent components remain:
\begin{align}
&C^{(3)}_{1, 1, 1}(\tau, \tau'), \quad C^{(3)}_{2, 2, 2}(\tau, \tau'),\\
&C^{(3)}_{1, 1, 2}(\tau, \tau')=C^{(3)}_{1, 2, 1}(\tau, \tau')=C^{(3)}_{2, 1, 1}(\tau, \tau'),\\
&C^{(3)}_{1, 2, 2}(\tau, \tau')=C^{(3)}_{2, 2, 1}(\tau, \tau')=C^{(3)}_{2, 1, 2}(\tau, \tau').
\end{align}
\par
Independent components of $C^{(3)}$ for a system with $R=5$, $k_{0}R=\pi/4 \mod2\pi$, and $\Delta=0$ are shown in the top panel of Figure \ref{fig: 3d_Order_Coh_res}. First, we note that the individual components of third-order coherence at $\tau, \tau'=0$ significantly exceed unity, signifying the bunching of phonons. This effect can be attributed to the fact that a single two-level system can only emit and absorb a single quantum of radiation at a time, which, in turn, significantly increases the probability of simultaneous detection of a pair of phonons in the waveguide. Moreover, this effect is more pronounced in those components of the correlation function, describing correlations with phonons in the channel of incident radiation $\mu=1$. This artifact again has to do with the fact of the enhancement of power extinction by an atom at zero detuning. 
\par
Another interesting feature of $C^{(3)}$ to be noticed is the presence of clear peak and downfall structures located on the lines $\tau'-\tau=nR, \ n\in\mathbb{Z}$. Physically $\delta\tau=\tau'-\tau$ corresponds to the average delay time between the second and third phonon detection events upon detection of the first one at zero time. The quantization of $\delta\tau$ in the units of deterministic time delay is the characteristic feature of the system under consideration and may be potentially observed in future experiments via the observation of enhancement/diminution of the conditional probability of arrival of the third particle. In fact, the peak and downfall structures discussed above are non-differentiable, as it was the case with the second-order coherence function, and again this phenomenon may be understood with the help of a simple picture of an effective cavity discussed in Sections \ref{2d_order_coh} and \ref{SPd}.  
\par
Beside the exact solution ${C}^{(3)}_{\mu, \mu', \mu''}(\tau, \tau')$, the solution based on the weak-correlation (WC) approximation $\tilde{C}^{(3)}_{\mu, \mu', \mu''}(\tau, \tau')$ as well as the difference between the exact solution and the WC one
\begin{align}
\label{eq: difference}
\Delta{C}^{(3)}_{\mu, \mu', \mu''}={C}^{(3)}_{\mu, \mu', \mu''}-\tilde{C}^{(3)}_{\mu, \mu', \mu''},
\end{align}
are shown in Figure \ref{fig: 3d_Order_Coh_res} in central and bottom pannels respectively. As one can anticipate, the WC approximation tends to notably overestimate the amplitude of the third-order coherence function. This effect is especially apparent in the vicinity of $\tau, \tau'=0$, where the WC approximation significantly overestimates the phononic bunching. Away from the temporal origin, though, the approximation becomes adequate and only slightly deviates from the exact result, as one may infer from the plots in the bottom panel of Figure \ref{fig: 3d_Order_Coh_res}. The discrepancy between the exact solution and its WC approximation is not hard to understand. Weak-correlation approximation ignores an infinite diagrammatic channel, consisting of exchange interaction diagrams between the phonons, which are, of course, of importance when one studies the statistical properties of particles.

\section{Conclusions and outlook}
In this paper, the diagrammatic theory of scattering and dynamics of multi-photon states in waveguide QED was developed. In particular, it was shown that the $N_{p}$-photon scattering matrices in single-qubit waveguide QED may be conveniently parametrized in terms of effective $N_{p}-1$-photon vertex functions and the equations satisfied by these vertex functions were established. Next, certain practical issues related to the direct sum representation of the $S$-matrix, separation of elastic contributions to effective vertices, as well as the generalized cluster decomposition were discussed. Further, a generalization to the waveguide QED systems with more than a single qubit was given. Specifically, in the case of the two-qubit systems, it was established that the equations governing multi-photon vertex functions remain the same as in the case of a single qubit, up to the inclusion of higher-order vertex corrections. Moreover, we have shown that once the integral equations governing $N_{q}$-photon scattering matrix in $N_{q}$ waveguide QED, these equations hold for any system of qubits and established the generic equations governing $2$ and $3$ photon scattering operators by considering $2$ and $3$ photon scattering on two and three qubits respectively.
\par
Next, the diagrammatic theory of scattering was applied to a problem of scattering of a weakly coherent pulse on the giant acoustic atom. Namely, by expanding a coherent state perturbatively in a coherence parameter up to third order in $|\alpha|$ and studying its scattering on the atom, we were able to establish the first, second, and third-order coherence functions of scattered radiation. Moreover, a set of approximation routines was suggested along with the exact method and the two were compared where appropriate. Further, the statistical properties of scattered surface acoustic waves were studied, and the effect of the non-Markovian nature of the setup on statistics was discussed.
\par
In our future work, we are going to present the generalization of the resummation approach enabling one to study real-time dynamics in waveguide QED systems. It would be of further interest to extend the present theory to study scattering in waveguide QED systems containing emitters with more complicated selection rules such as three and four-level systems. Another generalization of the theory presented in this paper of potential future interest is the study of particular 2 and 3-particle scattering problems on systems containing multiple distant qubits. Furthermore, it would be potentially interesting to assess the effects of counter-rotating terms, which were ignored throughout this work, which, however, will require the use of techniques other than the one discussed in this work. 
\section*{Acknowledgments}
We thank H. Schoeller for helpful instructions on diagrammatic methods in field theory applications and, in particular, for pointing to us the distributional Poincare-Bertrand identity. KP is grateful to A. Samson for enlightening discussions on numerical methods used throughout the paper. MP acknowledges the durable exchange of ideas on the subject of study with V. Gritsev and V. Yudson. This work was supported by the Deutsche Forschungsgemeinschaft (DFG) via the contract RTG 1995.

\begin{appendix}
\section{Cluster decomposition of three-photon $S$-matrix}
\label{3bodyScat}

The starting point of our analysis in this appendix goes back to the definition of the three-body component of the $S$-matrix entering its direct sum representation (throughout this appendix, for simplicity, it is assumed that $\omega_{\mu}(k)=\omega(k), \quad B_{\mu}=B, \quad \forall\mu\in\{1, ..., N_{c}\}$), namely 
\begin{widetext}
\begin{align}
\nonumber
\mathcal{S}_{3}=&\Bigg[\frac{1}{3!}\delta_{s_{1}', s_{1}}\delta_{s_{2}', s_{2}}\delta_{s_{3}', s_{3}}-2\pi{i}T^{(1, 1)}_{s_{1}', s_{1}}(\omega(k_{1}))\delta(\omega(k_{1}')-\omega(k_{1}))\frac{1}{2!}\delta_{s_{2}', s_{2}}\delta_{s_{3}', s_{3}}\\
\nonumber
&-2\pi{i}T^{(1, 2)}_{s_{1}'s_{2}', s_{1}s_{2}}(\omega(k_{1})+\omega(k_{2}))\delta(\omega(k_{1}')+\omega(k_{2}')-\omega(k_{1})-\omega(k_{2}))\frac{1}{1!}\delta_{s_{3}', s_{3}}\\
\nonumber
&- 2\pi{i}T^{(1, 3)}_{s_{1}'s_{2}'s_{3}', s_{1}s_{2}s_{3}}(\omega(k_{1})+\omega(k_{2})+\omega(k_{3}))\delta(\omega(k_{1}')+\omega(k_{2}')+\omega(k_{3}')-\omega(k_{1})-\omega(k_{2})-\omega(k_{3}))\Bigg]\\
\times&{a_{s_{1}'}^{\dagger}}a_{s_{2}'}^{\dagger}a_{s_{3}'}^{\dagger}a_{s_{3}}a_{s_{2}}a_{s_{1}}.
\end{align}
\end{widetext}
As before we write
\begin{align}
\nonumber
&T^{(1, 2)}_{s_{1}'s_{2}', s_{1}s_{2}}(\omega(k_{1})+\omega(k_{2}))=\\
\nonumber
&-\pi{i}\delta(\omega(k_{1}')-\omega(k_{1}))T^{(1, 1)}_{s_{1}', s_{1}}(\omega(k_{1}))T^{(1, 1)}_{s_{2}', s_{2}}(\omega(k_{2}))\\
&+T^{(1, 2, C)}_{s_{1}'s_{2}', s_{1}s_{2}}(\omega(k_{1})+\omega(k_{2})).
\end{align}
Here, as before, the equality symbol is understood in the sense of permutation equivalence and on-shell condition. Now let us start massaging the three-body transition operator. First of all, one has
\begin{widetext}
\begin{align}
& T^{(1, 3)}_{\mu_{1}'k_{1}', \mu_{2}'k_{2}', \mu_{3}'k_{3}'; \mu_{1}k_{1}, \mu_{2}k_{2}, \mu_{3}k_{3}}(\omega(k_{1})+\omega(k_{2})+\omega(k_{3}))
=g_{\mu_{1}}^{*}(k_{1})g_{\mu_{2}}^{*}(k_{2})g_{\mu_{3}}^{*}(k_{3})
g_{\mu_{1}'}(k_{1}')g_{\mu_{2}'}(k_{2}')g_{\mu_{3}'}(k_{3}')
\nonumber \\
&  \qquad \qquad \times \tilde{G}^{(1)}(\omega(k_{1}))\tilde{G}^{(1)}(\omega(k_{1}'))F^{(1, 2)}(k_{2}', k_{3}', k_{2}, k_{3}, \omega(k_{1})+\omega(k_{2})+\omega(k_{3})),
\end{align}
where $F^{(1, 2)}(k_{2}', k_{3}', k_{2}, k_{3}, \omega(k_{1})+\omega(k_{2})+\omega(k_{3}))$ is not an entirely connected object defined via
\begin{align}
    F^{(1, 2)}(k_{1}', k_{2}', k_{1}, k_{2}, \epsilon)=\frac{1}{g_{\mu_{1}'}(k_{1}')g_{\mu_{2}'}(k_{2}')g_{\mu_{1}}^{*}(k_{1})g_{\mu_{2}}^{*}(k_{2})}\Braket{g|W^{(1, 2)}_{\mu_{1}'k_{1}'\mu_{2}'k_{2}', \mu_{1}k_{1}\mu_{2}k_{2}}(\epsilon)|g}.
\end{align}

Separation of elastic contribution (\ref{eq: CII2_3}) translates into the following decomposition
\begin{align}
& F^{(1, 2)}(k_{2}', k_{3}', k_{2}, k_{3}, \omega(k_{1})+\omega(k_{2})+\omega(k_{3})) \nonumber \\
=&F^{(1, 1)}(k_{2}', k_{2}, \omega(k_{1}')+\omega(k_{2}')) \tilde{G}^{(1)}(\omega(k_{1})+\omega(k_{3})-\omega(k_{3}')) F^{(1, 1)}(k_{3}', k_{3}, \omega(k_{1})+\omega(k_{3}))\nonumber \\
 &+ \overline{F}^{(2, 2)}(k_{2}', k_{3}', k_{2}, k_{3}, \omega(k_{1})+\omega(k_{2})+\omega(k_{3})).
\end{align}
\end{widetext}
Bearing in mind that $\overline{F}^{(2, 2)}(k_{2}', k_{3}', k_{2}, k_{3}, \omega(k_{1})+\omega(k_{2})+\omega(k_{3}))$ is an analytic function we decompose the three photon $T$-matrix as follows
\begin{align}
& T^{(1, 3)}_{\mu_{1}'k_{1}', \mu_{2}'k_{2}', \mu_{3}'k_{3}'; \mu_{1}k_{1}, \mu_{2}k_{2}, \mu_{3}k_{3}}(\omega(k_{1})+\omega(k_{2})+\omega(k_{3}))
\nonumber \\
=& \overline{T}^{(1, 3)}_{\mu_{1}'k_{1}', \mu_{2}'k_{2}', \mu_{3}'k_{3}'; \mu_{1}k_{1}, \mu_{2}k_{2}, \mu_{3}k_{3}}(\omega(k_{1})+\omega(k_{2})+\omega(k_{3}))\nonumber \\
+& \hat{T}^{(1, 3)}_{\mu_{1}'k_{1}', \mu_{2}'k_{2}', \mu_{3}'k_{3}'; \mu_{1}k_{1}, \mu_{2}k_{2}, \mu_{3}k_{3}}(\omega(k_{1})+\omega(k_{2})+\omega(k_{3})),
\end{align}
where we have defined the following objects
\begin{align}
& \overline{T}^{(1, 3)}_{\mu_{1}'k_{1}', \mu_{2}'k_{2}', \mu_{3}'k_{3}'; \mu_{1}k_{1}, \mu_{2}k_{2}, \mu_{3}k_{3}}(\omega(k_{1})+\omega(k_{2})+\omega(k_{3})) \nonumber \\
=& g_{\mu_{1}}^{*}(k_{1})g_{\mu_{2}}^{*}(k_{2})g_{\mu_{3}}^{*}(k_{3}){g_{\mu_{1}'}(k_{1}')g_{\mu_{2}'}(k_{2}')g_{\mu_{3}'}(k_{3}')} \nonumber \\
& \times\tilde{G}(\omega(k_{1}))\tilde{G}(\omega(k_{1}')) \nonumber \\ 
& \times \overline{F}^{(1, 2)}(k_{2}', k_{3}', k_{2}, k_{3}, \omega(k_{1})+\omega(k_{2})+\omega(k_{3}))
\end{align}
and
\begin{align}
& \hat{T}^{(1, 3)}_{\mu_{1}'k_{1}', \mu_{2}'k_{2}', \mu_{3}'k_{3}'; \mu_{1}k_{1}, \mu_{2}k_{2}, \mu_{3}k_{3}}(\omega(k_{1})+\omega(k_{2})+\omega(k_{3})) \nonumber \\
=& g_{\mu_{1}}^{*}(k_{1})g_{\mu_{2}}^{*}(k_{2})g_{\mu_{3}}^{*}(k_{3}){g_{\mu_{1}'}(k_{1}')g_{\mu_{2}'}(k_{2}')g_{\mu_{3}'}(k_{3}')} \nonumber \\
& \times\tilde{G}(\omega(k_{1}))\tilde{G}(\omega(k_{1}'))F^{(1, 1)}(k_{2}', k_{2}, \omega(k_{1}')+\omega(k_{2}')) \nonumber \\
& \times \tilde{G}^{(1)}(\omega(k_{1})+\omega(k_{3})-\omega(k_{3}')) \nonumber \\
& \times F^{(1, 1)}(k_{3}', k_{3}, \omega(k_{1})+\omega(k_{3})).
\label{eq:T13G0}
\end{align}
In the last equation we have
\begin{align}
    & F^{(1, 1)}(k_{2}', k_{2}, \omega(k_{1}')+\omega(k_{2}')) = \hat{G}_{0}(\omega(k_{1}')-\omega(k_{2})) \nonumber \\ & \qquad +\overline{F}^{(1, 1)}(k_{2}', k_{2}, \omega(k_{1}')+\omega(k_{2}')), \\
    & F^{(1, 1)}(k_{3}', k_{3}, \omega(k_{1})+\omega(k_{3})) =  \hat{G}_{0}(\omega(k_{1})-\omega(k_{3}')) \nonumber \\
    & \qquad +  \overline{F}^{(1, 1)}(k_{3}', k_{3}, \omega(k_{1})+\omega(k_{3})) .
\end{align}

The terms in \eqref{eq:T13G0} containing $\hat{G}_{0}$ deserve a special attention since they can yield delta-functions determining additional conservation of frequencies. 

In particular, the terms with $\hat{G}_{0} \overline{F}^{(1, 1)}$ and $\overline{F}^{(1, 1)} \hat{G}_{0}$ together give
\begin{align}
& -2\pi{i}\delta(\omega(k_{3}')-\omega(k_{3}))T^{(1, 1)}_{s_{3}', s_{3}}(\omega(k_{3}))\nonumber \\
& \times \Bigg[T^{(1, 2, C)}_{s_{1}'s_{2}', s_{1}s_{2}}(\omega(k_{1})+\omega(k_{2})) \nonumber \\ 
&  -g_{\mu_{1}}^{*}(k_{1})g_{\mu_{2}}^{*}(k_{2}) g_{\mu_{1}'}(k_{1}')g_{\mu_{2}'}(k_{2}') \tilde{G}^{(1)}(\omega(k_{1}))\tilde{G}^{(1)}(\omega(k_{1}')) \nonumber \\
&\qquad \times P\Bigg(\frac{1}{\omega(k_{1})-\omega(k_{2}')}\Bigg)\Bigg] 
\label{eq:term1df} \\
&+g_{\mu_{1}}^{*}(k_{1})g_{\mu_{2}}^{*}(k_{2})g_{\mu_{3}}^{*}(k_{3}){g_{\mu_{1}'}(k_{1}')g_{\mu_{2}'}(k_{2}')g_{\mu_{3}'}(k_{3}')}\nonumber \\
& \quad \times \tilde{G}(\omega(k_{1}))\tilde{G}(\omega(k_{1}')) \tilde{G}^{(1)}(\omega(k_{1})+\omega(k_{3})-\omega(k_{3}')) \nonumber \\
& \times \Bigg[ P \Bigg(\frac{1}{\omega(k_{1}')-\omega(k_{2})}\Bigg) \overline{F}^{(1, 1)}(k_{3}', k_{3}, \omega(k_{1})+\omega(k_{3})) \nonumber \\
& \quad  + \overline{F}^{(1, 1)}(k_{2}', k_{2}, \omega(k_{1}')+\omega(k_{2}')) P\Bigg(\frac{1}{\omega(k_{1})-\omega(k_{3}')}\Bigg)\Bigg] .
\label{eq:term2df}
\end{align}
The term \eqref{eq:term1df} contributes to the $3=2+1$ cluster of the three-photon $S$-matrix. In turn, the term \eqref{eq:term2df} is completely connected and non-singular. This property becomes explicitly visible if we re-express it as
\begin{align}
& \frac{g_{\mu_{1}}^{*}(k_{1})g_{\mu_{2}}^{*}(k_{2})g_{\mu_{3}}^{*}(k_{3}) g_{\mu_{1}'}(k_{1}')g_{\mu_{2}'}(k_{2}')g_{\mu_{3}'}(k_{3}')}{\omega(k_{1}')-\omega(k_{1})} \nonumber \\
\times & \Bigg[\tilde{G}^{(1)}(\omega(k_{1}'))\tilde{G}^{(1)}(\omega(k_{3})) \tilde{G}^{(1)}(\omega(k_{3}')+\omega(k_{1}') -\omega(k_{1}))\nonumber \\
& \quad \times \overline{F}^{(1, 1)}(k_{2}', k_{2}, \omega(k_{1}')+\omega(k_{2}')+\omega(k_{3}')-\omega(k_{1})) \nonumber \\
&-\tilde{G}^{(1)}(\omega(k_{1})) \tilde{G}^{(1)}(\omega(k_{1})+\omega(k_{3})-\omega(k_{1}')) \tilde{G}^{(1)}(\omega(k_{3}')) \nonumber \\
& \quad \times \overline{F}^{(1, 1)}(k_{2}', k_{2}, \omega(k_{2}')+\omega(k_{3}'))\Bigg].
\end{align}
This representation makes it obvious that in the limit $k'_1 \to k_1$ this function is finite.

Now, let us consider the  contribution to \eqref{eq:T13G0}  containing $ \hat{G}_{0} \hat{G}_{0}$. It amounts to 
\begin{align}
& g_{\mu_{1}}^{*}(k_{1})g_{\mu_{2}}^{*}(k_{2})g_{\mu_{3}}^{*}(k_{3}) g_{\mu_{1}'}(k_{1}')g_{\mu_{2}'}(k_{2}')g_{\mu_{3}'}(k_{3}') \nonumber \\
& \times \tilde{G}(\omega(k_{1}))\tilde{G}(\omega(k_{3}')) \tilde{G}^{(1)}(\omega(k_{2})+\omega(k_{1})-\omega(k_{1}')) \nonumber \\
& \times\hat{G}_{0}(\omega(k_{3}')-\omega(k_{3}))\hat{G}_{0}(\omega(k_{1})-\omega(k_{1}'))
\nonumber \\
=& (-i\pi)^{2}T^{(1, 1)}_{s_{1}', s_{1}}(\omega(k_{1}))T^{(1, 1)}_{s_{2}', s_{2}}(\omega(k_{2}))T^{(1, 1)}_{s_{3}', s_{3}}(\omega(k_{3}))\nonumber \\
& \times \delta(\omega(k_{1}')-\omega(k_{1}))\delta(\omega(k_{2}')-\omega(k_{2})) \label{term_dg1} \\
-& 2\pi{i} g_{\mu_{1}}^{*}(k_{1})g_{\mu_{2}}^{*}(k_{2}) g_{\mu_{1}'}(k_{1}')g_{\mu_{2}'}(k_{2}') \tilde{G}^{(1)}(\omega(k_{1}))  \nonumber \\
& \times \tilde{G}^{(1)}(\omega(k_{1}')) T^{(1, 1)}_{s_{3}', s_{3}}(\omega(k_{3}))\delta(\omega(k_{3}')-\omega(k_{3})) \label{term_dg2} \\
+&g_{\mu_{1}}^{*}(k_{1})g_{\mu_{2}}^{*}(k_{2})g_{\mu_{3}}^{*}(k_{3}){g_{\mu_{1}'}(k_{1}')g_{\mu_{2}'}(k_{2}')g_{\mu_{3}'}(k_{3}')} \nonumber \\
& \times \tilde{G}(\omega(k_{3}')) \tilde{G}(\omega(k_{1})) \tilde{G}^{(1)}(\omega(k_{2})+\omega(k_{1})-\omega(k_{1}')) \nonumber \\
& \times {P}\Bigg(\frac{1}{\omega(k_{3}')-\omega(k_{3})}\Bigg) P\Bigg(\frac{1}{\omega(k_{1})-\omega(k_{1}')}\Bigg). \label{term_dg3}
\end{align}

The term \eqref{term_dg1} contributes to the $3=1+1+1$ cluster. The term \eqref{term_dg2} contributes to the $3=2+1$ cluster.

The term \eqref{term_dg3} requires a special consideration. Rewriting
\begin{align}
& \tilde{G}^{(1)}(\omega(k_{3}'))\tilde{G}^{(1)}(\omega(k_{2})+\omega(k_{1})-\omega(k_{1}'))\tilde{G}^{(1)}(\omega(k_{1}))\nonumber \\
&\times P\Bigg(\frac{1}{\omega(k_{3}')-\omega(k_{3})}\Bigg)  P\Bigg(\frac{1}{\omega(k_{1})-\omega(k_{1}')}\Bigg) \nonumber \\
=&\tilde{G}^{(1)}(\omega(k_{3}'))\tilde{G}^{(1)}(\omega(k_{1}))P\Bigg(\frac{1}{\omega(k_{3}')-\omega(k_{3})}\Bigg)\nonumber \\
& \times \frac{\tilde{G}^{(1)}(\omega(k_{2})+\omega(k_{1})-\omega(k_{1}'))-\tilde{G}^{(1)}(\omega(k_{2}))}{\omega(k_{1})-\omega(k_{1}')} \nonumber \\
+&\tilde{G}^{(1)}(\omega(k_{3}'))\tilde{G}^{(1)}(\omega(k_{2}))\tilde{G}^{(1)}(\omega(k_{1})) \nonumber \\
& \times P\Bigg(\frac{1}{\omega(k_{3}')-\omega(k_{3})}\Bigg)P\Bigg(\frac{1}{\omega(k_{1})-\omega(k_{1}')}\Bigg) \nonumber \\
=&\tilde{G}^{(1)}(\omega(k_{3}'))\tilde{G}^{(1)}(\omega(k_{1}))P\Bigg(\frac{1}{\omega(k_{3}')-\omega(k_{3})}\Bigg) \nonumber \\
& \times \frac{\tilde{G}^{(1)}(\omega(k_{2})+\omega(k_{1})-\omega(k_{1}'))-\tilde{G}^{(1)}(\omega(k_{2}))}{\omega(k_{1})-\omega(k_{1}')} \label{term_dl1} \\
 +&\tilde{G}^{(1)}(\omega(k_{2}))\tilde{G}^{(1)}(\omega(k_{1}))\nonumber \\
 & \times \frac{\tilde{G}^{(1)}(\omega(k_{3}'))-\tilde{G}^{(1)}(\omega(k_{3}))}{\omega(k_{3}')-\omega(k_{3})}P\Bigg(\frac{1}{\omega(k_{1})-\omega(k_{1}')}\Bigg) \label{term_dl2} \\
+&\tilde{G}^{(1)}(\omega(k_{3}))\tilde{G}^{(1)}(\omega(k_{2}))\tilde{G}^{(1)}(\omega(k_{1}))\nonumber \\ 
& \times P\Bigg(\frac{1}{\omega(k_{3}')-\omega(k_{3})}\Bigg)P\Bigg(\frac{1}{\omega(k_{1})-\omega(k_{1}')}\Bigg), \label{term_dl3}
\end{align}
we observe that
the terms \eqref{term_dl1} and \eqref{term_dl2} terms give together a non-singular contribution. In fact,
\begin{align}
& \tilde{G}^{(1)}(\omega(k_{3}'))\tilde{G}^{(1)}(\omega(k_{1}))
P\Bigg(\frac{1}{\omega(k_{3}')-\omega(k_{3})}\Bigg) \nonumber \\
& \times  \frac{\tilde{G}^{(1)}(\omega(k_{2})+\omega(k_{1})-\omega(k_{1}'))-\tilde{G}^{(1)}(\omega(k_{2}))}{\omega(k_{1})-\omega(k_{1}')} \nonumber \\
+&\tilde{G}^{(1)}(\omega(k_{2}))\tilde{G}^{(1)}(\omega(k_{1}))\frac{\tilde{G}^{(1)}(\omega(k_{3}'))-\tilde{G}^{(1)}(\omega(k_{3}))}{\omega(k_{3}')-\omega(k_{3})} \nonumber \\
& \times P\Bigg(\frac{1}{\omega(k_{1})-\omega(k_{1}')}\Bigg) \nonumber \\
&=\tilde{G}^{(1)}(\omega(k_{3}'))\tilde{G}^{(1)}(\omega(k_{1}))\tilde{G}^{(1)}(\omega(k_{2})) \nonumber \\
& \times \tilde{G}^{(1)}(\omega(k_{2}')+\omega(k_{3}')-\omega(k_{3}))\tilde{G}^{(1)}(\omega(k_{2}')) \nonumber \\
& \times\frac{(\tilde{G}^{(1)})^{-1}(\omega(k_{2}'))-(\tilde{G}^{(1)})^{-1}(\omega(k_{2}')+\omega(k_{3}')-\omega(k_{3}))}{\omega(k_{3}')-\omega(k_{3})} \nonumber \\
&\times\frac{(\tilde{G}^{(1)})^{-1}(\omega(k_{2}))-(\tilde{G}^{(1)})^{-1}(\omega(k_{2})+\omega(k_{1})-\omega(k_{1}'))}{\omega(k_{1})-\omega(k_{1}')}
\nonumber \\
&+\tilde{G}^{(1)}(\omega(k_{2}))\tilde{G}^{(1)}(\omega(k_{3}))\tilde{G}^{(1)}(\omega(k_{3}'))\frac{1}{\omega(k_{1}')-\omega(k_{1})} \nonumber \\
& \times\Bigg(\tilde{G}^{(1)}(\omega(k_{1}')) \nonumber \\
& \times \frac{(\tilde{G}^{(1)})^{-1}(\omega(k_{3}))-(\tilde{G}^{(1)})^{-1}(\omega(k_{1}')+\omega(k_{3}')-\omega(k_{1}))}{\omega(k_{1}')+\omega(k_{3}')-\omega(k_{1})-\omega(k_{3})} \nonumber \\
&-\tilde{G}^{(1)}(\omega(k_{1}))\frac{(\tilde{G}^{(1)})^{-1}(\omega(k_{3}))-(\tilde{G}^{(1)})^{-1}(\omega(k_{3}'))}{\omega(k_{3}')-\omega(k_{3})}\Bigg).
\end{align}

In contrast, the term \eqref{term_dl3} is singular and contributes to the $3=1+1+1$ cluster of the scattering matrix. To show this, we first fully symmetrize
\begin{align}
& \tilde{G}^{(1)}(\omega(k_{3}))\tilde{G}^{(1)}(\omega(k_{2}))\tilde{G}^{(1)}(\omega(k_{1})) \nonumber \\ 
& \times P\Bigg(\frac{1}{\omega(k_{3}')-\omega(k_{3})}\Bigg)P\Bigg(\frac{1}{\omega(k_{1})-\omega(k_{1}')}\Bigg) \nonumber \\
\to &\frac{1}{3}\tilde{G}^{(1)}(\omega(k_{3}))\tilde{G}^{(1)}(\omega(k_{2}))\tilde{G}^{(1)}(\omega(k_{1})) \nonumber \\
& \times \Bigg[P\Bigg(\frac{1}{\omega(k_{3}')-\omega(k_{3})}\Bigg)P\Bigg(\frac{1}{\omega(k_{1})-\omega(k_{1}')}\Bigg) \nonumber \\
& \quad +P\Bigg(\frac{1}{\omega(k_{2}')-\omega(k_{2})}\Bigg)P\Bigg(\frac{1}{\omega(k_{3})-\omega(k_{3}')}\Bigg) \nonumber \\
& \quad +P\Bigg(\frac{1}{\omega(k_{1}')-\omega(k_{1})}\Bigg)P\Bigg(\frac{1}{\omega(k_{2})-\omega(k_{2}')}\Bigg)\Bigg].
\label{symmP}
\end{align}
Owing to the Poincare-Bertrand distributional identity 
\begin{align}
P\Bigg(\frac{1}{x}\Bigg)P\Bigg(\frac{1}{y}\Bigg)&=P\Bigg(\frac{1}{y-x}\Bigg) \Bigg[P\Bigg(\frac{1}{x}\Bigg)-P\Bigg(\frac{1}{y}\Bigg)\Bigg] \nonumber \\
&+\pi^{2}\delta(x)\delta(y),
\end{align}
we establish the identity
\begin{align}
& P\Bigg(\frac{1}{\omega(k_{3}')-\omega(k_{3})}\Bigg)P\Bigg(\frac{1}{\omega(k_{1})-\omega(k_{1}')}\Bigg)\nonumber \\ 
+& P\Bigg(\frac{1}{\omega(k_{2}')-\omega(k_{2})}\Bigg)P\Bigg(\frac{1}{\omega(k_{3})-\omega(k_{3}')}\Bigg) \nonumber \\
+& P\Bigg(\frac{1}{\omega(k_{1}')-\omega(k_{1})}\Bigg)P\Bigg(\frac{1}{\omega(k_{2})-\omega(k_{2}')}\Bigg)\nonumber \\
=& \pi^{2}\delta(\omega(k_{3}')-\omega(k_{3}))\delta(\omega(k_{1}')-\omega(k_{1})),
\end{align}
leading us to the result
\begin{align}
  & \frac{\pi^2}{3} \tilde{G}^{(1)}(\omega(k_{3}))\tilde{G}^{(1)}(\omega(k_{2}))\tilde{G}^{(1)}(\omega(k_{1})) \\ & \times \delta(\omega(k_{3}')-\omega(k_{3}))\delta(\omega(k_{1}')-\omega(k_{1}))
\end{align}
in \eqref{symmP}.

Combining all of the above results, we arrive at the following decomposition of the three-photon $T$-matrix
\begin{widetext}
\begin{align}
& T^{(1, 3)}_{s_{1}'s_{2}'s_{3}', s_{1}s_{2}s_{3}}(\omega(k_{1})+\omega(k_{2})+\omega(k_{3})) \nonumber \\
=& \frac{(-2\pi{i})^{2}}{6}T^{(1)}_{s_{1}', s_{1}}(\omega(k_{1}))T^{(1)}_{s_{2}', s_{2}}(\omega(k_{2}))T^{(1)}_{s_{3}', s_{3}}(\omega(k_{3}))  \delta(\omega(k_{1}')-\omega(k_{1})) \delta(\omega(k_{2}')-\omega(k_{2}))\nonumber \\
 - & \, 2\pi{i}T^{(1, 2, C)}_{s_{1}'s_{2}', s_{1}s_{2}}(\omega(k_{1})+\omega(k_{2}))T^{(1)}_{s_{3}', s_{3}}(\omega(k_{3}))\delta(\omega(k_{3}')-\omega(k_{3})) + T^{(1, 3, C)}_{s_{1}'s_{2}'s_{3}', s_{1}s_{2}s_{3}}(\omega(k_{1})+\omega(k_{2})+\omega(k_{3})),
\end{align}
with the three-body connected part
\begin{align}
& T^{(1, 3, C)}_{s_{1}'s_{2}'s_{3}', s_{1}s_{2}s_{3}}(\omega(k_{1})+\omega(k_{2})+\omega(k_{3}))=g_{\mu_{1}}^{*}(k_{1})g_{\mu_{2}}^{*}(k_{2})g_{\mu_{3}}^{*}(k_{3}){g_{\mu_{1}'}(k_{1}')g_{\mu_{2}'}(k_{2}')g_{\mu_{3}'}(k_{3}')} \nonumber \\
& \times\Bigg\{\tilde{G}^{(1)}(\omega(k_{3}'))\tilde{G}^{(1)}(\omega(k_{1}))\tilde{G}^{(1)}(\omega(k_{2}))\tilde{G}^{(1)}(\omega(k_{2}')+\omega(k_{3}')-\omega(k_{3}))\tilde{G}^{(1)}(\omega(k_{2}')) \nonumber \\ 
& \qquad \times \frac{(\tilde{G}^{(1)})^{-1}(\omega(k_{2'}))-(\tilde{G}^{(1)})^{-1}(\omega(k_{2}')+\omega(k_{3}')-\omega(k_{3}))}{\omega(k_{3}')-\omega(k_{3})}\frac{(\tilde{G}^{(1)})^{-1}(\omega(k_{2}))-(\tilde{G}^{(1)})^{-1}(\omega(k_{2})+\omega(k_{1})-\omega(k_{1}'))}{\omega(k_{1})-\omega(k_{1}')}\nonumber \\
& \qquad + \frac{1}{\omega(k_{1}')-\omega(k_{1})} \Bigg[\tilde{G}^{(1)}(\omega(k_{2}))\tilde{G}^{(1)}(\omega(k_{3}))\tilde{G}^{(1)}(\omega(k_{3}')) \nonumber \\
& \qquad \qquad \times  \Bigg(\tilde{G}^{(1)}(\omega(k_{1}')) \frac{(\tilde{G}^{(1)})^{-1}(\omega(k_{3}))-(\tilde{G}^{(1)})^{-1}(\omega(k_{1}')+\omega(k_{3}')-\omega(k_{1}))}{\omega(k_{1}')+\omega(k_{3}')-\omega(k_{1})-\omega(k_{3})} \nonumber \\ & \qquad \qquad \qquad -\tilde{G}^{(1)}(\omega(k_{1}))\frac{(\tilde{G}^{(1)})^{-1}(\omega(k_{3}))-(\tilde{G}^{(1)})^{-1}(\omega(k_{3}'))}{\omega(k_{3}')-\omega(k_{3})}\Bigg) \nonumber \\
& \qquad \qquad +\tilde{G}^{(1)}(\omega(k_{1}'))\tilde{G}^{(1)}(\omega(k_{3}))\tilde{G}^{(1)}(\omega(k_{3}')+\omega(k_{1}')-\omega(k_{1})) \overline{F}^{(1, 1)}(k_{2}', k_{2}, \omega(k_{1}')+\omega(k_{2}')+\omega(k_{3}')-\omega(k_{1})) \nonumber \\ 
& \qquad \qquad -\tilde{G}^{(1)}(\omega(k_{1}))\tilde{G}^{(1)}(\omega(k_{3}'))\tilde{G}^{(1)}(\omega(k_{1})+\omega(k_{3})-\omega(k_{1}')) \overline{F}^{(1, 1)}(k_{2}', k_{2}, \omega(k_{2}')+\omega(k_{3}'))\Bigg] \nonumber \\
& \qquad +\tilde{G}(\omega(k_{1}))\tilde{G}(\omega(k_{1}'))\overline{F}^{(1, 1)}(k_{2}', k_{2}, \omega(k_{1}')+\omega(k_{2}'))\tilde{G}^{(1)}(\omega(k_{1})+\omega(k_{3})-\omega(k_{3}')) \overline{F}^{(1, 1)}(k_{3}', k_{3}, \omega(k_{1})+\omega(k_{3}))\Bigg\} \nonumber \\
&+\overline{T}^{(1, 3)}_{s_{1}'s_{2}'s_{3}', s_{1}s_{2}s_{3}}(\omega(k_{1})+\omega(k_{2})+\omega(k_{3})).
\end{align}
Plugging everything into the definition of the three-photon $S$-matrix we finally obtain
\begin{align}
\mathcal{S}_{3}& =\Bigg[\frac{1}{3!}S^{(1)}_{s_{1}', s_{1}}S^{(1)}_{s_{2}', s_{2}}S^{(1)}_{s_{3}', s_{3}}-2\pi{i}T^{(1, 2, C)}_{s_{1}'s_{2}', s_{1}s_{2}}(\omega(k_{1})+\omega(k_{2}))S^{(1)}_{s_{3}', s_{3}}\delta(\omega(k_{1}')+\omega(k_{2}')-\omega(k_{1})-\omega(k_{2})
\nonumber \\
\label{eq: S3-cluster}
& -2\pi{i}T^{(1, 3, C)}_{s_{1}'s_{2}'s_{3}', s_{1}s_{2}s_{3}}(\omega(k_{1})+\omega(k_{2})+\omega(k_{3}))\delta(\omega(k_{1}')+\omega(k_{2}')+\omega(k_{3}')-\omega(k_{1})-\omega(k_{2})-\omega(k_{3}))\Bigg] a_{s_{1}'}^{\dagger} a_{s_{2}'}^{\dagger}a_{s_{3}'}^{\dagger}a_{s_{3}}a_{s_{2}}a_{s_{1}}.
\end{align}
\end{widetext}
\section{Third-order coherence function}
\label{ap: 3ocf_exp}
In this appendix we present the formula for the third order coherence function of phonons in the giant atom model. Using the formula (\ref{eq: S3-cluster}) with $\omega(k)=k$ along with the notations introduced in the Section \ref{sec:GIANT_ATOM}, we obtain the following result upon contraction with the three-phonon Fock state
\begin{widetext}
\begin{align}
\nonumber
\mathcal{S}_{3}\ket{\Phi^{(3)}_{1}}&=\frac{1}{\sqrt{6}}\sum_{\{\mu'_{i}\}}\Bigg(\int_{k_{1}k_{2}k_{3}}\varphi(k_{1})\varphi(k_{2})\varphi(k_{3})S^{(1)}_{\mu_{1}', 1}(k_{1})S^{(1)}_{\mu_{2}', 1}(k_{2})S^{(1)}_{\mu_{3}', 1}(k_{3})a_{\mu_{1}'}^{\dagger}(k_{1})a_{\mu_{2}'}^{\dagger}(k_{2})a_{\mu_{3}'}^{\dagger}(k_{3})\ket{\Omega}\\
\nonumber
&-12\pi{i}\int_{k_{1}'k_{2}'k_{1}k_{2}k_{3}}\varphi(k_{1})\varphi(k_{2})\varphi(k_{3})T^{(2, C)}_{\mu_{1}'k_{1}', \mu_{2}'k_{2}', \mu_{1}k_{1}, \mu_{2}k_{2}}(k_{1}+k_{2})\delta(k_{1}'+k_{2}'-k_{1}-k_{2})S^{(1)}_{\mu_{3}'; 1}(k_{3})\\
\nonumber
&\times{a}_{\mu_{1}'}^{\dagger}(k_{1}')a_{\mu_{2}'}^{\dagger}(k_{2}')a_{\mu_{3}'}^{\dagger}(k_{3})\ket{\Omega}\\
\label{eq: SPE57}
&-12\pi{i}\Bigg(\frac{2\pi}{L}\Bigg)^{3/2}\int_{k_{1}'k_{2}'k_{3}'}Q(k_{1}', k_{2}', k_{3}')\delta(k_{1}'+k_{2}'+k_{3}')a_{\mu_{1}'}^{\dagger}(k_{1}')a_{\mu_{2}'}^{\dagger}(k_{2}')a_{\mu_{3}'}^{\dagger}(k_{3}')\ket{\Omega}\Bigg),
\end{align}
where we have introduced the following symmetrized version of the connected three-phonon transition operator
\begin{align}
\nonumber
Q(k_{1}', k_{2}', k_{3}')&=\frac{1}{6}\Big[T^{(3, C)}_{\mu_{1}'k_{1}', \mu_{2}'k_{2}', \mu_{3}'k_{3}', 10, 10, 10}(0)+T^{(3, C)}_{\mu_{1}'k_{1}', \mu_{3}'k_{3}', \mu_{2}'k_{2}', 10, 10, 10}(0)+T^{(3, C)}_{\mu_{3}'k_{3}', \mu_{2}'k_{2}', \mu_{1}'k_{1}', 10, 10, 10}(0)\\
\label{eq: SPE58}
&+T^{(3, C)}_{\mu_{3}'k_{3}', \mu_{1}'k_{1}', \mu_{2}'k_{2}', 10, 10, 10}(0)+T^{(3, C)}_{\mu_{2}'k_{2}', \mu_{1}'k_{1}', \mu_{3}'k_{3}', 10, 10, 10}(0)+T^{(3, C)}_{\mu_{2}'k_{2}', \mu_{3}'k_{3}', \mu_{1}'k_{1}', 10, 10, 10}(0)\Big].
\end{align}
\end{widetext}
Here we have suppressed the dependence of $Q(k_{1}', k_{2}', k_{3}')$ on $\{\mu'\}$ since in the giant atom model the coupling constants are independent of the channel index. Now we consider
\begin{widetext}
\begin{align}
\nonumber
a_{\mu''}(\tau_{3})a_{\mu'}(\tau_{2})a_{\mu}(\tau_{1})\mathcal{S}_{3}\ket{\Phi^{(3)}_{1}}&=\frac{\sqrt{6}}{L^{3/2}}S^{(1)}_{\mu'', 1}(0)S^{(1)}_{\mu', 1}(0)S^{(1)}_{\mu, 1}(0)\Bigg(1-4\pi{i}\Bigg[\frac{I^{(1)}_{\mu'', \mu'}(\tau_{3}-\tau_{2})}{S^{(1)}_{\mu'', 1}(0)S^{(1)}_{\mu', 1}(0)}\\
\label{eq: SPE62}
&+\frac{I^{(1)}_{\mu, \mu'}(\tau_{2}-\tau_{1})}{S^{(1)}_{\mu, 1}(0)S^{(1)}_{\mu', 1}(0)}+\frac{I^{(1)}_{\mu'', \mu}(\tau_{3}-\tau_{1})}{S^{(1)}_{\mu'', 1}(0)S^{(1)}_{\mu, 1}(0)}\Bigg]-12\pi{i}\frac{I^{(2)}(\tau_{3}-\tau_{1}, \tau_{2}-\tau_{1})}{S^{(1)}_{\mu'', 1}(0)S^{(1)}_{\mu', 1}(0)S^{(1)}_{\mu, 1}(0)}\Bigg),
\end{align}
\end{widetext}
where the following functions were defined
\begin{align}
\label{eq: SPE63}
I^{(1)}_{\mu', \mu}(t_{1})=&\int_{k}e^{ikt_{1}}M_{\mu', \mu}(k),\\
I^{(2)}(t_{2}, t_{1})=&\int_{k, q}e^{iqt_{2}}e^{ikt_{1}}Q(-k-q, k, q),
\end{align}
where $M_{\mu, \mu'}(k)=(T^{(2, C)}_{\mu' k, \mu-k, 10, 10}(0)+T^{(2, C)}_{\mu -k, \mu'k, 10, 10}(0))/2$, as before. By introducing the following variables $\tau'=\tau_{3}-\tau_{1}$, $\tau=\tau_{2}-\tau_{1}$, we can immediately write down the normalized third order coherence function to the lowest order in $\varphi$ as
\begin{widetext}
\begin{align}
C^{(3)}_{\mu'', \mu', \mu}(t', t)=\Bigg{|}1-4\pi{i}\Bigg[\frac{I^{(1)}_{\mu'', \mu'}(t'-t)}{S^{(1)}_{\mu'', 1}(0)S^{(1)}_{\mu', 1}(0)}+\frac{I^{(1)}_{\mu, \mu'}(t)}{S^{(1)}_{\mu, 1}(0)S^{(1)}_{\mu', 1}(0)}+\frac{I^{(1)}_{\mu'', \mu}(t')}{S^{(1)}_{\mu'', 1}(0)S^{(1)}_{\mu; 1}(0)}\Bigg]-12\pi{i}\frac{I^{(2)}(t', t)}{S^{(1)}_{\mu'', 1}(0)S^{(1)}_{\mu', 1}(0)S^{(1)}_{\mu, 1}(0)}\Bigg{|}^{2}.
\end{align}
\end{widetext}

\section{Diagrammatic representation of the generic three-body transition operator}
\label{ap: dressed}
In this Appendix we start our analysis with equation (\ref{eq: 3body_vertex_generator}). Taking matrix elements of (\ref{eq: 3body_vertex_generator}) in the three-particle subspace we arrive at the following integral equation
\begin{widetext}
\begin{align}
\nonumber
W^{(3, 2)}_{s_{1}'s_{2}', s_{1}s_{2}}(\epsilon)=&{D}^{(2)}_{s_{1}'s_{2}', s_{1}s_{2}}(\epsilon)\\
\nonumber
+&[{D}^{(2)}_{s_{1}'s_{2}', \bar{s}_{2}'\bar{s}_{1}'}(\epsilon)+{D}^{(2)}_{s_{1}'s_{2}', \bar{s}_{1}'\bar{s}_{2}'}(\epsilon)]{G}^{(1)}(\epsilon-\omega_{\bar{s}_{1}'}-\omega_{\bar{s}_{2}'})W^{(3, 2)}_{\bar{s}_{1}'\bar{s}_{2}', s_{1}s_{2}}(\epsilon)\\
\nonumber
+&{D}^{(2)}_{s_{1}'s_{2}', \bar{s}_{1}'\bar{s}_{2}}(\epsilon){G}^{(1)}(\epsilon-\omega_{\bar{s}_{1}'}-\omega_{\bar{s}_{2}})W^{(2, 1)}_{\bar{s}_{2}, \bar{s}_{2}'}(\epsilon){G}^{(1)}(\epsilon-\omega_{\bar{s}_{1}'}-\omega_{\bar{s}_{2}'})W^{(3, 2)}_{\bar{s}_{1}'\bar{s}_{2}', s_{1}s_{2}}(\epsilon)\\
\nonumber
+&D^{(2)}_{s_{1}'s_{2}', \bar{s}_{1}\bar{s}_{1}'}(\epsilon){G}^{(1)}(\epsilon-\omega_{\bar{s}_{1}}-\omega_{\bar{s}_{1}'})W^{(2, 1)}_{\bar{s}_{1}, \bar{s}_{2}'}(\epsilon){G}^{(1)}(\epsilon-\omega_{\bar{s}_{1}'}-\omega_{\bar{s}_{2}'})W^{(3, 2)}_{\bar{s}_{1}'\bar{s}_{2}', s_{1}s_{2}}(\epsilon)\\
\nonumber
+&{D}^{(2)}_{s_{1}'s_{2}', \bar{s}_{1}\bar{s}_{2}'}(\epsilon){G}^{(1)}(\epsilon-\omega_{\bar{s}_{1}}-\omega_{\bar{s}_{2}'})W^{(2, 1)}_{\bar{s}_{1}, \bar{s}_{1}'}(\epsilon){G}^{(1)}(\epsilon-\omega_{\bar{s}_{1}'}-\omega_{\bar{s}_{2}'})W^{(3, 2)}_{\bar{s}_{1}'\bar{s}_{2}', s_{1}s_{2}}(\epsilon)\\
\nonumber
+&{D}^{(2)}_{s_{1}'s_{2}', \bar{s}_{2}'\bar{s}_{2}}(\epsilon){G}^{(1)}(\epsilon-\omega_{\bar{s}_{2}'}-\omega_{\bar{s}_{2}})W^{(2, 1)}_{\bar{s}_{2}, \bar{s}_{1}'}(\epsilon){G}^{(1)}(\epsilon-\omega_{\bar{s}_{1}'}-\omega_{\bar{s}_{2}'})W^{(3, 2)}_{\bar{s}_{1}'\bar{s}_{2}', s_{1}s_{2}}(\epsilon)\\
\nonumber
+&{D}^{(2)}_{s_{1}'s_{2}', \bar{s}_{1}\bar{s}_{2}}(\epsilon){G}^{(1)}(\epsilon-\omega_{\bar{s}_{1}}-\omega_{\bar{s}_{2}})W^{(2, 2)}_{\bar{s}_{2}\bar{s}_{1}, \bar{s}_{2}'\bar{s}_{1}'}(\epsilon){G}^{(1)}(\epsilon-\omega_{\bar{s}_{1}'}-\omega_{\bar{s}_{2}'})W^{(3, 2)}_{\bar{s}_{1}'\bar{s}_{2}', s_{1}s_{2}}(\epsilon)\\
\nonumber
+&{D}^{(2)}_{s_{1}'s_{2}', \bar{s}_{1}\bar{s}_{2}}(\epsilon){G}^{(1)}(\epsilon-\omega_{\bar{s}_{1}}-\omega_{\bar{s}_{2}})W^{(2, 2)}_{\bar{s}_{1}\bar{s}_{2}, \bar{s}_{2}'\bar{s}_{1}'}(\epsilon){G}^{(1)}(\epsilon-\omega_{\bar{s}_{1}'}-\omega_{\bar{s}_{2}'})W^{(3, 2)}_{\bar{s}_{1}'\bar{s}_{2}', s_{1}s_{2}}(\epsilon)\\
\nonumber
+&{D}^{(2)}_{s_{1}'s_{2}', \bar{s}_{1}\bar{s}_{2}}(\epsilon){G}^{(1)}(\epsilon-\omega_{\bar{s}_{1}}-\omega_{\bar{s}_{2}})W^{(2, 2)}_{\bar{s}_{1}\bar{s}_{2}, \bar{s}_{1}'\bar{s}_{2}'}(\epsilon){G}^{(1)}(\epsilon-\omega_{\bar{s}_{1}'}-\omega_{\bar{s}_{2}'})W^{(3, 2)}_{\bar{s}_{1}'\bar{s}_{2}', s_{1}s_{2}}(\epsilon)\\
+&{D}^{(2)}_{s_{1}'s_{2}', \bar{s}_{1}\bar{s}_{2}}(\epsilon){G}^{(1)}(\epsilon-\omega_{\bar{s}_{1}}-\omega_{\bar{s}_{2}})W^{(2, 2)}_{\bar{s}_{2}'\bar{s}_{1}, \bar{s}_{1}'\bar{s}_{2}'}(\epsilon){G}^{(1)}(\epsilon-\omega_{\bar{s}_{1}'}-\omega_{\bar{s}_{2}'})W^{(3, 2)}_{\bar{s}_{1}'\bar{s}_{2}', s_{1}s_{2}}(\epsilon),\end{align}
\end{widetext}
where the projection of $\mathcal{D}$ onto the $2$-particle subspace is given by
\begin{align}
\nonumber
    &D^{(2)}_{s_{1}'s_{2}', s_{1}s_{2}}(\epsilon)\\
    &=v_{s_{1}'}G_{0}(\epsilon-\omega_{s_{2}'})v_{s_{2}'}G^{(1)}(\epsilon)v_{s_{2}}^{\dagger}G_{0}(\epsilon-\omega_{s_{2}})v_{s_{1}}^{\dagger}.
\end{align}
Baring this in mind we make the following ansatz
\begin{align}
\nonumber
    &W^{(3, 2)}_{s_{1}'s_{2}', s_{1}s_{2}}(\epsilon)\\
    &=v_{s_{1}'}G_{0}(\epsilon-\omega_{s_{2}'})v_{s_{2}'}G^{(3, 0)}(\epsilon)v_{s_{2}}^{\dagger}G_{0}(\epsilon-\omega_{s_{2}})v_{s_{1}}^{\dagger}.
\end{align}
This, in turn, leads to the following Dyson equation
\begin{align}
G^{(3, 0)}(\epsilon)=G^{(1)}(\epsilon)+G^{(1)}(\epsilon)\Sigma^{(3, 0)}(\epsilon)G^{(3, 0)}(\epsilon),
\end{align}
where the self-energy in the three-excitation subspace is given by
\begin{widetext}
\begin{align}
\nonumber
\Sigma^{(3, 0)}(\epsilon)=&[v_{s_{1}'}^{\dagger}G_{0}(\epsilon-\omega_{s_{1}'})v_{s_{2}'}^{\dagger}+v_{s_{2}'}^{\dagger}G_{0}(\epsilon-\omega_{s_{2}'})v_{s_{1}'}^{\dagger}]{G}^{(1)}(\epsilon-\omega_{s_{1}'}-\omega_{s_{2}'})v_{s_{1}'}G_{0}(\epsilon-\omega_{s_{2}'})v_{s_{2}'}\\
\nonumber
+&v_{s_{2}}^{\dagger}G_{0}(\epsilon-\omega_{s_{2}})v_{s_{1}'}^{\dagger}{G}^{(1)}(\epsilon-\omega_{s_{1}'}-\omega_{s_{2}})W^{(2, 1)}_{s_{2}, s_{2}'}(\epsilon){G}^{(1)}(\epsilon-\omega_{s_{1}'}-\omega_{s_{2}'})v_{s_{1}'}G_{0}(\epsilon-\omega_{s_{2}'})v_{s_{2}'}\\
\nonumber
+&v_{s_{1}'}^{\dagger}G_{0}(\epsilon-\omega_{s_{1}'})v_{s_{1}}^{\dagger}{G}^{(1)}(\epsilon-\omega_{s_{1}}-\omega_{s_{1}'})W^{(2, 1)}_{s_{1}, s_{2}'}(\epsilon){G}^{(1)}(\epsilon-\omega_{s_{1}'}-\omega_{s_{2}'})v_{s_{1}'}G_{0}(\epsilon-\omega_{s_{2}'})v_{s_{2}'}\\
\nonumber
+&v_{s_{2}'}^{\dagger}G_{0}(\epsilon-\omega_{s_{2}'})v_{s_{1}}^{\dagger}{G}^{(1)}(\epsilon-\omega_{s_{1}}-\omega_{s_{2}'})W^{(2, 1)}_{s_{1}, s_{1}'}(\epsilon){G}^{(1)}(\epsilon-\omega_{s_{1}'}-\omega_{s_{2}'})v_{s_{1}'}G_{0}(\epsilon-\omega_{s_{2}'})v_{s_{2}'}\\
\nonumber
+&v_{s_{2}}^{\dagger}G_{0}(\epsilon-\omega_{s_{2}})v_{s_{2}'}^{\dagger}{G}^{(1)}(\epsilon-\omega_{s_{2}'}-\omega_{s_{2}})W^{(2, 1)}_{s_{2}, s_{1}'}(\epsilon){G}^{(1)}(\epsilon-\omega_{s_{1}'}-\omega_{s_{2}'})v_{s_{1}'}G_{0}(\epsilon-\omega_{s_{2}'})v_{s_{2}'}\\
\nonumber
+&v_{s_{2}}^{\dagger}G_{0}(\epsilon-\omega_{s_{2}})v_{s_{1}}^{\dagger}{G}^{(1)}(\epsilon-\omega_{s_{1}}-\omega_{s_{2}})W^{(2, 2)}_{s_{2}s_{1}, s_{2}'s_{1}'}(\epsilon){G}^{(1)}(\epsilon-\omega_{s_{1}'}-\omega_{s_{2}'})v_{s_{1}'}G_{0}(\epsilon-\omega_{s_{2}'})v_{s_{2}'}\\
\nonumber
+&v_{s_{2}}^{\dagger}G_{0}(\epsilon-\omega_{s_{2}})v_{s_{1}}^{\dagger}{G}^{(1)}(\epsilon-\omega_{s_{1}}-\omega_{s_{2}})W^{(2, 2)}_{s_{1}s_{2}, s_{2}'s_{1}'}(\epsilon){G}^{(1)}(\epsilon-\omega_{s_{1}'}-\omega_{s_{2}'})v_{s_{1}'}G_{0}(\epsilon-\omega_{s_{2}'})v_{s_{2}'}\\
\nonumber
+&v_{s_{2}}^{\dagger}G_{0}(\epsilon-\omega_{s_{2}})v_{s_{1}}^{\dagger}{G}^{(1)}(\epsilon-\omega_{s_{1}}-\omega_{s_{2}})W^{(2, 2)}_{s_{1}s_{2}, s_{1}'s_{2}'}(\epsilon){G}^{(1)}(\epsilon-\omega_{s_{1}'}-\omega_{s_{2}'})v_{s_{1}'}G_{0}(\epsilon-\omega_{s_{2}'})v_{s_{2}'}\\
+&v_{s_{2}}^{\dagger}G_{0}(\epsilon-\omega_{s_{2}})v_{s_{1}}^{\dagger}{G}^{(1)}(\epsilon-\omega_{s_{1}}-\omega_{s_{2}})W^{(2, 2)}_{s_{2}s_{1}, s_{1}'s_{2}'}(\epsilon){G}^{(1)}(\epsilon-\omega_{s_{1}'}-\omega_{s_{2}'})v_{s_{1}'}G_{0}(\epsilon-\omega_{s_{2}'})v_{s_{2}'}.
\end{align}
\end{widetext}
Let us now consider the expression (\ref{eq: Tphoton_Transition_operator}) for the transition operator. Concentrating on the three-photon subspace we obtain
\begin{widetext}
\begin{align}
    T^{(3, 3)}_{s_{1}'s_{2}'s_{3}', s_{1}s_{2}s_{3}}(\epsilon)=&\Braket{g|v_{s_{1}'}{G}^{(1)}(\epsilon-\omega_{s_{2}'}-\omega_{s_{3}'})[W^{(2, 2)}_{s_{2}'s_{3}', s_{2}s_{3}}(\epsilon)+V^{(3, 2)}_{s_{2}'s_{3}'}(\epsilon)G^{(3, 0)}(\epsilon)\overline{V}^{(3, 2)}_{s_{2}s_{3}}(\epsilon)]{G}^{(1)}(\epsilon-\omega_{s_{2}}-\omega_{s_{3}})v^{\dagger}_{s_{1}}|g},\\
    \nonumber
    V^{(3, 2)}_{s_{1}'s_{2}'}=&v_{s_{1}'}G_{0}(\epsilon)v_{s_{2}'}+W^{(2, 1)}_{s_{1}', s}(\epsilon)G^{(1)}(\epsilon-\omega_{s})v_{s}G_{0}(\epsilon)v_{s_{2}'}+W^{(2, 1)}_{s_{1}', s}(\epsilon)G^{(1)}(\epsilon-\omega_{s})v_{s_{2}'}G_{0}(\epsilon-\omega_{s})v_{s}\\
        \nonumber
    +&W^{(2, 2)}_{s_{1}'s_{2}',
    \bar{s}_{1}'\bar{s}_{2}'}(\epsilon)G^{(1)}(\epsilon-\omega_{\bar{s}_{1}'}-\omega_{\bar{s}_{2}'})v_{\bar{s}_{1}'}G_{0}(\epsilon-\omega_{\bar{s}_{2}'})v_{\bar{s}_{2}'}\\
    +&W^{(2, 2)}_{s_{1}'s_{2}', \bar{s}_{2}'\bar{s}_{1}'}(\epsilon)G^{(1)}(\epsilon-\omega_{\bar{s}_{1}'}-\omega_{\bar{s}_{2}'})v_{\bar{s}_{1}'}G_{0}(\epsilon-\omega_{\bar{s}_{2}'})v_{\bar{s}_{2}'},\\
            \nonumber
    \overline{V}^{(3, 2)}_{s_{1}s_{2}}=&v_{s_{1}}G_{0}(\epsilon)v_{s_{2}}+v_{s_{1}}^{\dagger}G_{0}(\epsilon)v_{s'}^{\dagger}G^{(1)}(\epsilon-\omega_{s'})W^{(2, 1)}_{s', s_{2}}(\epsilon)+v_{s'}^{\dagger}G_{0}(\epsilon-\omega_{s'})v_{s_{1}}^{\dagger}G^{(1)}(\epsilon-\omega_{s'})W^{(2, 1)}_{s', s_{2}}(\epsilon)\\
            \nonumber
    +&v^{\dagger}_{\bar{s}_{2}}G_{0}(\epsilon-\omega_{\bar{s}_{2}})v^{\dagger}_{\bar{s}_{1}}G^{(1)}(\epsilon-\omega_{\bar{s}_{1}}-\omega_{\bar{s}_{2}})W^{(2, 2)}_{\bar{s}_{1}\bar{s}_{2}, s_{1}s_{2}}(\epsilon)\\
     +&v^{\dagger}_{\bar{s}_{2}}G_{0}(\epsilon-\omega_{\bar{s}_{2}})v^{\dagger}_{\bar{s}_{1}}G^{(1)}(\epsilon-\omega_{\bar{s}_{1}}-\omega_{\bar{s}_{2}})W^{(2, 2)}_{\bar{s}_{2}\bar{s}_{1}, s_{1}s_{2}}(\epsilon).
\end{align}
\end{widetext}
Now, our goal is to rewrite the renormalized two-particle emission $V_{s_{1}'s_{2}'}$ and absorption $\overline{V}_{s_{1}s_{2}}$ vertices, as well as the self-energy in three-excitation subspace $\Sigma^{(3, 3)}$ in terms of full Green's and vertex functions, as it is presented in the main text. 
\par
First, we note the following identity 
\begin{widetext}
\begin{align}
\nonumber
    &v_{s_{1}'}G_{0}(\epsilon)v_{s_{2}'}+W^{(2, 1)}_{s_{1}', s}(\epsilon)G^{(1)}(\epsilon-\omega_{s})v_{s}G_{0}(\epsilon)v_{s_{2}'}=v_{s'_1} G_0(\epsilon)v_{s'_2}   + W_{s'_1, s}^{(2, 1, i)}(\epsilon) G^{(1)}(\epsilon-\omega_{s})  v_{s} G_0(\epsilon)    v_{s'_2}\\
    \nonumber
    &+V_{s'_1}^{(2, 1)}(\epsilon) G^{(2, 0)}(\epsilon) \overline{V}_s^{(2, 1)}(\epsilon) G^{(1)}(\epsilon-\omega_{s})  v_{s} G_0(\epsilon)    v_{s'_2} =  v_{s'_1} G_0(\epsilon)    v_{s'_2}   + W_{s'_1, s}^{(2, 1, i)} G^{(1)}(\epsilon-\omega_{s})  v_{s} G_0(\epsilon)    v_{s'_2}\\
    &+V_{s'_1}^{(2, 1)}(\epsilon) G^{(2, 0)}(\epsilon) \Sigma^{(2, 0)} G_0(\epsilon) 
        \nonumber
    v_{s'_2}=v_{s'_1} G_0(\epsilon)    v_{s'_2}   + W_{s'_1, s}^{(2, 1, i)}(\epsilon) G^{(1)}(\epsilon-\omega_{s})  v_{s} G_0(\epsilon)    v_{s'_2} \\
        \label{eq:ident_1}
    &+V_{s'_1}^{(2, 1)}(\epsilon) [ G^{(2, 0)}(\epsilon) -G_0(\epsilon)   ] v_{s'_2}= V_{s'_1}^{(2, 1)}(\epsilon)  G^{(2, 0)}(\epsilon)  v_{s'_2}.
\end{align}
\end{widetext}
Analogously
\begin{align}
\nonumber
    &v_{s_1}^{\dagger} G_0(\epsilon)  v_{s_2}^{\dagger} + v_{s_1}^{\dagger} G_0(\epsilon)  v_{s'}^{\dagger}  G^{(1)}(\epsilon-\omega_{s'})  W_{s',s_2}^{(2,1)}(\epsilon)\\
    \label{eq:ident_2}
    &=v_{s_1}^{\dagger} G^{(2,0)}(\epsilon)  \overline{V}_{s_2}^{(2, 1)}(\epsilon).
\end{align}
Analysing the structure of equations satisfied by $W^{(2, 1)}$ and $W^{(2, 2)}$ one easily concludes that
\begin{align}
\nonumber
\label{eq: 2body_right}
    W^{(2, 2)}_{s_{1}'s_{2}', s_{1}s_{2}}(\epsilon)&=W^{(2, 1)}_{s_{1}', s_{1}}(\epsilon)G^{(1)}(\epsilon)W^{(2, 1)}_{s_{2}', s_{2}}(\epsilon)\\
    &+W^{(2, 1)}_{s_{1}', s}(\epsilon)G^{(1)}(\epsilon-\omega_{s})W^{(2, 2)}_{s_{2}'s, s_{1}s_{2}}(\epsilon),
    \end{align}
    \begin{align}
    \nonumber
\label{eq: 2body_left}
    W^{(2, 2)}_{s_{1}'s_{2}', s_{1}s_{2}}(\epsilon)&= W^{(2, 1)}_{s_{1}', s_{1}}(\epsilon)G^{(1)}(\epsilon)W^{(2, 1)}_{s_{2}', s_{2}}(\epsilon)\\
    &+W^{(2, 2)}_{s_{1}'s_{2}', ss_{1}}(\epsilon)G^{(1)}(\epsilon-\omega_{s})W^{(2, 1)}_{s,s_{2}}(\epsilon).
\end{align}
Multiplying (\ref{eq: 2body_right}) and (\ref{eq: 2body_left}) by $G^{(1)}(\epsilon)v_{s_{2}}G_{0}(\epsilon)v_{s_{1}}$ from the right and by $v_{s_{2}'}^{\dagger}G_{0}(\epsilon)v_{s_{1}'}^{\dagger}G^{(1)}(\epsilon)$ from the left respectively and contracting the relevant indices, we obtain
\begin{widetext}
\begin{align}
\nonumber
    &  W_{s'_1 s'_2, s_1 s_2}^{(2,2)}(\epsilon)   G^{(1)}(\epsilon-\omega_{s_{1}}-\omega_{s_{2}}) v_{s_2} G_0(\epsilon-\omega_{s_{2}}) v_{s_1}= W_{s'_1 , s_1}^{(2,1)}(\epsilon)  G^{(1)}(\epsilon-\omega_{s_{1}})  [V_{ s'_2}^{(2, 1)} (\epsilon)G^{(2, 0)}(\epsilon-\omega_{s_{1}}) v_{s_1} \\
    &- v_{ s'_2} G_0(\epsilon-\omega_{s_{1}}) v_{s_1}]+ W_{s'_1 , s}^{(2,1)}(\epsilon)  G^{(1)}(\epsilon-\omega_{s})    W_{s'_2 s, s_1 s_2}^{(2,2)}(\epsilon) G^{(1)}(\epsilon-\omega_{s_{1}}-\omega_{s_{2}}) v_{s_2} G_0(\epsilon-\omega_{s_{1}}) v_{s_1},\\
    \nonumber
    & v_{s'_2}^{\dagger}  G_0(\epsilon-\omega_{s_{2}'}) v_{s'_1}^{\dagger}  G^{(1)}(\epsilon-\omega_{s_{1}'}-\omega_{s_{2}'}) W_{s'_1 s'_2, s_1 s_2}^{(2, 2)}(\epsilon)=  [ v_{s'_2}^{\dagger}  G^{(2, 0)}(\epsilon-\omega_{s_{2}'}) \overline{V}_{s_1}^{(2, 1)}(\epsilon) -  v_{s'_2}^{\dagger}  G_0(\epsilon-\omega_{s_{2}'}) v_{s_1}^{\dagger}  ]  \\
 &  \times G^{(1)}(\epsilon-\omega_{s_{2}'})   W_{s'_2, s_2}^{(2,1)}  + v_{s'_2}^{\dagger}  G_0(\epsilon-\omega_{s_{2}'}) v_{s'_1}^{\dagger}  G^{(1)}(\epsilon-\omega_{s_{1}'}-\omega_{s_{2}'}) W_{s'_1 s'_2, s s_1 }^{(2,2)}    G^{(1)}(\epsilon-\omega_{s}) W_{s , s_2}^{(2,1)}(\epsilon) .
\end{align}
\end{widetext}
Now, defining the following objects
\begin{align}
\nonumber
    &K^{(d)}_{s'_1 s'_2}(\epsilon)\\
    \nonumber
    &=  W_{s'_1 s'_2, s_1 s_2}^{(2,2)}(\epsilon)   G^{(1)}(\epsilon-\omega_{s_{2}}-\omega_{s_{1}}) v_{s_2} G_0(\epsilon-\omega_{s_{1}}) v_{s_1} \\
    &+ W_{s'_1 , s_1}^{(2,1)} (\epsilon) G^{(1)}(\epsilon-\omega_{s_{1}}) v_{ s'_2} G_0(\epsilon-\omega_{s_{1}}) v_{s_1},
\end{align}
\begin{align}
        \nonumber
    &\overline{K}^{(d)}_{s_1 s_2}(\epsilon)\\
        \nonumber
    &= v_{s'_2}^{\dagger}  G_0(\epsilon-\omega_{s_{2}'}) v_{s'_1}^{\dagger}  G^{(1)}(\epsilon-\omega_{s_{1}'}-\omega_{s_{2}'}) W_{s'_1 s'_2, s_1 s_2}^{(2,2)}(\epsilon) \\
    &+  v_{s'_2}^{\dagger}  G_0(\epsilon-\omega_{s_{2}'}) v_{s_1}^{\dagger}   G^{(1)}(\epsilon-\omega_{s_{2}'})   W_{s'_2, s_2}^{(2,1)}(\epsilon),
\end{align}
we establish the following equations 
\begin{widetext}
\begin{align}
\nonumber
    K^{(d)}_{s'_1 s'_2}(\epsilon) &= W_{s'_1 , s_1}^{(2,1)}(\epsilon)G^{(1)}(\epsilon-\omega_{s_{1}})V_{ s'_2}^{(2, 1)}(\epsilon-\omega_{s_{1}}) G^{(2, 0)}(\epsilon-\omega_{s_{1}}) v_{s_1}\\
    &+ W_{s'_1 , s}^{(2,1)}(\epsilon)  G^{(1)}(\epsilon-\omega_{s})    [K^{(d)}_{s'_2 s}(\epsilon)  - W_{s'_2 , s_1}^{(2,1)}(\epsilon-\omega_{s})  G^{(1)}(\epsilon-\omega_{s_{1}}-\omega_{s}) v_{ s} G_0(\epsilon-\omega_{s_{1}}) v_{s_1}],
    \end{align}
    \begin{align}
    \nonumber
    \overline{K}^{(d)}_{s_1 s_2}(\epsilon) &=  v_{s'_2}^{\dagger}  G^{(2, 0)}(\epsilon-\omega_{s_{2}'}) \overline{V}_{s_1}^{(2, 1)}(\epsilon-\omega_{s_{2}'})  G^{(1)}(\epsilon-\omega_{s_{2}'})   W_{s'_2, s_2}^{(1,0)} (\epsilon)\\
    &+   [\overline{K}^{(d)}_{s s_1}(\epsilon) - v_{s'_2}^{\dagger}  G_0(\epsilon-\omega_{s_{2}'}) v_{s}^{\dagger}   G^{(1)}(\epsilon-\omega_{s_{2}'}-\omega_{s})   W_{s'_2, s_1}^{(2,1)}(\epsilon-\omega_{s}) ] G^{(1)}(\epsilon-\omega_{s}) W_{s , s_2}^{(2,1)}(\epsilon) .
\end{align}
\end{widetext}
Analogously multiplying (\ref{eq: 2body_right}) and (\ref{eq: 2body_left}) by $G^{(1)}(\epsilon)v_{s_{1}}G_{0}(\epsilon)v_{s_{2}}$ from the right and by $v_{s_{1}'}^{\dagger}G_{0}(\epsilon)v_{s_{2}'}^{\dagger}G^{(1)}(\epsilon)$ from the left respectively, contracting the $s_{1, 2}$ indices, and defining
\begin{align}
\nonumber
    &K^{(e)}_{s_{1}'s_{2}'}(\epsilon)\\
    &=W^{(2, 2)}_{s_{1}'s_{2}', s_{1}s_{2}}(\epsilon)G^{(1)}(\epsilon-\omega_{s_{1}}-\omega_{s_{2}})v_{s_{1}}G_{0}(\epsilon-\omega_{s_{2}})v_{s_{2}},
\end{align}
\begin{align}
\nonumber
        &\overline{K}^{(e)}_{s_{1}s_{2}}(\epsilon)\\
        &=v_{s_{1}'}^{\dagger}G_{0}(\epsilon-\omega_{s_{1}'})v_{s_{2}'}^{\dagger}G^{(1)}(\epsilon-\omega_{s_{1}'}-\omega_{s_{2}'})W^{(2, 2)}_{s_{1}'s_{2}', s_{1}s_{2}}(\epsilon),
\end{align}
we deduce
\begin{widetext}
\begin{align}
\nonumber
    K^{(e)}_{s'_1 s'_2}(\epsilon)&= W_{s'_1 , s_1}^{(2,1)} (\epsilon)G^{(1)}(\epsilon-\omega_{s_{1}}) W_{s'_2, s_2}^{(2,1)}(\epsilon-\omega_{s_{1}}) G^{(1)}(\epsilon-\omega_{s_{1}}-\omega_{s_{2}}) v_{s_1} G_0(\epsilon-\omega_{s_{2}})  v_{s_2}\\
    &+ W_{s'_1 , s}^{(2,1)}(\epsilon)  G^{(1)}(\epsilon-\omega_{s})   K^{(e)}_{s'_2 s}(\epsilon)  ,\\
    \nonumber
    \overline{K}^{(e)}_{s_1 s_2}(\epsilon) &= v_{s'_1}^{\dagger} G_0(\epsilon-\omega_{s_{1}'}) v_{s'_2}^{\dagger}  G^{(1)}(\epsilon-\omega_{s_{1}'}-\omega_{s_{2}'}) W_{s'_1 , s_1}^{(2,1)}(\epsilon-\omega_{s_{2}'})  G^{(1)}(\epsilon-\omega_{s_{2}'})   W_{s'_2, s_2}^{(2,1)}(\epsilon)\\
    &+ \overline{K}^{(e)}_{s s_1}(\epsilon)   G^{(1)}(\epsilon-\omega_{s}) W_{s , s_2}^{(2,1)}(\epsilon).
\end{align}
\end{widetext}
Further we define
\begin{align}
    &K_{s_{1}'s_{2}'}(\epsilon)=K_{s_{1}'s_{2}'}^{(d)}(\epsilon)+K_{s_{1}'s_{2}'}^{(e)}(\epsilon),\\
    &\overline{K}_{s_{1}s_{2}}(\epsilon)=\overline{K}_{s_{1}s_{2}}^{(d)}(\epsilon)+\overline{K}_{s_{1}s_{2}}^{(e)}(\epsilon).
\end{align}
With these definitions it is easy to show that the two-particle emission/absorption vertices are given by
\begin{align}
\label{eq: C25}
    &V^{(3, 2)}_{s_{1}'s_{2}'}(\epsilon)=V^{(2, 1)}_{s_{1}'}(\epsilon)G^{(2, 0)}(\epsilon)V^{(2, 1)}_{s_{2}'}(\epsilon)+\tilde{K}_{s_{1}'s_{2}'}(\epsilon),\\
    \label{eq: C26}
    &\overline{V}^{(3, 2)}_{s_{1}s_{2}}(\epsilon)=\overline{V}^{(2, 1)}_{s_{1}}(\epsilon)G^{(2, 0)}(\epsilon)\overline{V}^{(2, 1)}_{s_{2}}(\epsilon)+\tilde{\overline{K}}_{s_{1}s_{2}}(\epsilon),
\end{align}
where
\begin{align}
\nonumber
    &\tilde{K}_{s_{1}'s_{2}'}(\epsilon)={K}_{s_{1}'s_{2}'}(\epsilon)\\
    &-V_{s_{1}'}^{(2, 1)}(\epsilon)G^{(2, 0)}(\epsilon)W^{(2, 1, i)}_{s_{2}', s_{1}}(\epsilon)G^{(1)}(\epsilon-\omega_{s_{1}})v_{s_{1}},\\
\nonumber    
    &\tilde{\overline{K}}_{s_{1}s_{2}}(\epsilon)={\overline{K}}_{s_{1}s_{2}}(\epsilon)\\
    &-v_{s_{2}'}^{\dagger}G^{(1)}(\epsilon-\omega_{s_{2}'})W^{(2, 1, i)}_{s_{2}', s_{1}}(\epsilon)G^{(2, 0)}(\epsilon)\overline{V}_{s_{2}}^{(2, 1)}(\epsilon),
\end{align}
obey the following integral equations
\begin{widetext}
\begin{align}
\nonumber
    \tilde{K}_{s_{1}'s_{2}'}(\epsilon)=&W^{(2, 1)}_{s_{1}', s}(\epsilon)G^{(1)}(\epsilon-\omega_{s})V^{(2, 1)}_{s_{2}'}(\epsilon-\omega_{s})G^{(2, 0)}(\epsilon-\omega_{s})V^{(2, 1)}_{s}(\epsilon)-V_{s_{1}'}^{(2, 1)}(\epsilon)G^{(2, 0)}(\epsilon)W^{(2, 1, i)}_{s_{2}', s}(\epsilon)G^{(1)}(\epsilon-\omega_{s})v_{s}\\
    \label{eq: C29}
    +&W^{(2, 1)}_{s_{1}, s}(\epsilon)G^{(1)}(\epsilon-\omega_{s})\tilde{K}_{s_{2}'s}(\epsilon),\\
    \nonumber
    \tilde{\overline{K}}_{s_{1}s_{2}}(\epsilon)=&\overline{V}^{(2, 1)}_{s}(\epsilon)G^{(2, 0)}(\epsilon-\omega_{s})V^{(2, 1)}_{s_{1}}(\epsilon-\omega_{s})G^{(1)}(\epsilon-\omega_{s})W^{(2, 1)}_{s, s_{2}}(\epsilon)-v_{s}^{\dagger}G^{(1)}(\epsilon-\omega_{s})W^{(2, 1, i)}_{s, s_{1}}(\epsilon)G^{(2, 0)}(\epsilon)\overline{V}^{(2, 1)}_{s_{2}}(\epsilon)\\
    \label{eq: C30}
    +&\tilde{\overline{K}}_{ss_{1}}(\epsilon)G^{(1)}(\epsilon-\omega_{s})W^{(2, 1)}_{s, s_{2}}(\epsilon).
\end{align}
\end{widetext}
Using equations (\ref{eq: C25}), (\ref{eq: C26}) together with (\ref{eq: C29}), (\ref{eq: C30}), we finally arrive at the following equations
\begin{align}
\nonumber
    V^{(3, 2)}_{s_{1}'s_{2}'}(\epsilon)&=V_{s_{1}'}^{(2, 1)}(\epsilon)G^{(2, 0)}(\epsilon)v_{s_{2}'}\\
    &+W^{(2, 1)}_{s_{1}', s}(\epsilon)G^{(1)}(\epsilon-\omega_{s})V^{(3, 2)}_{s_{2}'s}(\epsilon),
    \end{align}
\begin{align}
\nonumber
    \overline{V}^{(3, 2)}_{s_{1}s_{2}}(\epsilon)&=v_{s_{1}}G^{(2, 1)}(\epsilon)\overline{V}_{s_{2}}(\epsilon)\\
    &+\overline{V}^{(3, 2)}_{ss_{1}}(\epsilon)G^{(1)}(\epsilon-\omega_{s})W^{(2, 1)}_{s, s_{2}}(\epsilon),
\end{align}
which are precisely the equations (\ref{eq: twophotonabsorb}) and (\ref{eq: twophotonemis}) stated in the main text.
\par
Now, we turn our attention to the self-energy bubble. Let us show that equations (\ref{eq:3photon_self1}) and (\ref{eq:3photon_self2}) hold via direct substitution.  One has
\begin{widetext}
\begin{align}
\nonumber
\Sigma^{(3, 0)} (\epsilon) &=  ( v_{s'_1}^{\dagger} G_0(\epsilon-\omega_{s_{1}'})  v_{s'_2}^{\dagger} +  v_{s'_2}^{\dagger} G_0(\epsilon-\omega_{s_{2}'})  v_{s'_1}^{\dagger} ) G^{(1)}(\epsilon-\omega_{s_{1}'}-\omega_{s_{2}'}) V_{s'_1 s'_2}^{(3, 2)}(\epsilon) \\
\nonumber
&=    v_{s'_2}^{\dagger} [ G^{(2, 0)}(\epsilon-\omega_{s_{2}'}) - G_0(\epsilon-\omega_{s_{2}'})] v_{s'_2} - v_{s'_1}^{\dagger} G_0(\epsilon-\omega_{s_{1}'}) v_s^{\dagger} G^{(1)}   (\epsilon-\omega_{s_{1}'}-\omega_{s})v_{s'_1} G_0(\epsilon-\omega_{s})   v_{s} \\
\nonumber
&+ v_{s'_1}^{\dagger} G_0(\epsilon-\omega_{s_{1}'})  v_{s'_2}^{\dagger}  G^{(1)}(\epsilon-\omega_{s_{1}'}-\omega_{s_{2}'}) V_{s'_1} ^{(2, 1)}(\epsilon-\omega_{s_{2}'})G^{(2, 0)}(\epsilon-\omega_{s_{2}'}) v_{s'_2}\\
\nonumber
&+  v_{s'_1}^{\dagger} G^{(2, 0)}(\epsilon-\omega_{s_{1}'}) \overline{V}_s^{(2, 1)}(\epsilon-\omega_{s_{1}'})G^{(1)}(\epsilon-\omega_{s}-\omega_{s_{1}'})v_{s'_1} G_0(\epsilon-\omega_{s})   v_{s}\\
\nonumber
&+ v_{s'_2}^{\dagger} G_0(\epsilon-\omega_{s_{2}'})  v_{s'_1}^{\dagger}  G^{(1)}(\epsilon-\omega_{s_{1}'}-\omega_{s_{2}'}) W_{s'_2 , s}^{(2,1)}(\epsilon-\omega_{s_{1}'}) G^{(1)}(\epsilon-\omega_{s_{1}'}-\omega_{s})   v_{s'_1} G_0(\epsilon-\omega_{s})   v_{s} \\
\nonumber
&+ v_{s'_1}^{\dagger} G_0(\epsilon-\omega_{s_{1}'})  v_{s'_2}^{\dagger}  G^{(1)}(\epsilon-\omega_{s_{1}'}-\omega_{s_{2}'}) W_{s'_1 s'_2 , s' s}^{(2,2)}(\epsilon) G^{(1)}  (\epsilon-\omega_{s'}-\omega_{s})v_{s'} G_0(\epsilon-\omega_{s})    v_{s} \\
\nonumber
&+v_{s'_1}^{\dagger} G_0(\epsilon-\omega_{s_{1}'})  v_{s'_2}^{\dagger}  G^{(1)}(\epsilon-\omega_{s_{1}'}-\omega_{s_{2}'}) W_{s'_1 s'_2 , s s'}^{(2,2)} (\epsilon)G^{(1)}(\epsilon-\omega_{s'}-\omega_{s})  v_{s'} G_0(\epsilon-\omega_{s})    v_{s} \\
\nonumber
&+v_{s'_2}^{\dagger} G_0(\epsilon-\omega_{s_{2}'})  v_{s'_1}^{\dagger}  G^{(1)}(\epsilon-\omega_{s_{1}'}-\omega_{s_{2}'})W_{s'_1 s'_2 , s' s}^{(2,2)}(\epsilon)G^{(1)}(\epsilon-\omega_{s'}-\omega_{s})  v_{s'} G_0(\epsilon-\omega_{s})    v_{s}\\
&+v_{s'_2}^{\dagger} G_0(\epsilon-\omega_{s_{2}'})  v_{s'_1}^{\dagger}  G^{(1)}(\epsilon-\omega_{s_{1}'}-\omega_{s_{2}'})W_{s'_1 s'_2 , s s'}^{(2,2)}(\epsilon)G^{(1)}(\epsilon-\omega_{s'}-\omega_{s})  v_{s'} G_0(\epsilon-\omega_{s})    v_{s},\\
\nonumber
\Sigma^{(3, 0)} (\epsilon) &=\overline{V}_{s_1 s_2}^{(3, 2)} G^{(1)}(\epsilon-\omega_{s_{1}}-\omega_{s_{2}})  (v_{s_1} G_0(\epsilon-\omega_{s_{2}})    v_{s_2}   + v_{s_2} G_0    (\epsilon-\omega_{s_{1}})v_{s_1}  ) \\
\nonumber
&=   v_{s_2}^{\dagger} [G^{(2, 0)}(\epsilon-\omega_{s_{2}})  - G_0(\epsilon-\omega_{s_{2}}) ]   v_{s_2}   - v_{s'}^{\dagger} G_0(\epsilon-\omega_{s'})  v_{s_1}^{\dagger}  G^{(1)}(\epsilon-\omega_{s_{1}}-\omega_{s'})  v_{s'} G_0 (\epsilon-\omega_{s_{1}})v_{s_1}    \\
\nonumber
&+  v_{s'}^{\dagger} G_0(\epsilon-\omega_{s'})  v_{s_1}^{\dagger}  G^{(1)}(\epsilon-\omega_{s_{1}}-\omega_{s'})V_{s'}^{(2, 1)}(\epsilon-\omega_{s_{1}}) G^{(2, 0)}(\epsilon-\omega_{s_{1}}) v_{s_1}    \\
\nonumber
&+  v_{s_2}^{\dagger} G^{(2, 0)}(\epsilon-\omega_{s_{2}})  \overline{V}_{s_1}^{(2, 1)}(\epsilon-\omega_{s_{2}})  G^{(1)}(\epsilon-\omega_{s_{1}}-\omega_{s_{2}})  v_{s_2} G_0(\epsilon-\omega_{s_{1}})    v_{s_1} \\
\nonumber
&+  v_{s'}^{\dagger} G_0(\epsilon-\omega_{s'})  v_{s_1}^{\dagger}  G^{(1)}  (\epsilon-\omega_{s'}-\omega_{s_{1}})W_{s',s_2}^{(2,1)}(\epsilon-\omega_{s_{1}})   G^{(1)}(\epsilon-\omega_{s_{1}}-\omega_{s_{2}})  v_{s_1} G_0(\epsilon-\omega_{s_{2}})    v_{s_2}  \\
\nonumber
&+ v_{s'}^{\dagger} G_0(\epsilon-\omega_{s'})  v_{s}^{\dagger}G^{(1)}(\epsilon-\omega_{s}-\omega_{s'})  W_{s s' , s_1 s_2}^{(2,2)}(\epsilon)  G^{(1)}(\epsilon-\omega_{s_{1}}-\omega_{s_{2}})  v_{s_1} G_0(\epsilon-\omega_{s_{2}})    v_{s_2} \\
\nonumber
&+ v_{s'}^{\dagger} G_0(\epsilon-\omega_{s'})  v_{s}^{\dagger}G^{(1)}(\epsilon-\omega_{s}-\omega_{s'}) W_{s' s , s_1 s_2}^{(2,2)}(\epsilon)  G^{(1)}(\epsilon-\omega_{s_{1}}-\omega_{s_{2}})  v_{s_1} G_0(\epsilon-\omega_{s_{2}})    v_{s_2}  \\
\nonumber
&+ v_{s'}^{\dagger} G_0(\epsilon-\omega_{s'})  v_{s}^{\dagger}G^{(1)}(\epsilon-\omega_{s}-\omega_{s'}) W_{s s' , s_1 s_2}^{(2,2)} (\epsilon)  G^{(1)}(\epsilon-\omega_{s_{1}}-\omega_{s_{2}})  v_{s_2} G_0(\epsilon-\omega_{s_{1}})    v_{s_1} \\
&+ v_{s'}^{\dagger} G_0(\epsilon-\omega_{s'})  v_{s}^{\dagger}G^{(1)}(\epsilon-\omega_{s}-\omega_{s'}) W_{s' s , s_1 s_2}^{(2,2)} (\epsilon)  G^{(1)}(\epsilon-\omega_{s_{1}}-\omega_{s_{2}})  v_{s_2} G_0(\epsilon-\omega_{s_{1}})v_{s_1}.
\end{align}
\end{widetext}
This, together with identities (\ref{eq:ident_1}) and (\ref{eq:ident_2}) justifies the proposed representation of the self-energy.

\section{Effect of the non-zero detuning}
\label{ap:detun}
\begin{figure}[t]
\includegraphics[width=0.9\columnwidth]{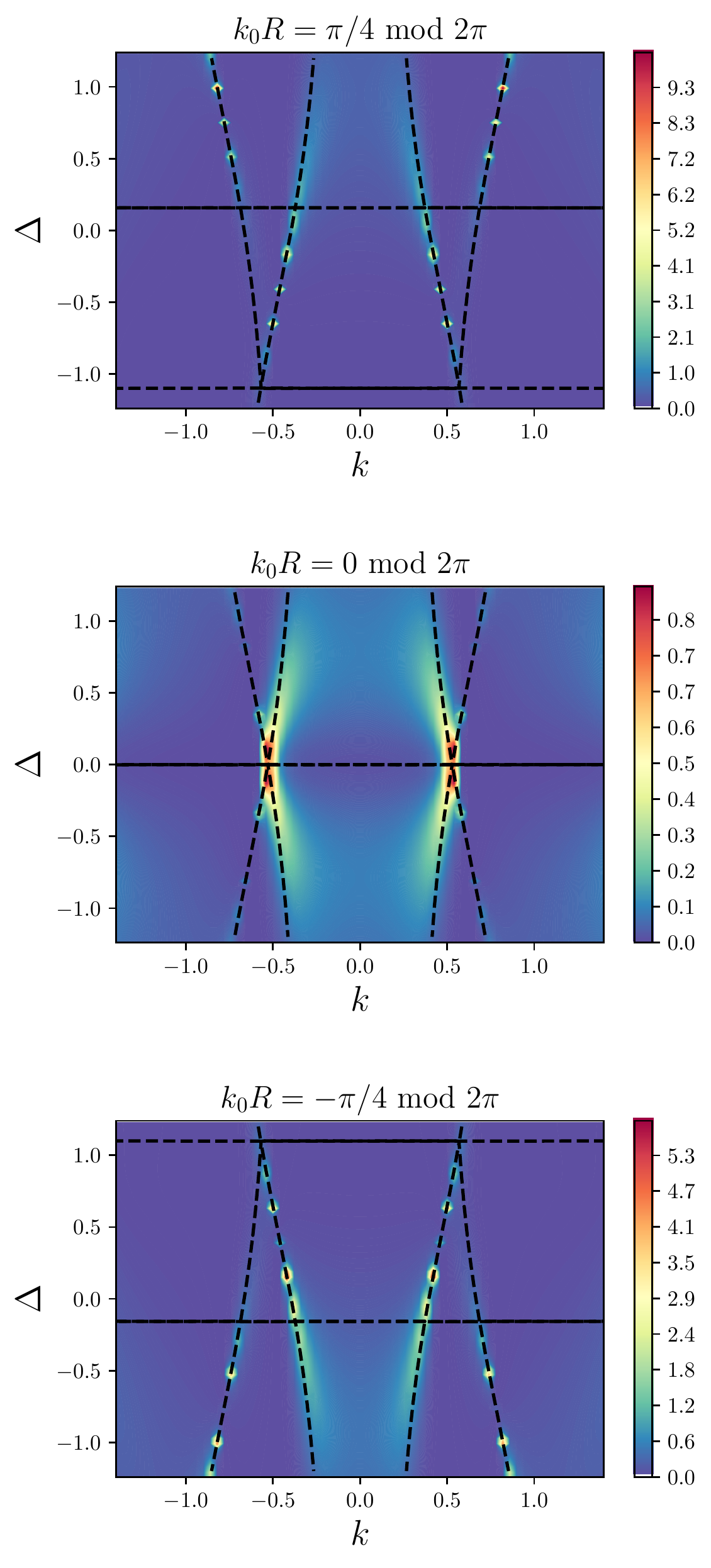} 
\caption{Spectral power density (scaled by $1/\Phi^{2}$) for of the giant atom model as a function of $\Delta, \ k$ for various values of $k_{0}R$ and $\gamma R=5$. Here the dashed black lines indicate the pole position of the dressed Green's function in the single excitation subspace.}
        \label{fig: S_nz}
\end{figure}
\begin{figure}[t]
\includegraphics[width=0.9\columnwidth]{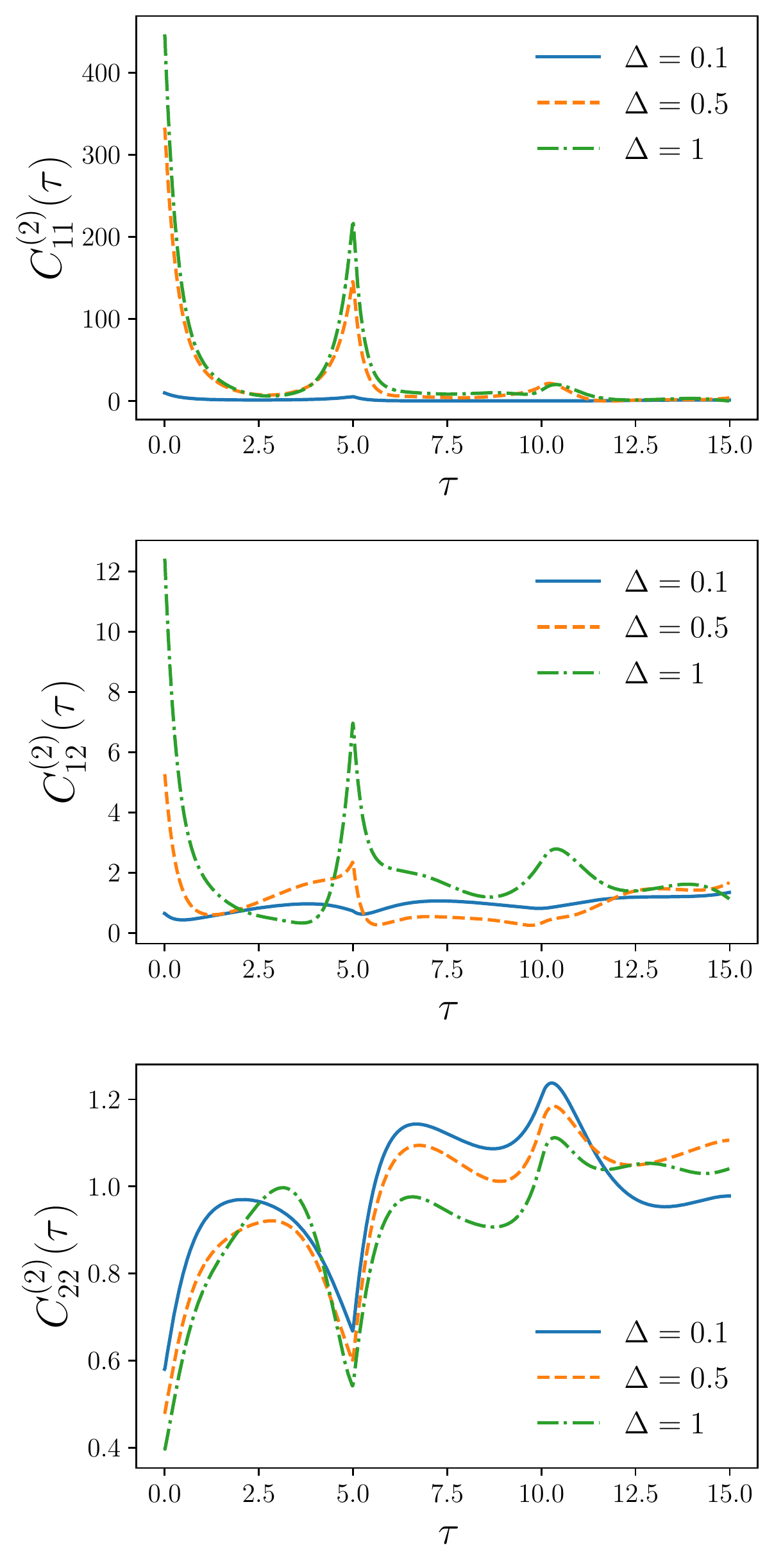} 
\caption{Second order coherence function for the system with $k_{0}R=3\pi/11, \ \gamma R=5, \ \Delta=0.1, \ 0.5, \ 1.0$ (as before, the the second order coherence function is dimensionless).}
        \label{fig: g2d}
\end{figure}
In this appendix we analyse the effect on non-zero detuning of the atom from radiation on the observable quantities. In particular, we focus on the spectral power density and the second order coherence function. As it was discussed in Section \ref{SPd}, the sharp bound-state-like peaks in the line-shape of spectral density may be understood with a simple physical picture of an effective cavity. When one increases (or decreases) the detuning from zero value, one effectively changes the modes supported by the cavity and thus one expects the position of the peaks to be shifted. The precise location of the resonances in the spectral density as  function of $\Delta$ and $k$ for various values of $k_{0}R$ is shown in Figure $\ref{fig: S_nz}$. As we can see, for non-zero dephasing $k_{0}R\neq0$ the spectrum is not a symmetric function of $\Delta$. For negative dephasing we find that emission into the zero modes is enhanced for a negatively detuned atom $\Delta<0$, whereas the picture is opposite for positive $k_{0}R$. In general, we can clearly resolve a pair of sharp peaks which eventually merge together at certain values of parameters (e.g. $k_{0}R=\Delta=0$). It is interesting to note that the location of this peaks is almost entirely determined by the poles of the dressed propagator in the single-excitation subspace $(G^{(1)}(k))^{-1}=k+\Delta+i\gamma(1+e^{i(k+k_{0})R})=0$. This equation is solved by 
\begin{align}
    k=\pm i\frac{R\gamma-iR\Delta-W_{n}(-\gamma Re^{i(k_{0}-\Delta-i\gamma)R})}{R},
    \label{pole}
\end{align}
where $W_{n}(z)$ is the $n^{\text{th}}$ branch of the Lambert $W$-function, also known as the product logarithm. The real part of (\ref{pole}) with  $n=0, \pm1$ is plotted as black dashed lines in Figure \ref{fig: S_nz}. The vertical lines in Figure \ref{fig: S_nz} represent the discontinuous jumps of the Lambert function across the branch cut. As one may notice, the formula (\ref{pole}) is indeed in perfect agreement with numerical results.
\par
Let us now consider the second order coherence function at non-zero detuning. Second order coherence for the system with $k_{0}R=3\pi/11, \ \gamma R=5, \ \Delta=0.1, \ 0.5, \ 1.0$ is shown in Figure \ref{fig: g2d}. We note that the presence of detuning does not affect the general trend of strong photon bunching in the first channel and their corresponding anti-bunching in the second one, as was discussed in Section \ref{SPd}. Neither the detuning affects the presence of non-differentiable peaks occurring at integer multiplies of delay time $R$.

\end{appendix}

\end{document}